\documentclass[12pt,amsmath,amssymb,nofootinbib,superscriptaddress]{revtex4-1}
\usepackage{amsmath} %
\usepackage{pstricks}
\usepackage{color} 
\usepackage{amssymb}
\usepackage{slashed}
\usepackage{graphicx,amsfonts,layout,appendix,subfigure}
\input{epsf} 
\def\beq{\begin{equation}} \def\eeq{\end{equation}}
\def\beqn{\begin{eqnarray}} \def\eeqn{\end{eqnarray}}
 \def\to{\rightarrow}

\begin{document} 

\newcommand\sss{\scriptscriptstyle}
\newcommand\IDC{\textbf{\textit{{\rm Id}}}_C}
\newcommand\SUNT{\textit{\textbf{T}}}
\newcommand\pslashed{\slashed{p}}
\newcommand\DST{D_{\rm ST}}
\newcommand\DDirac{D_{\rm Dirac}}
\newcommand\factor{\rm Factor}
\newcommand\IC{\textbf{\textit{I}}}
\newcommand\Spmatrix{\textit{\textbf{Sp}}}
\newcommand\Spelement{{\rm Sp}}
\newcommand\Spfunction{{\rm Split}}
\newcommand\NC{N_C}
\newcommand\CG{c_{\Gamma}}
\newcommand\DR{D_{\rm R}}
\newcommand\mug{\mu_\gamma} \newcommand\mue{\mu_e}
\newcommand\mui{\mu_{\sss I}} \newcommand\muf{\mu_{\sss F}}
\newcommand\mur{\mu_{\sss R}} \newcommand\muo{\mu_0} \newcommand\me{m_e}
\newcommand\as{\alpha_{\sss S}} \newcommand\ep{\epsilon}
\newcommand\Th{\theta} \newcommand\epb{\overline{\epsilon}}
\newcommand\aem{\alpha_{\rm em}} \newcommand\refq[1]{$^{[#1]}$}
\newcommand\avr[1]{\left\langle #1 \right\rangle}
\newcommand\lambdamsb{\Lambda_5^{\rm \sss \overline{MS}}}
\newcommand\qqb{{q\overline{q}}} \newcommand\qb{\overline{q}} %
\newcommand\MSB{{\rm \overline{MS}}} \newcommand\MS{{\rm MS}}
\newcommand\DIG{{\rm DIS}_\gamma} \newcommand\CA{C_{\sss A}}
\newcommand\DA{D_{\sss A}} \newcommand\CF{C_{\sss F}}
\newcommand\TF{T_{\sss F}} % \newcommand\Jetlist{\{J_l\}_{1,2}}
\newcommand\qeps{q^2_{\epsilon}} % \newcommand\Jetlist{\{J_l\}_{1,2}}

%********************************************************************************************************
\begin{titlepage}
\renewcommand{\thefootnote}{\fnsymbol{footnote}}
\begin{flushright}
     LPN13-069\\ IFIC/13-73
     \end{flushright}
\par \vspace{10mm}

\begin{center}
{\Large \bf
Double collinear splitting amplitudes\\[1ex] 
at next-to-leading order
}
\end{center}

\par \vspace{2mm}
\begin{center}
{\bf Germ\'an F. R. Sborlini}~$^{(a) (b)}$\footnote{{\tt gfsborlini@df.uba.ar}},
{\bf Daniel de Florian}~$^{(a)}$\footnote{{\tt deflo@df.uba.ar}} and 
{\bf Germ\'an Rodrigo}~$^{(b)}$\footnote{{\tt german.rodrigo@csic.es}}

\vspace{5mm}

${}^{(a)}$Departamento de F\'\i sica and IFIBA, FCEyN, Universidad de Buenos Aires, \\
(1428) Pabell\'on 1 Ciudad Universitaria, 
Capital Federal, Argentina \\
\vspace*{2mm}
${}^{(b)}$Instituto de F\'{\i}sica Corpuscular, 
Universitat de Val\`encia - Consejo Superior de Investigaciones Cient\'{\i}ficas,
Parc Cient\'{\i}fic, E-46980 Paterna (Valencia), Spain \\

\vspace{5mm}

\end{center}

\par \vspace{2mm}
\begin{center} {\large \bf Abstract} \end{center}
\begin{quote}
\pretolerance 10000

We compute the next-to-leading order (NLO) QCD corrections to the $1 \to 2$ splitting amplitudes in different dimensional regularization (DREG) schemes. Besides recovering previously known results, we explore new DREG schemes and analyze their consistency by comparing the divergent structure with the expected behavior predicted by Catani's formula. Through the introduction of scalar-gluons, we show the relation among splittings matrices computed using different schemes. Also, we extended this analysis to cover the double collinear limit of scattering amplitudes in the context of QCD+QED.

\end{quote}

\vspace*{\fill}
\begin{flushleft}

January 2014

\end{flushleft}
\end{titlepage}

\setcounter{footnote}{0}

%********************************************************************************************************
%********************************************************************************************************

%%%%%%%%%%%%%%%%%%%%%%%%%%%%%%%%%%%%%%%%%%%%%%%%%%%%%%%%%%%%%%%%
\section{Introduction}
%%%%%%%%%%%%%%%%%%%%%%%%%%%%%%%%%%%%%%%%%%%%%%%%%%%%%%%%%%%%%%%%

%%%%%%%%%%%%%%%%%%%%%%%%%%%%%%%%%%%%%%

Reaching higher-orders in the context of perturbative QCD implies a great challenge, but it becomes crucial to achieve the level of accuracy required by nowadays experiments. LHC results need to be compared to high precision theoretical predictions to unveil novel high-energy physics phenomena. In order to achieve that goal, it is necessary to understand the infrared (IR) divergent structure of QCD amplitudes. Within the framework of dimensional regularization \cite{Bollini:1972ui,'tHooft:1972fi}, a lot of work has been performed at one-loop, two-loop and higher-order loops \cite{Catani:1998bh, Sterman:2002qn, Aybat:2006wq, Aybat:2006mz, Dixon:2008gr, Becher:2009cu, Gardi:2009qi, Becher:2009qa, Dixon:2009ur}. Many methods rely on the properties of the collinear/soft limit to perform the analytical subtraction of IR divergences, which allows to obtain finite cross-sections at colliders (for instance, see Ref. \cite{Catani:1996vz}).

Centering in the collinear limit, it is known that scattering amplitudes and squared amplitudes take a rather simple form in this class of kinematical configurations. Moreover, there are proven factorization properties\footnote{See \cite{Collins:1989gx} and references therein.} which show that IR collinear divergences exhibit a universal process-independent behavior, although strict collinear factorization is violated in the space-like region\cite{Catani:2011st, Forshaw:2012bi, Catani:2012iw}. 

At the squared amplitude level, this universal behaviour is captured by the Altarelli-Parisi (AP) kernels (also known as \textit{splitting functions}), which were first introduced in Ref. \cite{Altarelli:1977zs} for the double-collinear limit at tree-level, and at the amplitude level by the splitting amplitudes \cite{Berends:1987me, Mangano:1990by}. Since then, splitting amplitudes and Altarelli-Parisi kernels have been studied at one-loop \cite{Bern:1998sc, Bern:1999ry, Bern:1993qk, Bern:1994zx, Bern:1995ix,Kosower:1999rx} and two-loops \cite{Bern:2004cz, Badger:2004uk} in the double collinear limit. A higher-order loops analysis has been performed at the amplitude level in Ref. \cite{Kosower:1999xi}. In the multiple collinear limit, splitting functions were studied at tree-level \cite{Campbell:1997hg, Catani:1998nv, Catani:1999ss, DelDuca:1999ha,Birthwright:2005ak, Birthwright:2005vi} and there are some results for the triple-collinear limit at one-loop order \cite{Catani:2003vu}.

Dimensional regularization (DREG) can be implemented in various ways when performing an explicit computation. This gives rise to different DREG schemes. Since theoretical results have to be compared with experiments, it is expected that they do not depend on the scheme being used. However, since splitting functions and splitting amplitudes are not physical observables, they can exhibit some scheme dependence. For this reason, it is necessary to understand how to relate the results obtained with different schemes. In the double collinear limit, this topic has been discussed in Ref. \cite{Catani:1996pk}. At one-loop level, computations were performed using several schemes choices, for both splitting amplitudes and AP kernels. In particular, in Ref. \cite{Kunszt:1993sd}, the scheme dependence for $2\to 2$ processes was studied at one-loop level and the authors also suggested a way to relate one-loop AP kernels in some usual DREG schemes. 

In this paper, we discuss the scheme dependence of splittings amplitudes at NLO. Starting from the QCD Lagrangian in $(4-2\epsilon)$ dimensions, we decompose the gluon field and define scalar-gluons associated with the extra-dimensional degrees of freedom\footnote{For this reason, some authors also call them \textit{$\epsilon$-scalars}.} (see Ref. \cite{Harlander:2006rj, Kilgore:2011ta}). Using these artificial particles we establish a link among results in several schemes, besides exploring novel DREG parameters' configuration. It is also important to mention that scalar-gluons were useful to solve some theoretical issues related with factorization in QCD when working in DREG \cite{Signer:2005iu, Signer:2008va}.

%%%%%%%%%%%%%%%%%%%%%%%%%%%%%%%%%%%%%%%%%%%%

This article is organized as follows. In Section 2, we briefly define the different schemes in DREG and introduce scalar-gluons in QCD. We motivate the effective Feynman rules for these objects, starting from a Lagrangian-level decomposition. In Section 3 we discuss the kinematics of the collinear limit and introduce the splitting matrices. In Section 4 we present results for the $q\to g q$ splitting at NLO. We recover known expressions, compared them with Catani's formula for the IR-divergent structure and analyze the scalar-gluon contributions. In the last part of that section, we calculate the AP kernels. In Section 5, we tackle the $g\to q \bar{q}$ splitting and put more emphasis in the study of the scalar contributions. The $g \to gg$ splitting amplitude is discussed in Section 6. Since photons play a crucial role in today's collider physics (Higgs boson background, study of quark-gluon plasma and jet quenching, etc.), we extend our results to cover the $q \to \gamma q$ and $\gamma \to q \bar{q}$ splittings in Section 7. Finally, conclusions are presented in Section 8.

%%%%%%%%%%%%%%%%%%%%%%%%%%%%%%%%%%%%%%%%%%%%%%%%%%%%%%%%%%%%%%%%
\section{Dimensional regularization and QCD}
%%%%%%%%%%%%%%%%%%%%%%%%%%%%%%%%%%%%%%%%%%%%%%%%%%%%%%%%%%%%%%%%
In the context of perturbative QCD, when we want to compute higher-order contributions we have to face Feynman integrals. Generally, these are ill-defined because they involve non-integrable expressions. So, it is mandatory to introduce a regularization method to give sense to the theory. Moreover, QCD has both ultraviolet (UV) and IR singularities, so we need prescriptions to treat them. Due to the simultaneous treatment of UV-IR divergences and gauge-invariance preserving formalism, DREG is one of the most used method to regularize QCD.

Introduced in the late sixties \cite{Bollini:1972ui, 'tHooft:1972fi}, the main idea of DREG is to modify the space-time dimension. If $D_{\rm ST}$ is the new dimension of the space-time, then divergences appear as poles in $D_{\rm ST}-4$. To keep the coupling constant dimensionless, one has to introduce a factor $\mu^{4-D_{\rm ST}}$ multiplying the Lagrangian density. Also, one should extend vectors, spinors and tensors to a $D_{\rm ST}$-dimensional space. 

Depending on the DREG scheme, it is possible to keep some quantities living in a $4$-dimensional space. In some sense, this is equivalent to specify the symmetries of the extended theory. We know that QCD is a quantum field theory on a $4$-dimensional space-time which is invariant under the action of the $4$-dimensional Poincar\`e group. When we extend the theory to a $D_{\rm ST}$-dimensional space-time, it is possible to force a $D_{\rm ST}$-dimensional Poincar\`e invariance or just retain the physical $4$-dimensional invariance. In this work we will play with these options and explore their consequences in the final results. 

%********************************************************************************************************
\subsection{DREG schemes definition}
%********************************************************************************************************
Let's start with a general four-dimensional quantum field theory (QFT). We know that any one-loop scattering amplitude can be written in the general form
\beqn
{\cal M}^{(1)} &=& \sum_{k} \, A^{\mu_1 \ldots \mu_{n_k}}_{k} \, \int \frac{d^4 q}{(2 \pi)^4} \, \frac{q_{\mu_1} \ldots q_{\mu_{n_k}}}{D^{\alpha(k,1)}_{\sigma(k,1)} \ldots D^{\alpha(k,n'_k)}_{\sigma(k,n'_k)}} \, ,
\eeqn
where $A_{k}$ are coefficients depending on external momenta, color configuration and the kind of particles involved in the process (both internal and external). The DREG changes the space-time dimension to $D_{\rm ST}$ in order to allow for convergence of loop integrals. Usually, we take $D_{\rm ST}=4-2\epsilon$ (with $\epsilon$ a complex number) and perform the replacement
\beqn
\int \frac{d^4 q}{(2 \pi)^4}  &\to& \int \frac{d^{D_{\rm ST}} q}{(2 \pi)^{D_{\rm ST}}} \equiv \imath \int_q \ .
\eeqn
Using Passarino-Veltman decomposition or any other reduction method, we can solve tensor-type integrals and write them as
\beqn
\int_q \, \frac{q^{\mu_1} q^{\mu_2} \ldots q^{\mu_m}}{D^{\alpha_1}_1 D^{\alpha_2}_2 \ldots D^{\alpha_n}_n} &=& \sum_A \, F^{\mu_1 \ldots \mu_m}_A(\left\{p_i\right\},\left\{\alpha_i\right\},{\eta^{\DST}}) I^{\rm scalar}_A (\left\{p_i \cdot p_j\right\}, D_{\rm ST}) \, ,
\eeqn
where $\left\{D_i\right\}$ and $\left\{p_i\right\}$ denotes the set of possible denominators (with $\alpha_i \geq 0$) and physical $4$-vectors defined in the $4$-dimensional unregularized theory, respectively. Note that we are using the $D_{\rm ST}$-dimensional metric tensor in this expansion. This is an important point since we can not take the limit $\epsilon \to 0$ while computing integrals, so it is not allowed to make the replacement $\eta^{D_{\rm ST}}\to \eta^4$ until we finish the calculation. 

On the other hand, DREG does not impose any specific treatment of the objects that appear in the coefficients $A_k$. Since $A_k$ depend on the Dirac's algebra dimension and the number of fermion and boson polarizations, this means that we can change them and set an specific convention for our computations: this is called a DREG scheme. The parameters used to define a DREG scheme in the context of massless QCD (or massless QCD+QED, as we discuss in the last part of this article) are
\begin{itemize}
	\item $n_g$: number of external gluon polarizations,
	\item $h_g$: number of internal gluon polarizations,
	\item $n_q$: number of external quark polarizations,
	\item $h_q$: number of internal quark polarizations, and
	\item $D_{\rm Dirac}$: dimension of the Dirac's algebra.
\end{itemize}
There is another subtlety related with the dimensionality of particle's momenta. DREG forces loop momenta to be $D_{\rm ST}$-dimensional to ensure convergence, but there is no restriction over external momenta. For that reason, we could use them in $D_{\rm ST}$ or $4$ dimensions in the context of different schemes. At the amplitude level, we usually consider external particles to be physical, so their momenta are naturally $4$-dimensional. However, when we compute squared matrix elements and perform phase space integrals, we can face IR singularities again. This time divergences originate in some regions of phase-space where particles have soft, collinear or soft-collinear kinematics. DREG can be used to regularize phase space integrals \cite{Catani:1996pk}, which implies that unobserved external momenta have to be extended to a $D_{\rm ST}$-dimensional space-time. For that reason, it is also possible to consider external momenta being $D_{\rm ST}$-dimensional. 
 
Following with DREG schemes definition, we first study how to relate the previously mentioned parameters with the way that computations are performed. First of all, note that the $D_{\rm ST}$-dimensional space-time metric can be written as a direct product of a 4-dimensional and a $D_{\rm ST}-4$-dimensional contribution. So, if $\eta^{\DST}_{\mu \nu} = \eta^{\rm 4}_{\mu \nu} \oplus \eta^{\DST-4}_{\mu \nu}$, with $\eta^{\rm 4}_{\mu \nu}$ the usual $4$-dimensional Minkowski metric, then $\eta^{\DST}_{\mu \nu} {\left(\eta^{\DST-4}\right)}^{\mu \nu}=D_{\rm ST}-4$. On the other hand, we can introduce vectors and spinors in this space\footnote{In the context of smooth manifolds, vector fields are defined as sections to the tangent bundle and spinor fields arise as representations of a Clifford algebra induced by the metric over the tangent space. So, $D_{\rm Dirac}$ refers to the dimension of the tangent space and it must be $D_{\rm Dirac}=D_{\rm ST}$ by definition. However, in the context of DREG we can treat them independently, since we are ultimately interested in taking the limit $\epsilon\to 0$ to recover physical results.}. We write $\DDirac=4-2\delta \epsilon$, with $\delta=0$ or $\delta=1$ to work with a $4$-dimensional or a $\DST$-dimensional algebra, respectively. Spinors are defined starting from a representation $R$ of Dirac's algebra; that is, we have a set of objects $\left\{\gamma^{\mu_k}\right\}_{k=0\ldots D_{\rm Dirac}-1} \in R$ which verifies
\beqn
\left\{\gamma^{\mu},\gamma^{\nu}\right\} &=& 2 (\eta^{D_{\rm Dirac}})^{\mu\nu} {\rm Id} \, ,
\eeqn
where ${\rm Id}$ refers to the identity in the space where representation $R$ is defined. Since fermions are described by spinors, the number of polarizations of a massless fermion is related to ${\rm Tr}({\rm Id})$. In particular, we can define
\beqn
{\rm Tr}^{\rm Ext}({\rm Id}) &=& 2n_q \, ,
\\ {\rm Tr}^{\rm Int}({\rm Id}) &=& 2h_q\, ,
\eeqn
where ${\rm Tr}^{\rm Ext}$ and ${\rm Tr}^{\rm Int}$ denote the trace over external and internal fermionic states, respectively, since we are treating internal and external fermions in an independent way. It is interesting to appreciate that changing $h_q$ or $n_q$ only modifies contributions which involve Dirac's traces, because using Dirac algebra and cyclic-invariance of traces, it turns out that traces are always proportional to ${\rm Tr}({\rm Id})$. We introduced the parameters $\beta$ and $\beta_{\rm R}$ to write
\beqn
n_q &=& 2- 2 \beta_{\rm R} \epsilon \ \to \ {\rm Tr}^{\rm Ext}({\rm Id}) = 4- 4\beta_{\rm R} \epsilon \, ,
\\ h_q &=& 2- 2 \beta \epsilon \ \to \ {\rm Tr}^{\rm Int}({\rm Id}) = 4- 4\beta \epsilon \, ,
\eeqn
and control the number of fermion polarizations when performing the computations.

Now let's turn to the parameter $h_g$ which is related to the gluon propagator. Working in an axial gauge, we write the sum over internal gluon's polarizations as
\beqn
d_{\mu \nu}(p,n)&=& - \left(\eta^{\rm 4}_{\mu \nu}+\alpha_R\eta^{D_{\rm Dirac}-4}_{\mu \nu}\right) + \frac{p_{\mu}n_{\nu}+n_{\mu}p_{\nu}}{p \cdot n} \ ,
\label{GluonPolarizationsSUMINTERNAL}
\eeqn
where $n$ is a light-like vector which verifies $n^2=0$ and $n\cdot p \neq 0$. Here we introduced $\alpha_R$ to take into account the number of polarization associated with internal gluons. In particular, we know that
\beq
h_g = d_{\mu \nu}(p,n) (\eta^{D_{\rm ST}})^{\mu\nu} \, .
\eeq
Using Eq. \ref{GluonPolarizationsSUMINTERNAL}, we see that if $\alpha_R=0$ then $h_g=2$ independently of the value of $D_{\rm Dirac}$, while if we choose $\alpha_R=1$ then $h_g=D_{\rm Dirac}-2$. It is important to note that this result relies in the fact that $n$ is the $\DST$-dimensional null-extension of a four-vector and the metric tensor is diagonal even in $D_{\rm ST}$-dimensions. Also, we can appreciate that choosing $\delta=0$ (i.e $\DDirac=4$) removes the dependence on $\alpha_R$ in Eq. \ref{GluonPolarizationsSUMINTERNAL}.

To control the number of external gluon's polarizations we define
\beqn
d^{\rm Ext}_{\mu \nu}(p,Q) &=& \sum_{\rm phys. pol.} \, \epsilon^{*}_{\mu}(p)\epsilon_{\nu}(p) =  - \left(\eta^{\rm 4}_{\mu \nu}+\alpha \eta^{D_{\rm ST}-4}_{\mu \nu}\right) + \frac{p_{\mu}Q_{\nu}+Q_{\mu}p_{\nu}}{p \cdot Q} \, ,
\label{GluonPolarizationsSUMEXTERNAL}
\eeqn
where $Q$ is an arbitrary null-vector which fulfills $Q^2=0$ and $Q \cdot p \neq 0$. When performing the explicit computation we will take $Q=n$ with the sake of simplifying the intermediate steps. Again, using Eq. \ref{GluonPolarizationsSUMEXTERNAL}
\beq
n_g= d^{\rm Ext}_{\mu \nu}(p,n) (\eta^{D_{\rm ST}})^{\mu\nu} = 2- 2 \alpha \epsilon \, ,
\eeq
where we express the result explicitly in terms of $\alpha$. 

\begin{table}

	\centering

		\begin{tabular}{|l|| c| c| c| c| c|}
		  \hline

		  Scheme & $n_g$ & $h_g$  & $\hphantom{a}$ $\delta$ $\hphantom{a}$ & $\hphantom{a}$ $\alpha_R$ $\hphantom{a}$ & $\hphantom{a}$ $\alpha$ $\hphantom{a}$ \\
			\hline
			CDR & $2-2\epsilon$ & $2-2\epsilon$ & 1 & 1 & 1 \\
			HV & 2 & $2-2\epsilon$ & 1 & 1 & 0 \\
			FDH & 2 & 2 & 0 & 1 & 0 \\
			\hline
			HSA & $2-2\epsilon$ & 2 & 1 & 0 & 1 \\
			HSB & 2 & 2 & 1 & 0 & 0 \\
			\hline

		\end{tabular}

	\caption{Table of DREG schemes used in this paper. All these schemes set the number of internal and external fermion's polarizations to $2$ (i.e $\beta=0=\beta_R$).}

	\label{tab:TableDREGSCHEMES}

\end{table}

At this point, it is important to recall some properties of the gluon's polarization tensors $d_{\mu \nu}(p,n)$ and $d^{\rm Ext}_{\mu \nu}(p,n)$. Since we are working in the light-cone gauge (LCG), these objects should fulfill the following identities:
\begin{itemize}

	\item $d_{\mu \nu}(p,n) n^{\mu} = 0 = d^{\rm Ext}_{\mu \nu}(p,n) n^{\mu}$ (orthogonality to $n$),
	\item $d^{\rm Ext}_{\mu \nu}(p,n) p^{\mu} = 0$ (orthogonality to external momenta $p$), and
	\item $d_{\mu \nu}(p,n) p^{\mu} \propto p^2$.

\end{itemize}

These requirements are related with some physical restrictions. The first condition is due to the gauge choice, while the fact that external gluons are massless guarantees the validity of the second identity. The last condition is imposed in order to recover orthogonality with external momenta when the virtual particle is on-shell. Having introduced a parametrization for polarization tensors in Eqs. \ref{GluonPolarizationsSUMINTERNAL} and \ref{GluonPolarizationsSUMEXTERNAL}, we show that
\beqn
d_{\mu \nu}(p,n) n^{\mu} &=& - n_{\nu} + \frac{(p\cdot n) \, n_{\nu}+n^2 \, p_{\nu}}{p \cdot n} =0= d^{\rm Ext}_{\mu \nu}(p,n) n^{\mu} \, ,
\eeqn
where we use strongly that $n$ is a light-like $4$-vector (and $n^{\mu}\eta^{D_{\rm Dirac}-4}_{\mu \nu}=0$). Something similar happens if we consider external momenta as the null-extension of a light-like $4$-vector, i.e. $p=p^{(4)}\oplus \vec{0}$. In this case, we obtain
\beqn
d^{\rm Ext}_{\mu \nu}p^{\mu} &=& - p_{\nu} + \frac{p^2 \, n_{\nu}+(p \cdot n) \, p_{\nu}}{p \cdot n} =0 \, ,
\eeqn
which shows that all the requirement are fulfilled for external momenta, independently of the dimension of the space in which external momenta live. However in all DREG schemes, internal momenta have to be expressed as $p=p^{(4)}\oplus p^{(\DST-4)}$ (i.e. with a non trivial component in the transverse space) and when we contract with $d_{\mu \nu}(p,n)$ we get
\beqn
\nonumber d_{\mu \nu}(p,n) p^{\mu} &=& - \left(p^{(4)}_{\nu}+\delta \alpha_R p^{(\DST-4)}_{\nu}\right) + \frac{p^2 \, n_{\nu}+(p\cdot n)\, p_{\nu}}{p \cdot n} =
\\  &=& p^2 \frac{n_{\nu}}{p \cdot n} + (1- \delta \alpha_R)  p^{(\DST-4)}_{\nu} \, ,
\eeqn
which shows that, for certain combinations of parameters, propagators fail to fulfill some physical consistence conditions. In the following, we discuss deeply about this fact, performing some explicit computations and showing that failing to verify this conditions could lead to some unexpected IR divergences.

Having introduced the possible parameters that we can modify in the context of DREG, let's explain how to recover some of the most frequently used schemes. In conventional dimensional regularization (CDR) \cite{Bollini:1972ui, 'tHooft:1972fi, Gastmans:1973uv}, internal and external particles are treated in the same way. We consider that particle's momenta is $\DST$-dimensional, gluons have $2-2\epsilon$ polarizations and massless fermions only have $2$ polarizations. In other words, CDR uses $n_g=h_g=2-2\epsilon$ and $h_q=n_q=2$, with $\DDirac=4-2\epsilon$. Equivalently, we can work in this scheme setting $\alpha=1$, $\alpha_R=1$ and $\delta=1$, according to our definitions.

On the other hand, we can set external particles in four-dimensions while keeping internal ones in $D_{\rm ST}$-dimensions. This is the 't Hooft-Veltman scheme (HV), first introduced in Ref. \cite{'tHooft:1972fi}. External momenta are four-dimensional and external massless fermions and gluons have only $2$ physical polarization states (i.e. $n_g=2=n_q$). However, internal gluons have $2-2\epsilon$ polarizations ($h_g=2-2\epsilon$) while internal fermions only have $2$ ($h_q=2$). Using our parameters, we can settle in this scheme by choosing $\delta=1$, $\alpha=0$ and $\alpha_R=1$.

Sometimes it is preferable to treat all the particles as $4$-dimensional objects, although internal ones have their momenta extended to a $\DST$-dimensional space. In the four-dimensional helicity scheme (FDH) \cite{Signer:2008va, Siegel:1979wq}, massless fermions and gluons have $2$ polarizations states, independently of being internal or external particles. It means that FDH scheme is defined by setting $n_g=h_g=2$, $n_q=h_q=2$ and $\DDirac=4$, or $\delta=0$ and $\alpha=0$ in our convention. The main advantage of this configuration is the possibility of using many identities and properties derived from the helicity method, which can be used to obtain very compact expressions. 

Closely related with FDH, there is other choice called dimensional reduction (DRED) \cite{Siegel:1979wq, Capper:1979ns}. In this scheme, both external and internal particles have $4$-dimensional polarization vectors, but external momenta are continued to $\DST$-dimensions. This forces us to include scalar-gluons in order to achive consistency. A deep discussion about dimensional reduction schemes and its implementation can be found in Ref. \cite{Signer:2008va}\footnote{In this reference, the authors group DRED and FDH into the category of \textit{dimensional reduction schemes}. Moreover, they are called \textit{old} and \textit{new} dimensional reduction, respectively. It is important to note that these schemes are not equivalent, but they could lead to the same results.}.

Beyond these three well-established schemes, changing the parameters $\alpha_R$, $\alpha$, $\delta$ and the value of ${\rm Tr}({\rm Id})$, we can construct more configurations. Variations of CDR, HV and FDH with ${\rm Tr}({\rm Id})=4-4\epsilon$ were studied in Ref. \cite{Catani:1996pk}. In particular, in that paper, the authors analyzed the consequences of choosing those \textit{toy-models} when performing a full NLO computation with the subtraction method. Since we can easily modify the values of $\beta$ and $\beta_R$ in our codes, we give most of our results for an arbitrary number of fermion polarizations. In the last part of this article, we also discuss the results computed in the toy-scheme (TSC) defined by $n_g=n_q=2-2\epsilon=h_q=h_g$ which was introduced in Ref. \cite{Catani:1996pk}. Specifically, we show that this scheme preserves the supersymmetric Ward identity at one-loop level.

Aside from allowing different values of $h_q$ and $n_q$, here we also discuss \textit{hybrid-schemes} that use $\DDirac=4-2\epsilon$ ($\delta=1$) and $h_g=2$ ($\alpha_R=0$), with the possibility of setting $n_g=2-2\epsilon$ ($\alpha=1$) or $n_g=2$ ($\alpha=0$). To make the discussion easier, we call them hybrid-scheme A (HSA) and hybrid-scheme B (HSB), respectively. A summary of all the schemes treated in this paper is displayed in Table \ref{tab:TableDREGSCHEMES}. Although these new schemes seem to a valid choice, we can anticipate that they are inconsistent unless we add some scalar-gluon contributions.

%********************************************************************************************************
\subsection{Scalar-gluons: Lagrangian level decomposition}
%********************************************************************************************************
An interesting fact is related to the appearance of new scalar-type particles when we use certain DREG schemes in a $D$-dimensional space. We can decompose $D$-vectors into $4$-vectors plus $D-4$ scalar particles \cite{Capper:1979ns}; this forces us to introduce new Feynman rules for these particles and, of course, new diagrams contribute to the scattering-amplitudes. Note that this decomposition suggests that non-physical degrees of freedom can be absorbed into a \textit{certain amount}\footnote{Specifically, since there are $D-4$ transverse dimensions, there should be $D-4$ scalar particles in the problem. Note that we are using the signature $\left(+---\right)$ for the space-time metric.} of scalar-particles.

To get the Feynman rules for scalar-gluons, let's start with the usual $4$-dimensional massless-QCD Lagrangian density,
\beqn
{\cal L}_{\rm QCD} &=& \sum_f \, \bar{\Psi}^{i}_f(\imath \gamma^{\mu} D_{\mu,ij})\Psi^j_f - \frac{1}{4} G^a_{\mu \nu} G^{a \, \mu \nu} \, ,
\label{QCD4DimensionalLagrangian}
\eeqn
where $\left\{i,j\right\}$ are color indices, $G_{\mu \nu} = \partial_{\mu}A^a_{\nu}-\partial_{\nu}A^a_{\mu}-g_s f_{a b c} A^b_{\mu} A^c_{\nu}$ is the gauge-field strength tensor, $D_{\mu,ij} = \partial_{\mu} \delta_{ij}+\imath g_s A^a_{\mu} T^a_{ij}$ is the covariant derivative and we are summing over the possible quark flavors. Knowing that kinetic terms are associated with propagators, let's expand the interaction component which can be written as
\beqn
{\cal L}^{\rm Int}_{\rm QCD} &=& {\cal L}^{\rm \bar{f}gf}_{\rm QCD} + {\cal L}^{\rm 3g}_{\rm QCD} + {\cal L}^{\rm 4g}_{\rm QCD} \, ,
\eeqn
with
\beqn
{\cal L}^{\rm \bar{f}gf}_{\rm QCD} &=& - \sum_f g_s\mu^{\epsilon} T_{ij}^a \, \bar{\Psi}^i_f.\gamma^{\mu}.\Psi^j_f \, A^a_{\mu} \, ,
\\ {\cal L}^{\rm 3g}_{\rm QCD} &=& \frac{g_s\mu^{\epsilon}f_{a b c}}{2} \, \left(\partial_{\mu}A^a_{\nu}-\partial_{\nu}A^a_{\mu}\right) A^{b \, \mu} A^{c \, \nu} \, ,
\\ {\cal L}^{\rm 4g}_{\rm QCD} &=& -\frac{g^2_s\mu^{2 \epsilon}}{4} \, f_{a b c} f_{a d e}  A^b_{\mu} A^c_{\nu} A^{d \, \mu} A^{e \, \nu} \, ,
\eeqn
which are associated to the fermion-gluon, three-gluon and four-gluon vertices, respectively. Note that we take $D=4-2\epsilon$ as the space dimension.

Now, let's consider that the space-time metric $\eta^{D}_{\mu\nu}$ is diagonal and can be expressed as a direct product. We can perform the decomposition
\beqn
A^a_{\mu} &=& \hat{A}^a_{\mu} + \tilde{A}^a_{\mu} \, ,
\\ {\gamma}^{\mu} &=& \hat{\gamma}^{\mu} + \tilde{\gamma}^{\mu} \, ,
\eeqn 
with $\hat{A}$, $\hat{\gamma}$ living in a $4$-dimensional space and $\tilde{A}$, $\tilde{\gamma}$ in the unphysical $D-4$-dimensional transverse space. Also, we have
\beqn
\left\{\tilde{\gamma}^{\mu},\hat{\gamma}^{\nu} \right\}&=&0 \, ,
\\ \eta^{D}_{\mu\nu} \, \tilde{A}^{\mu} \hat{A}^{\nu} &=& 0 \, ,
\eeqn
due to the orthogonality of the $4$-dimensional and the $(D-4)$-dimensional subspaces. The validity of this decomposition is general. However it is suitable when we consider that the theory only retains $4$-dimensional Poincar\`e invariance. In this case, $\tilde{A}$ behaves like a $4$-vector while $\hat{A}$ does not transform under the $4$-dimensional Poincar\`e group. This fact motivates that $\hat{A}$ is called scalar-gluon. On the other hand, under the assumption of $D$-dimensional Poincar\`e invariance, a general Lorentz transformation might mix $\tilde{A}$ and $\hat{A}$ although $A$ transforms as a $D$-vector. 

\begin{figure}[htb]
	\centering
		\includegraphics[width=0.60\textwidth]{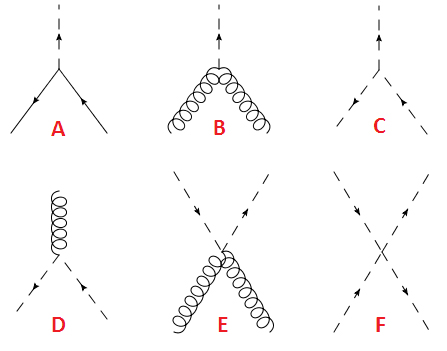}
	\caption{Available vertices involving scalar-gluons. Expanding QCD Lagrangian, we find six kind of vertices: \textbf{A} fermion-scalar-fermion, \textbf{B} gluon-scalar-gluon, \textbf{C} triple-scalar interaction, \textbf{D} scalar-gluon-scalar, \textbf{E} 2scalar-2gluon, and \textbf{F} 4scalar. Momenta associated with gluons and scalar-gluons are considered outgoing.}
	\label{fig:ScalarVerteces}
\end{figure}

Using the expressions for the interaction terms in the Lagrangian Eq. \ref{QCD4DimensionalLagrangian} we get
\beqn
{\cal L}^{\rm \bar{f}gf}_{\rm QCD} &=& - \mu^{\epsilon} g_s T_{ij}^a \, \sum_f\,  \left(\bar{\Psi}^i_f\hat{\gamma}^{\mu}\Psi^j_f \, \hat{A}^a_{\mu} + \bar{\Psi}^i_f\tilde{\gamma}^{\mu}\Psi^j_f \, \tilde{A}^a_{\mu} \right) \, ,
\\ {\cal L}^{\rm 3g}_{\rm QCD} &=& \mu^{\epsilon} g_s f_{a b c} \, \left[ (\partial_{\mu}\hat{A}^a_{\nu}) \hat{A}^{b \, \mu} \hat{A}^{c \, \nu}+(\partial_{\mu}\hat{A}^a_{\nu}) \tilde{A}^{b \, \mu} \hat{A}^{c \, \nu} +(\partial_{\mu}\tilde{A}^a_{\nu}) \hat{A}^{b \, \mu} \tilde{A}^{c \, \nu} +(\partial_{\mu}\tilde{A}^a_{\nu}) \tilde{A}^{b \, \mu} \tilde{A}^{c \, \nu}\right] \, ,
\\ {\cal L}^{\rm 4g}_{\rm QCD} &=& -\frac{\mu^{2 \epsilon} g^2_s}{4} \, f_{a b c} f_{a d e} \left[\hat{A}^b_{\mu} \hat{A}^c_{\nu} \hat{A}^{d \, \mu} \hat{A}^{e \, \nu} + 2 \hat{A}^b_{\mu} \tilde{A}^c_{\nu} \hat{A}^{d \, \mu} \tilde{A}^{e \, \nu}+  \tilde{A}^b_{\mu} \tilde{A}^c_{\nu} \tilde{A}^{d \, \mu} \tilde{A}^{e \, \nu}\right]\,,
\label{ExpansionLagrangiano}
\eeqn
where we must take into account that some indices live in the $4$-dimensional physical space, while others stay only in the transverse space. In Fig. \ref{fig:ScalarVerteces} the available vertices are drawn. We have six possible interaction vertices which involves scalar-gluons: 2fermions-scalar, 2gluons-scalar, 2scalars-gluon, 3scalars, 2scalars-2gluons and 4scalars.

\begin{figure}[htb]
	\centering
		\includegraphics[width=0.99\textwidth]{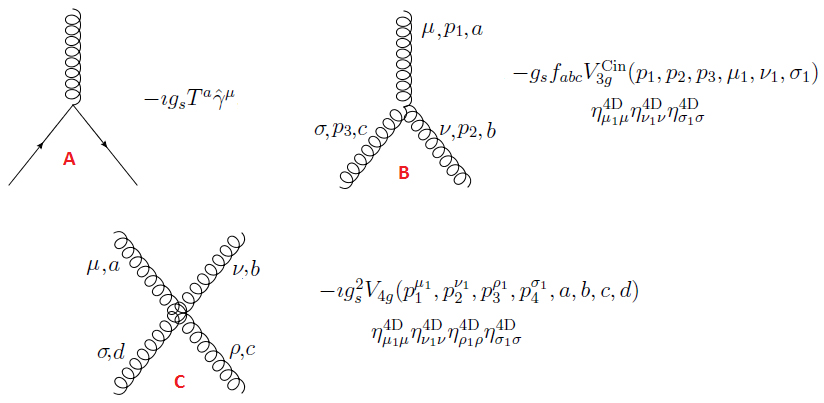}
	\caption{Usual $4$-dimensional QCD interaction vertices: \textbf{A} fermion-gluon-fermion, \textbf{B} triple-gluon and \textbf{C} quadruple-gluon vertex. Momenta associated with gluons are considered outgoing.}
	\label{fig:StandardRulesQCD}
\end{figure}

After identifying the Lagrangian terms that originate the possible scalar interactions, we are able to deduce the corresponding Feynman rules. Introducing the functions
\beqn
V_{3g}\left(p^{\mu}_1,p^{\nu}_2,p^{\sigma}_3,a,b,c \right) &=& f_{abc} \left[ {\left(p_2-p_1\right)}_{\sigma} \eta^{\DST}_{\mu \nu} + {\left(p_3-p_2\right)}_{\mu} \eta^{\DST}_{\nu \sigma} + {\left(p_1-p_3\right)}_{\nu} \eta^{\DST}_{\sigma \mu}\right]
\label{TripleGluonVertex}
\\ &=& f_{abc} V^{\rm Cin}_{3g}\left(p_1,p_2,p_3;\mu,\nu,\sigma \right) \, ,
\eeqn
and
\beqn
\nonumber V_{4g}(p^{\mu}_1,p^{\nu}_2,p^{\rho}_3,p^{\sigma}_4,a,b,c,d) &=& f_{a b e} f_{c d e} \left(\eta^{\DST}_{\mu \rho}\eta^{\DST}_{\nu \sigma}-\eta^{\DST}_{\mu \sigma}\eta^{\DST}_{\nu \rho} \right) + f_{a c e} f_{b d e} \left(\eta^{\DST}_{\mu \nu}\eta^{\DST}_{\rho\sigma}-\eta^{\DST}_{\mu \sigma}\eta^{\DST}_{\nu \rho} \right)
\\ &+& f_{a d e} f_{c b e} \left(\eta^{\DST}_{\mu \nu}\eta^{\DST}_{\rho \sigma}-\eta^{\DST}_{\mu \rho}\eta^{\DST}_{\nu \sigma} \right) \, ,
\label{QuadrupleGluonVertex}
\eeqn
the usual Feynman rules for $4$-dimensional QCD reads
\begin{itemize}
	\item fermion-gluon-fermion vertex $- \imath g_s \SUNT^a \hat{\gamma}^{\mu}$,
	\item triple-gluon vertex $- g_s V_{3g}(p^{\mu_1}_1,p^{\nu_1}_2,p^{\sigma_1}_3,a,b,c) \eta^{\rm 4}_{\mu_1 \mu} \eta^{\rm 4}_{\nu_1 \nu} \eta^{\rm 4}_{\sigma_1 \sigma}$,
	\item and quadruple-gluon $- \imath g^2_s V_{4g}(p^{\mu_1}_1,p^{\nu_1}_2,p^{\rho_1}_3,p^{\sigma_1}_4,a,b,c,d) \eta^{\rm 4}_{\mu_1 \mu} \eta^{\rm 4}_{\nu_1 \nu} \eta^{\rm 4}_{\rho_1 \rho} \eta^{\rm 4}_{\sigma_1 \sigma}$,
\end{itemize}
where we are projecting Lorentz index to the $4$-dimensional space through the contraction with $\eta^{\rm 4}$. In Fig. \ref{fig:StandardRulesQCD} we write explicitly the usual QCD rules, to clarify momentum sign and ordering conventions. From these rules and conventions, we get the following associated Feynman rules for scalar interactions;
\begin{itemize}
	\item fermion-scalar-fermion vertex $- \imath g^{\rm scalar}_s \mu^{\epsilon}\SUNT^a \tilde{\gamma}^{\mu}$,
	\item gluon-scalar-gluon vertex $- g^{\rm scalar}_s \mu^{\epsilon} V_{3g}(p^{\mu_1}_1,p^{\nu_1}_2,p^{\sigma_1}_3,a,b,c) \eta^{\rm \epsilon}_{\mu_1 \mu} \eta^{\rm 4}_{\nu_1 \nu} \eta^{\rm 4}_{\sigma_1 \sigma}$,
	\item scalar-gluon-scalar vertex $- g^{\rm scalar}_s \mu^{\epsilon} V_{3g}(p^{\mu_1}_1,p^{\nu_1}_2,p^{\sigma_1}_3,a,b,c) \eta^{\rm 4}_{\mu_1 \mu} \eta^{\rm \epsilon}_{\nu_1 \nu} \eta^{\rm \epsilon}_{\sigma_1 \sigma}$,
	\item triple-scalar vertex $- g^{\rm scalar}_s \mu^{\epsilon} V_{3g}(p^{\mu_1}_1,p^{\nu_1}_2,p^{\sigma_1}_3,a,b,c) \eta^{\epsilon}_{\mu_1 \mu} \eta^{\rm \epsilon}_{\nu_1 \nu} \eta^{\rm \epsilon}_{\sigma_1 \sigma}$,
	\item 2scalar-2gluon $- 2 \imath {\left(g^{\rm scalar}_s\right)}^2 \mu^{2 \epsilon} V_{4g}(p^{\mu_1}_1,p^{\nu_1}_2,p^{\rho_1}_3,p^{\sigma_1}_4,a,b,c,d) \eta^{\rm \epsilon}_{\mu_1 \mu} \eta^{\rm \epsilon}_{\nu_1 \nu} \eta^{\rm 4}_{\rho_1 \rho} \eta^{\rm 4}_{\sigma_1 \sigma}$,
	\item and 4scalar $- \imath {\left(g^{\rm scalar}_s\right)}^2 \mu^{2 \epsilon} V_{4g}(p^{\mu_1}_1,p^{\nu_1}_2,p^{\rho_1}_3,p^{\sigma_1}_4,a,b,c,d) \eta^{\rm \epsilon}_{\mu_1 \mu} \eta^{\rm \epsilon}_{\nu_1 \nu} \eta^{\rm \epsilon}_{\rho_1 \rho} \eta^{\rm \epsilon}_{\sigma_1 \sigma}$,
\end{itemize}
where we used the labeling introduced in Fig. \ref{fig:ScalarVerteces} and the conventions shown in Fig. \ref{fig:StandardRulesQCD}. The additional factor 2 in the 2scalar-2gluon vertex comes from the expansion of the Lagrangian in Eq. \ref{ExpansionLagrangiano}.

As a final remark, note that when scalar-gluons appear we make the replacement $g_s \to  g^{\rm scalar}_s$. This is done because we consider scalar-gluons as a new kind of particles which are not necessarily related with vector-gluons. So, following Landau's principle, we have to write the most general Lagrangian compatible with certain reasonable requirements. But these requirements do not exclude the possibility of having different coupling constants, so we introduce $g^{\rm scalar}_s$ and treat it independently of $g_s$. Although $g_s=g^{\rm scalar}_s$ at leading order in $g_s$, these couplings do not evolve in the same way and, in consequence, can differ at higher-orders (see \cite{Harlander:2006rj, Harlander:2007ws}).

%********************************************************************************************************
\subsection{Effective Feynman rules and other considerations for scalar-gluons}
%********************************************************************************************************
Working with scalar-gluons involves taking into account some technical details. In order to be able to write scattering amplitudes that include scalar-particles, let's motivate some effective Feynman rules and explain other useful points.

Let's start with the gluon propagator, in an axial gauge
\beqn
D_G\left(p,\mu,\nu\right)&=&\imath \frac{d_{\mu \nu}(p,n)}{p^2 + \imath 0} \ ,
\label{GluonPropagatorGeneral}
\eeqn
where $d_{\mu\nu}(p,n)$ is given by Eq. \ref{GluonPolarizationsSUMINTERNAL}, using a null-vector $n$. Here we explicitly indicate that we are using the Feynman prescription to compute propagators. However, in the following we will omit the term $+\imath 0$ in propagator denominators, although its presence is always understood. As we shown previously, the number of gluon polarizations can be modified changing $\alpha_R$ and $\delta$. Working in a $D_{\rm ST}$-dimensional space-time, gluons can be treated as $D_{\rm ST}$-vectors setting $\eta = \eta^{\DST}$ inside the definition of the propagator (or $\alpha_R=1$ and $\delta=1$ with our parametrization), in which case $h_g = d_{\mu \nu}(p,n) {(\eta^{\DST})}^{\mu \nu} = D_{\rm ST} -2$. Also it is possible to decompose them in a $4$-dimensional gluon plus scalar-gluons, by setting $\eta = \eta^{4}$ in $D_G$ and using the propagator
\beqn
D_S\left(p,\mu,\nu\right)&=&-\imath \frac{\eta^{\rm \epsilon}_{\mu \nu}}{p^2+ \imath 0} \, ,
\label{PropagadorGluonEscalar}
\eeqn
for the scalar-gluon component. If we count the number of polarizations in this case, vector-gluons contribute with $d^{\rm 4D}_{\mu \nu}(p,n) {(\eta^{\DST})}^{\mu \nu} = 2$ while scalar-gluons add $\eta^{\rm \epsilon}_{\mu \nu} {(\eta^{\DST})}^{\mu \nu} = D_{\rm ST}-4$ polarizations. For this reason, scalar-gluons have to be included when we set the number of polarizations of internal gluons to $2-2\epsilon$ while working with a $4$-dimensional Dirac's algebra. In other words, HV results could be recovered adding to the FDH computation the corresponding scalar-gluon contribution. 

It is worth noting that a completely similar analysis can be performed with external gluons. If they are treated as $D_{\rm ST}$-vectors, we can decompose them as $4$-dimensional gluons plus $D_{\rm ST}-4$ scalar particles. This implies that we can also use scalar-gluons as external legs to compensate the number of degrees of freedom of the system when working with $D_{\rm Dirac}=4$. Explicitly, 
\beqn
\sum_{\rm pol. \in \DST} \, \epsilon^{*}_{\mu}(p)\epsilon_{\nu}(p) &=&  \left(-\eta^{\rm 4}_{\mu \nu} + \frac{p^{\mu}Q^{\nu}+Q^{\mu}p^{\nu}}{p \cdot Q} \right) + \left(-\eta^{\rm \epsilon}_{\mu \nu}\right)
\\ &=& \sum_{\rm pol. \in 4D} \, \epsilon^{*}_{\mu}(p)\epsilon_{\nu}(p) \, + \hat{\epsilon}^{*}_{\hat{\mu}}(p)\hat{\epsilon}_{\hat{\nu}}(p) \, ,
\label{SumaPolarizacionesSCALARES}
\eeqn
where we are using a diagonal extension of space-time metric and we interpret $\left\{\hat{\mu},\hat{\nu} \right\} \in (\DST-4)$ as real (or complex) numbers. 
 
Scalar-vector decomposition is performed with the aim of being able to use the well-known Dirac algebra properties in integer-dimensional spaces, and, in some sense, forget about the transverse $\epsilon$-dimensional components artificially introduced during the regularization process. Being more explicit, when we retain only $4$-dimensional Poincar\`e invariance, we are setting fermions in $4$-dimensions. So, we have to consider $D_{\rm Dirac}=4$ which simplifies a lot the treatment of spinor chains.

Now, let's tackle the associated effective Feynman rules for scalar-gluons. Before doing that, we reinterpretate the meaning of extra-dimensions and additional gluon polarizations. As a staring point, let's remark that DREG is a particular dimension extension of a $4$-dimensional QFT. So, additional gluon polarizations are related to additional space-time dimensions. Now, we have two options: we can retain only the original invariance under the action of the $4$-dimensional Poincar\`e group or we can force a $D$-dimensional invariance. In the latest choice, we are going to have $D$-dimensional vector type gluons, while in the first one we will be able to separate $4$-vector type gluons from $(D-4)$-flavors of scalar type fields (which we called scalar-gluons here). Note that, in this step, we have used that $\eta^{\DST}$ is a flat-diagonal extension of usual Minkowski metric, which allows us to \textit{convert extra-dimensions into flavors of scalar particles}. And this is the key point to write the Feynman rules.

Using the definitions of $V_{3g}$ and the induced rules for scalar-gluons vertices at Lagrangian level, we can get some effective rules to work with these particles. To get them we will modify the expressions given before, using the fact that $\eta^{\DST}$ is diagonal (i.e. it does not mix physical and transverse contributions). So, starting with the triple vertex we have:

\begin{itemize}

	\item $g^{\rm scalar}_s \mu^{\epsilon} f_{a b c} \eta^{\rm 4}_{\nu \sigma} (p_2-p_3)_{\hat{\mu}}$ for the 2gluon-scalar vertex;
	\item $g^{\rm scalar}_s \mu^{\epsilon} f_{a b c} \eta^{\epsilon}_{\hat{\nu} \hat{\sigma}} (p_2-p_3)_{\mu}$ for the 2scalar-gluon vertex;
	\item and $g^{\rm scalar}_s \mu^{\epsilon} V_{3g}^{\rm Cin}(p_1^{\hat{\mu}},p_2^{\hat{\nu}},p_3^{\hat{\rho}},a,b,c)$ for the 3-scalar vertex

\end{itemize}

Note that these rules agree with the usual form of Feynman rules for vector-scalar interactions. Also, here $\eta^{\epsilon}_{\rho \sigma}$ can be interpreted as a delta function whose value is $1$ if scalar-particles have the same index and $0$ otherwise.

Following the same ideas, we can simplify quadruple interactions and we get these rules:
\begin{itemize}
	\item $- 2 \imath {\left(g^{\rm scalar}_s\right)}^2 \mu^{2 \epsilon} (f_{a c e}f_{b d e}+f_{a d e}f_{b c e}) \eta^{\rm 4}_{\sigma \rho} \eta^{\epsilon}_{\mu \nu}$ for the 2gluon-2scalar vertex;
	\item and $- \imath {\left(g^{\rm scalar}_s\right)}^2 \mu^{2\epsilon} V_{4g}(\hat{\mu},\hat{\nu},\hat{\sigma},\hat{\rho},a,b,c,d)$ for the 4scalar vertex;
\end{itemize}
where, again, we see agreement among these expressions and the ones associated with standard quadruple scalar-vector interactions.

Finally, let's make a comment about the fermion-scalar interaction. This is the only vertex which involves dealing with $\hat{\gamma}$ matrices. Since $D_{\rm Dirac}=4$, these extra-gamma matrices act trivially over spinors, so we do not have to include them inside spinor chains: this leads to helicity-violation interactions. Moreover, if we have two $\hat{\gamma}$ matrices inside a chain, using the fact that $\left\{\hat{\gamma}^{\mu},\tilde{\gamma}^{\nu} \right\}=0$ and $\left\{\hat{\gamma}_{\mu},\hat{\gamma}_{\nu} \right\}=2 \eta^{\epsilon}_{\mu \nu} \, {\rm Id}$ we can get ride of the transverse-dimensional indices. We give an explicit example when computing the $q \to g q$ splitting amplitude at NLO.

%****************************************************************************************************%
\subsection{Computational implementation}
%****************************************************************************************************%
We implemented the computation in {\sc Mathematica} and we used {\sc FeynCalc} (version 8.2) \cite{Mertig:1990an} to handle Dirac's algebra. {\sc FeynCalc} used $D$ as the dimension of Dirac's algebra; in particular, since it was used to perform Dirac and Lorentz algebra, then $\DDirac=D$. To compute integrals we used the results shown in the literature (for example, see Ref. \cite{Kosower:1999rx}) and the integration by parts method (IBP) \cite{Chetyrkin:1981qh}, implemented through the {\sc Mathematica}'s package {\sc FIRE} \cite{Smirnov:2008iw, Smirnov:2013dia}. We set $d=4-2\epsilon$ as the space-time dimension in this package.

%%%%%%%%%%%%%%%%%%%%%%%%%%%%%%%%%%%%%%%%%%%%%%%%%%%%%%%%%%%%%%%%
\section{Collinear limits of scattering amplitudes in QCD}
%%%%%%%%%%%%%%%%%%%%%%%%%%%%%%%%%%%%%%%%%%%%%%%%%%%%%%%%%%%%%%%%

%********************************************************************************************************
%\subsection{Kinematics of the double-collinear}
%********************************************************************************************************
To study the double collinear limit of scattering amplitudes, the first step consists in identifying the relevant kinematical variables. We describe the momenta of the external particles using the vectors $p^{\mu}_1$ and $p^{\mu}_2$, which refer to the outgoing particles. Here it is important to note that $\mu$ is a Lorentz index which runs over the space-time dimension $D_{\rm ST}=4-2\epsilon$. Since they refer to external particles, we assume that components along the additional dimensions are zero, so $p^{\mu}_1$ and $p^{\mu}_2$ behave like usual four-vectors. The momenta of the incoming particle can be obtained from momentum-conservation rules, $p_{12}^{\mu}=\left(p^{\mu}_1+p^{\mu}_2\right)$, and its virtuality is given by
\beqn
s_{12}&=&p_{12}^{\mu} p_{12}^{\nu} \eta^{\DST}_{\mu \nu} = p_{12}^{\mu} p_{12}^{\nu} \eta^{4}_{\mu\nu} \ .
\eeqn
Since we are defining $\eta^{\DST}_{\mu \nu}$ as a flat-extension of the usual four-dimensional Minkowski metric, inner product with four-vectors behaves like a projection. Then, due to the fact that we are working with massless QCD, $p^2_1=0=p^2_2$.

To describe the collinear limit, we introduce a null-vector $n^{\mu}$ ($n^2=0$) that satisfies $n\cdot p_{12} \neq 0$ and that is the zero-extension of a usual four-vector (we call them with the same name to reduce the notation). Choosing that vector is equivalent to introduce a preferred direction in space-time, which allows us to get rid of unphysical degrees of freedom. In other words, we use $n$ to settle in the light-cone gauge. Working in the light-cone gauge has advantages (for example, we do not have to consider diagrams with ghosts), but it introduces an extra-denominator in loop-integrals which makes them harder to compute. However the most important benefit of choosing a physical gauge is the possibility to exploit collinear factorization properties in an easier way.

Returning to the kinematics of the double collinear limit, we can introduce a collinear null-vector, $\tilde{P}$, which satisfies ${\tilde{P}}^2=0$, $n \cdot \tilde{P} \neq 0$ and $p_{12}^{\mu} \to \tilde{P}^{\mu}$ when $s_{12} \to 0$. This allows us to define the momentum fraction of particle $i$ as
\beqn
z_i &=& \frac{n \cdot p_i}{n \cdot \tilde{P}} \ \ \ i \in \left\{1,2\right\} \, ,
\eeqn
where we have the constraint $z_1+z_2=1$ and therefore we can describe the collinear limit using only the scalar variable $z_1$. (Strictly speaking, we need to use also $s_{12}$ and $n \cdot \tilde{P}$, but since they are dimensionful we can guess their scaling properties and factorize them). We can think about $z_i$ as a measure of the contribution of particle $i$ to the longitudinal total momentum relative to $n$. In other words, $n$ is used here to parametrize the approach to the collinear limit.

On the other hand, it is necessary to take into account the transverse component of the outgoing particles relative to the longitudinal component proportional to $\tilde{P}$. To do this we define $k_{\bot}^{\mu}$, which verifies $n \cdot k_{\bot} = 0 = k_{\bot} \cdot \tilde{P}$. Due to the relations among $n$, $k_{\bot}$ and $\tilde{P}$ we can use them to parametrize the momentum of the outgoing particles \cite{Catani:1999ss} as
\beqn
p^{\mu}_1 &=& z_1 \, \tilde{P}^{\mu} + k_{\bot}^{\mu} - \frac{{k_{\bot}}^2}{2 z_1 n\cdot \tilde{P}} \, n^{\mu} \ \, ,
\label{k1Decomposition}
\\ p^{\mu}_2 &=& (1-z_1) \, \tilde{P}^{\mu} - k_{\bot}^{\mu} - \frac{{k_{\bot}}^2}{2 (1-z_1) n\cdot \tilde{P}} \, n^{\mu} \ \, ,
\label{k2Decomposition}
\eeqn
where $z_i$ is the momentum fraction associated with particle $i$ and ${k_{\bot}}^2=-z_1(1-z_1)s_{12}$. Note that this parametrization is consistent with the fact that both outgoing particles are on-shell and massless. On the other hand, when performing the explicit computation we do not need to express external momenta in terms of $\tilde{P}$, $n$ and $k_{\bot}$: this decomposition is relevant to simplify spinor chains or scalar products that appear in matrix elements. Also, we use $n\cdot p_{12} = n \cdot \tilde{P}$ because
\beqn
p_{12}^{\mu}&=&\tilde{P}^{\mu}+\frac{s_{12}}{2 n\cdot \tilde{P}} \, n^{\mu} \ ,
\label{PDecomposition}
\eeqn
with $\pslashed_{12}u(n)=\slashed{\tilde{P}}u(n)$.

%****************************************************************************************************%
%\subsection{Collinear limit of scattering amplitudes}
%****************************************************************************************************%
After describing collinear kinematics, let's settle some conventions to write scattering amplitudes in the context of massless-QCD with photons. Due to the presence of color charges, we will express matrix elements in color$+$spin space \cite{Catani:1996jh, Catani:1996vz}. A general $n$-particle matrix element can be written as ${\cal M}^{c_1,c_2,\ldots,c_n ;s_1,s_2,\ldots,s_n}_{a_1,a_2,\ldots,a_n}\left(p_1,p_2,\ldots,p_n\right)$, where $\left\{c_1,c_2,\ldots,c_n\right\}$, $\left\{s_1,s_2,\ldots,s_n\right\}$ and $\left\{a_1,a_2,\ldots,a_n\right\}$ are respectively color, spin and flavor indices. Of course, $\left\{p_1,p_2,\ldots,p_n\right\}$ are particle's momenta. To expand color$+$spin space we can introduce an orthonormal basis $\left\{ \left|c_1,c_2,\ldots,c_n\right\rangle \otimes\left|s_1,s_2,\ldots,s_n\right\rangle \right\}$, whose dual basis allows us to express matrix elements as
\beq
{\cal M}^{c_1,\ldots,c_n ;s_1,\ldots,s_n}_{a_1,\ldots,a_n}\left(p_1,\ldots,p_n\right) = \left( \left\langle c_1,\ldots,c_n\right| \otimes\left\langle s_1,\ldots,s_n\right| \right) \left|{\cal M}_{a_1,\ldots,a_n}\left(p_1,\ldots,p_n\right) \right\rangle \,
\eeq
where $\left|{\cal M}_{a_1,\ldots,a_n}\left(p_1,p_2,\ldots,p_n\right) \right\rangle$ is a vector in color$+$spin space. We need to remark that external legs are being considered as on-shell particles (and, moreover, QCD partons are massless). 

\begin{figure}[htb]
	\centering
		\includegraphics[width=0.80\textwidth]{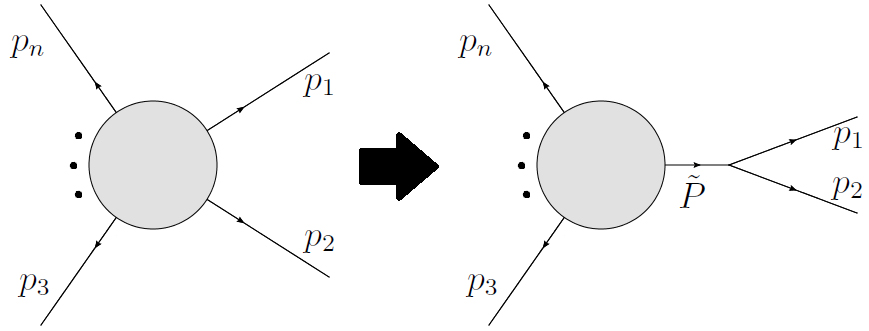}
	\caption{Typical contribution to the most divergent part of an $n$-particle scattering amplitude in the double collinear limit.}
	\label{fig:CollinearSplitting}
\end{figure}

Let's consider an $n$-particle scattering amplitude and assume that two particles, labeled as $1$ and $2$, become collinear. Since we are interested in studying the most divergent part of this kinematic limit, we will only consider diagrams in which $1$ and $2$ come from a parent leg $P$, as shown in Fig. \ref{fig:CollinearSplitting}. It is important to note that, in order to simplify factorization properties, we have to perform the computation in a physical gauge (for example, see Ref. \cite{Pritchard:1978ts}).

Following Fig. \ref{fig:CollinearSplitting}, and using the kinematical variables introduced in the previous section, we can write this contribution as
\beqn
\nonumber {\cal M}^{c_1,c_2\ldots;s_1,s_2\ldots}_{a_1,a_2\ldots}\left(p_1,p_2,\ldots\right) &\approx& - \imath \, \sum_P {\cal A}^{c_{P'},c_1,c_2;s_1,s_2}_{P;a_1,a_2}\left(p_{12},p_1,p_2\right) \, {\rm Prop}(P;{p_{12}})_{c_P c_{P'}}
\\ &\times&  {\cal A}^{c_P,c_3\ldots;s_3\ldots}_{P;a_3\ldots}\left(p_{12},p_3,\ldots\right) \, ,
\label{DescomposicionMatrixElement}
\eeqn
where we have introduced the amputated amplitudes $\cal A$ and a general propagator $\rm Prop$, which depends on the kinematics and the class of particles involved in the process. It is important to note two facts: we are summing over all possible flavors of particle $P$, and $p_{12}$ is the momentum vector associated with the intermediate particle.

Since we are working in massless-QCD, $P$ can be a gluon or a quark. If $P$ is a quark, $a_1$ and $a_2$ have to be a quark and a gluon. On the other hand, if $P$ is a gluon, $a_1$ and $a_2$ can be a quark-antiquark pair or a gluon-gluon pair. It is important to notice that a quark-antiquark pair can become collinear because we are considering them as massless particles, but gluons can always be collinear or soft. 

Let's analyze what happens with each possible choice of $P$. If $P$ is a quark, then its propagator is
\beq
\frac{\imath \delta_{i j} }{\pslashed_{12}} = \frac{\imath \delta_{i j}}{s_{12}}\left(\slashed{\tilde{P}}+\frac{s_{12}}{2 n\cdot\tilde{P}} \slashed{n} \right) = \frac{\imath \delta_{i j} \slashed{\tilde{P}}}{s_{12}} + {\cal O}(s_{12}^0) \, ,
\label{LimiteFermionPropagador}
\eeq
where we used the definition of the light-like vector $\tilde{P}$ and we keep only the most divergent contributions in the limit $s_{12}\to 0$. Here, $\left\{i,j\right\}$ are color indices associated to the fundamental representation of ${\rm SU}(N)$. Since $\tilde{P}$ is a null-vector, it is possible to consider $\tilde{P}$ as the momenta of a massless quark. So, using the completeness relation of massless spinors, we are able to use the expressions
\beqn
\slashed{\tilde{P}}&=&\sum_{\lambda \, {\rm phys.pol.}} \, u_{\lambda}(\tilde{P})\bar{u}_{\lambda}(\tilde{P}) = \sum_{\lambda \, {\rm phys.pol.}} \, v_{\lambda}(\tilde{P})\bar{v}_{\lambda}(\tilde{P}) \, ,
\eeqn
with $\lambda$ being a label for possible physical polarizations of intermediate quark and antiquark, respectively. These considerations leads us to rewrite Eq. \ref{LimiteFermionPropagador} as
\beqn
\frac{\imath \delta_{i j} }{\pslashed_{12}} &=& \sum_{\lambda \, {\rm phys.pol.}} \delta_{i j} \, \frac{\imath u_{\lambda}(\tilde{P})}{s_{12}} \, \bar{u}_{\lambda}(\tilde{P}) \,  + {\cal O}(s_{12}^0) \, .
\eeqn
Now, going back to Eq. \ref{DescomposicionMatrixElement}, we obtain
\beqn
\nonumber {\cal M}^{c_1,c_2\ldots;s_1,s_2\ldots}_{a_1,a_2\ldots}\left(p_1,p_2,\ldots\right) &\approx& \sum_{\lambda \, {\rm phys.pol.}} \, \frac{1}{s_{12}} {\cal A}^{c_P,c_1,c_2;s_1,s_2}_{P;a_1,a_2}\left(p_{12},p_1,p_2\right) u_{\lambda}(\tilde{P}) 
\\ \nonumber &\times& \left(\bar{u}_{\lambda}(\tilde{P}) {\cal A}^{c_P,c_3\ldots;s_3\ldots}_{P;a_3\ldots}\left(p_{12},p_3,\ldots\right) \right)
\\ \nonumber &\equiv&  \sum_{\lambda \, {\rm phys.pol.}} \left(\frac{1}{s_{12}} {\cal A}^{c_P,c_1,c_2;s_1,s_2}_{P;a_1,a_2}\left(p_{12},p_1,p_2\right) u_{\lambda}(\tilde{P}) \right)  
\\ &\times& {\cal M}^{c_P,c_3\ldots;\lambda,s_3\ldots}_{P,a_3\ldots}\left(\tilde{P},p_3,\ldots\right) \, ,
\label{SplittingCasoQ}
\eeqn
where, in the last line, we rearranged the factors to form an $n-1$ matrix element associated with a process which replaces legs $1$ and $2$ with a unique on-shell massless particle $P$. This can be done because we are working in a kinematical region where $1 \parallel 2$, so $s_{12} \to 0$ and we put the divergent factors to the left-side of the propagator. In other words, the replacement $p_{12} \to \tilde{P}$ is possible in ${\cal A}^{c_P,c_3\ldots;s_3\ldots}_{P;a_3\ldots}\left(p_{12},p_3,\ldots\right)$ because it is finite in the collinear limit.

If we consider now the case in which $P$ is a gluon, we have to write the propagator as
\beq
\frac{\imath d_{\mu \nu}(p_{12},n)}{s_{12}} = \frac{\imath }{s_{12}}\left(- \eta_{\mu \nu} + \frac{{p_{12}}_{\mu}n_{\nu}+{p_{12}}_{\nu}n_{\mu}}{n \cdot p_{12}}\right) \, ,
\label{LimiteGluonPropagador1}
\eeq
where $\eta$ is a metric tensor which depends on the number of polarizations of gluons. As done for the quark case, we can use the definition of $\tilde{P}$ and perform the expansion
\beqn
d_{\mu \nu}(p_{12},n) &=& - \eta_{\mu \nu} + \frac{\tilde{P}_{\mu} n_{\nu}+\tilde{P}_{\nu} n_{\mu}}{n \cdot \tilde{P}} + s_{12} \frac{n_{\mu} n_{\nu}}{{(n \cdot \tilde{P})}^2} \approx - \eta_{\mu \nu} + \frac{\tilde{P}_{\mu} n_{\nu}+\tilde{P}_{\nu} n_{\mu}}{n \cdot \tilde{P}} = d_{\mu \nu}(\tilde{P},n) \, ,
\eeqn
and together with the completeness relation
\beq
d_{\mu \nu}(\tilde{P},n) = \sum_{\lambda \, {\rm phys. pol.}} \, \epsilon^{*}_{\mu}(\tilde{P},\lambda) \epsilon_{\nu}(\tilde{P},\lambda) \, ,
\eeq
leads us to the expression
\beqn
\frac{\imath d_{\mu \nu}(p_{12},n)}{s_{12}} &\approx& \sum_{\lambda \, {\rm phys.pol.}} \, \frac{\epsilon_{\mu}(\tilde{P},\lambda)}{s_{12}} \epsilon^{*}_{\nu}(\tilde{P},\lambda) + {\cal O}(s_{12}^0) \, ,
\eeqn
which is a valid approximation in the collinear limit. Applying these results to Eq. \ref{DescomposicionMatrixElement} we get
\beqn
\nonumber {\cal M}^{c_1,c_2\ldots;s_1,s_2\ldots}_{a_1,a_2\ldots}\left(p_1,p_2,\ldots\right) &\approx& \sum_{\lambda \, {\rm phys.pol.}} \, \frac{1}{s_{12}} {\cal A}^{c_P,c_1,c_2;s_1,s_2;\mu}_{P;a_1,a_2}\left(p_{12},p_1,p_2\right) \epsilon_{\mu}(\tilde{P},\lambda) 
\\ \nonumber &\times& \left(\epsilon^{*}_{\nu}(\tilde{P},\lambda) {\cal A}^{c_P,c_3\ldots;s_3\ldots;\nu}_{P;a_3\ldots}\left(p_{12},p_3,\ldots\right) \right)
\\ \nonumber &\equiv&  \sum_{\lambda \, {\rm phys.pol.}} \left(\frac{1}{s_{12}} {\cal A}^{c_P,c_1,c_2;s_1,s_2;\mu}_{P;a_1,a_2}\left(p_{12},p_1,p_2\right) \epsilon_{\mu}(\tilde{P},\lambda) \right)  
\\ &\times& {\cal M}^{c_P,c_3\ldots;\lambda,s_3\ldots}_{P,a_3\ldots}\left(\tilde{P},p_3,\ldots\right) \, ,
\label{SplittingCasoG}
\eeqn
where, again, we are able to rearrange the expression in such a way that the first factor contains all the divergent contributions and the second one is a reduced-matrix element for a $n-1$-particle process.

From Eqs. \ref{SplittingCasoQ} and \ref{SplittingCasoG}, we can motivate the definition of splitting matrices and amplitudes. Working in the double-collinear limit, the quark initiated splitting matrix can be written as
\beqn
\Spmatrix_{q\to a_1 a_2} &=& \frac{1}{s_{12}} \left|{\cal A}_{q,a_1,a_2}\left(p_{12},p_1,p_2\right) \right\rangle u(\tilde{P}) \, ,
\eeqn
with $a_1$ and $a_2$ being a gluon and a quark, respectively. When the parent particle is a gluon, we get
\beqn
\Spmatrix_{g\to a_1 a_2} &=& \frac{1}{s_{12}} \left|{\cal A}^{\mu}_{g,a_1,a_2}\left(p_{12},p_1,p_2\right) \right\rangle  \epsilon_{\mu}(\tilde{P}) \, ,
\eeqn
being $a_1$ and $a_2$ a quark-antiquark or a gluon pair. In both cases, $\left|{\cal A}_{P,a_1,a_2}\left(p_{12},p_1,p_2\right) \right\rangle$ is the amputated scattering amplitude associated to the process $P \to a_1 a_2$, without being projected over the color$+$spin space. If we project $\Spmatrix$ over color$+$spin space we get the so-called splitting amplitudes. To recover splitting functions (as defined in Ref. \cite{Kosower:1999rx}), we just have to remove color information from splitting amplitudes.

Now it is important to note that we left $p_{12}$ as incoming momenta in the amputated amplitude, instead of using $\tilde{P}$. This is related to the fact that the presence of divergences in the definition of splitting matrices forces us to regularize them and keep the $s_{12}$ dependence explicitly. For that reason we must consider that the incoming particle is slightly off-shell and include all possible Feynman diagrams, also those which include self-energy corrections to the parent leg. We will emphasize this fact when computing explicitly some scattering amplitudes at NLO. 

Finally, we have to remark that it is possible to get the divergent contribution to the splitting matrices at NLO without performing a full computation. For the double-collinear limit, according to Ref. \cite{Catani:2011st} we can write
\beqn
\Spmatrix^{(1)}\left(p_1,p_2;\tilde{P}\right) &=& \Spmatrix^{(1)}_{H} \left(p_1,p_2;\tilde{P}\right) + \IC_{C} \left(p_1,p_2;\tilde{P} \right) \Spmatrix^{(0)} \left(p_1,p_2;\tilde{P}\right) \, ,
\label{FormulaCatani1}
\eeqn
with $\Spmatrix^{(1)}_{H}$ containing only the rational dependence on the momenta $p_1$, $p_2$ and $\tilde{P}$, and
\beqn
\nonumber \IC_{C} \left(p_1,p_2;\tilde{P} \right) &=& \CG g_s^{2} \left(\frac{-s_{12}-\imath 0}{\mu^2} \right)^{-\epsilon}
\\ \nonumber &\times& \left\{ \frac{1}{\epsilon^2}\left(C_{12}-C_1-C_2 \right) + \frac{1}{\epsilon} \left(\gamma_{12}-\gamma_1-\gamma_2 + b_0 \right) \right.
\\ &-& \left. \frac{1}{\epsilon} \left[\left(C_{12}+C_1-C_2\right)f(\epsilon,z_1)+\left(C_{12}+C_2-C_1\right)f(\epsilon,1-z_1) \right] \right\} \, ,
\label{FormulaCatani2}
\eeqn
which contains all the divergent contributions and non-rational functions of $z_1$. Here $C_i$ are the Casimir factors associated with the parton $a_i$ ($C_i=C_A$ for gluons and $C_i=C_F$ for quarks), $\gamma_i$ depend on the flavor of $a_i$ and $c_{\Gamma}$ is the $D$-dimensional volume factor associated with one-loop integrals, i.e.
\beqn
\CG &=& \frac{\Gamma(1+\epsilon)\Gamma(1-\epsilon)^2}{(4\pi )^{2-\epsilon }\Gamma(1-2\epsilon)} \, .
\eeqn
If $N_f$ is the number of quark flavors, then
\beq
\gamma_q=\gamma_{\bar{q}}=\frac{3}{2}C_F \ \ , \ \ \gamma_g=\frac{11\,C_A-2\,N_f}{6} \, ,
\eeq
and $b_0=\gamma_g$ is the first perturbative coefficient of the QCD $\beta$ function, according to our normalization. Besides that, the function $f(\epsilon,z)$ is given by
\beqn
f(\epsilon,z)&=&\frac{1}{\epsilon}\left( _2 F_1(1,-\epsilon,1-\epsilon,1-z^{-1})-1 \right) \, ,
\eeqn
and it is associated with the kinematical behavior of matrix elements in the collinear limit.

%%%%%%%%%%%%%%%%%%%%%%%%%%%%%%%%%%%%%%%%%%%%%%%%%%%%%%%%%%%%%%%%
\section{The $q\to q g$ splitting matrix}
%%%%%%%%%%%%%%%%%%%%%%%%%%%%%%%%%%%%%%%%%%%%%%%%%%%%%%%%%%%%%%%%

When working in the LCG, the presence of internal gluons makes more difficult to perform an explicit computation. So, we start with the $q \to g q$ splitting and we explain the differences among schemes. At NLO we can write the corresponding splitting matrix as
\beq
\Spmatrix_{q \to g q} = \Spmatrix^{(0)}_{q \to g q} + \Spmatrix^{(1)}_{q \to g q} \ ,
\eeq
where the LO contribution is
\beqn
\Spmatrix^{(0)}_{q \to g q} &=& \frac{g_s \mu^{\epsilon}}{s_{12}} \SUNT^a \bar{u}(p_2)\slashed{\epsilon}{\left(p_1\right)}u(\tilde{P}) \ .
\eeqn
Even at LO, we can decompose $\gamma^{\mu}=\tilde{\gamma}^{\mu}+\hat{\gamma}^{\mu}$ when considering $\DDirac=4-2\epsilon$. This leads to the expression
\beqn
\Spmatrix^{(0)}_{q \to g q} &=& \frac{g_s \mu^{\epsilon}}{s_{12}} \SUNT^a \left[\bar{u}(p_2)\tilde{\gamma}^{\mu}u(\tilde{P})+\bar{u}(p_2)\hat{\gamma}^{\mu}u(\tilde{P}) \right] \epsilon_{\mu}(p_1)\, ,
\label{SplittingLOqgqSEPARACION}
\eeqn
which includes an helicity-violating term that contributes only in CDR or HSA schemes. However, since gluons are treated as $D$-dimensional vectors in CDR, it is not required to separate explicitly the helicity-violating term. The situation is going to be different in HSA scheme because the presence of both $4$ and $\DST$-dimensional metrics leads to a non-equal mixing between $\bar{u}(p_2){\gamma}^{\mu}u(\tilde{P})$ and $\bar{u}(p_2)\tilde{\gamma}^{\mu}u(\tilde{P})$.

\begin{figure}[htb]
	\centering
		\includegraphics[width=0.60\textwidth]{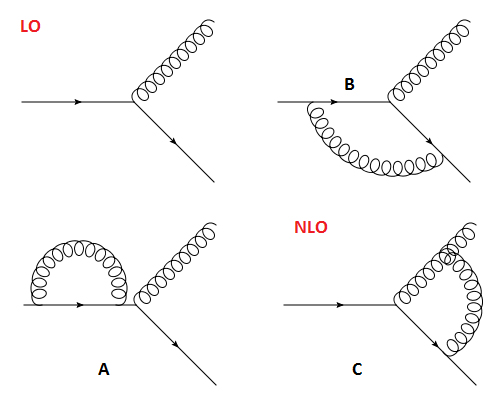}
	\caption{Feynman diagrams associated with $q(\tilde{P}) \to g(p_1) q(p_2)$ at NLO, including the self-energy correction to the parent parton. We show all the standard QCD contributions up to ${\cal O}(g_s^3)$.}
	\label{fig:NLODiagrams1}
\end{figure}

It is important to appreciate that we are starting from the amputated amplitude related with $q(p_{12}) \to g(p_1) q (p_2)$. This explains why we must take into account self-energy corrections to the incoming particle. In other words, to calculate the NLO corrections to the splitting matrix $\Spmatrix_{q \to g q}$ we need to include all the diagrams shown in Fig. \ref{fig:NLODiagrams1}. However, it is necessary to take into account other kind of contributions to explore consistently the different schemes. As we mentioned in Section 2, when we treat QCD in the context of DREG, it is possible to decompose $\DST$-dimensional gluons into $4$-dimensional vectors and scalar particles. Due to the fact that we can make that differentiation when drawing Feynman diagrams, it is useful to introduce the following classification of diagrams:
\begin{itemize}

	\item \textit{standard QCD contributions} (STD);
	\item \textit{helicity preserving interactions mediated by scalar gluons} (SCA-nHV);
	\item and \textit{helicity-violating interactions} (SCA-HV).

\end{itemize}

To compute STD contributions, we start from $4$-dimensional QCD and draw the associated Feynman diagrams, using only gluons and quarks to do this. Conversely, SCA contributions include scalar-gluons as internal or external particles. SCA-nHV only allows the presence of internal gluons with the additional requirement that external particles do not violate helicity conservation. To be more explicit, let's center in the $q \to g q$ process. In $4$-dimensional QCD, incoming and outgoing quarks have the same helicity because quark-gluon interaction is represented by a vector-like vertex. So, SCA-nHV only takes into account that kind of diagrams. Instead of that, SCA-HV contributions only allow helicity configurations that are forbidden by usual $4$-dimensional QCD interactions.

Let's start describing the standard NLO QCD contribution. It can be expressed as
\beqn
\Spmatrix^{(1,{\rm STD})}_{q \to g q} &=& \Spmatrix^{(1,A)}_{q \to g q} + \Spmatrix^{(1,B)}_{q \to g q}+\Spmatrix^{(1,C)}_{q \to g q} \, ,
\eeqn
where $\Spmatrix^{(1,i)}_{q \to g q}$ refers to the diagram $i\in \left\{A,B,C \right\}$, as shown in Fig. \ref{fig:NLODiagrams1}. Writing each contribution we have,
\beqn
\nonumber \Spmatrix^{(1,A)}_{q \to g q} &=& -\frac{g^3_s \mu^{3 \epsilon} C_F}{s_{12}^2} \, \SUNT^a \,  \bar{u}(p_2)\slashed{\epsilon}{\left(p_1\right)}\pslashed_{12}\gamma^{\nu}\gamma^{\alpha}\gamma^{\rho}u(\tilde{P})
\\ &\times&  \int_q \, \frac{{\left(p_{12}-q\right)}_{\alpha} d_{\rho \nu}\left(q,n\right)}{q^2 t_{12q}} \, ,
\\ \nonumber \Spmatrix^{(1,B)}_{q \to g q} &=& \frac{g^3_s \mu^{3 \epsilon}(C_A-2C_F)}{2s_{12}} \, \SUNT^a \, \bar{u}(p_2)\gamma^{\rho}\gamma^{\alpha}\slashed{\epsilon}{\left(p_1\right)}\gamma^{\beta}\gamma^{\nu}u(\tilde{P})
\\ &\times& \int_q \, \frac{{\left(p_{12}-q\right)}_{\beta} {\left(p_2-q\right)}_{\alpha} d_{\rho \nu}\left(q,n\right)}{q^2 t_{2q} t_{12q}} \, ,
\\ \nonumber \Spmatrix^{(1,C)}_{q \to g q} &=& - \frac{g^3_s \mu^{3 \epsilon}C_A}{2s_{12}} \, \SUNT^a \, \epsilon_{\mu}{\left(p_1\right)} \bar{u}(p_2)\gamma^{\nu}\gamma^{\alpha}\gamma^{\beta}u(\tilde{P})
\\ &\times& \int_q \, \frac{{\left(p_2-q\right)}_{\alpha} V^{\rm Cin}_{3g}\left(p_1,q,-p_1-q;\mu,\nu_1,\mu_1\right) d_{\beta \mu_1}\left(p_1+q,n\right) d_{\nu \nu_1}\left(q,n\right)}{q^2 s_{1q} t_{2q}} \, ,
\eeqn
where we are not making any distinction between $4$ and $\DST$-dimensional gluons. In particular, in the context of HSA scheme, we should interpret 
\beqn
\slashed{\epsilon}(p_1) &=& \tilde{\gamma}^{\mu}\epsilon_{\mu}(p_1) + \hat{\gamma}^{\mu} \hat{\epsilon}_{\mu}(p_1) \, ,
\eeqn
because there are $2-2\epsilon$ gluon's degrees of freedom but vector-gluons have only two polarizations (setting $\alpha=0$ and $\alpha_R=0$) while the remaining polarizations are associated with scalar-gluons. Also, it is useful to note that $\Spmatrix^{(1,A)}_{q \to g q}$ can be rewritten as
\beqn
\Spmatrix^{(1,A)}_{q \to g q} &=& \Sigma(p^2_{12}) \,  \Spmatrix^{(0)}_{q \to g q} \, ,
\label{SplittingqgqSELFENERGY}
\eeqn
where $\Sigma(p^2_{12})$ is the NLO correction to quark self-energy. Except for this diagram, all the others correspond to the ones that appear when computing $q\to g q$ with massless on-shell particles.

\begin{figure}[htb]
	\centering
		\includegraphics[width=0.80\textwidth]{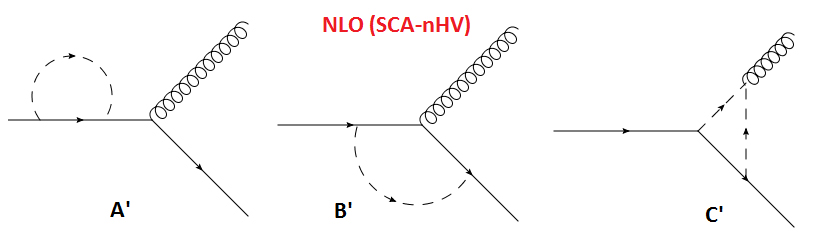}
	\caption{Feynman diagrams associated with the scalar-gluon contributions to $q(\tilde{P}) \to g(p_1) q(p_2)$ at NLO. We show only SCA-nHV configurations.}
	\label{fig:NLODiagramsScalar1}
\end{figure}

Now let's turn to the scalar-gluon contributions to the splitting matrix. We only consider diagrams associated with helicity configurations that are allowed by $4$-dimensional QCD interactions. Since scalar contributions are related with the fermion-gluon-fermion and triple-gluon vertices, and we have to keep external physical particles only, diagrams containing internal scalar-gluons start at NLO. As we can see in Fig. \ref{fig:NLODiagramsScalar1}, we have corrections to the three standard QCD diagrams (see Fig. \ref{fig:NLODiagrams1}). Explicitly, the associated contributions are
\beqn
\Spmatrix^{(1,{\rm SCA-nHV})}_{q \to g q}  &=& \Spmatrix^{(1,A')}_{q \to g q}  + \Spmatrix^{(1,B')}_{q \to g q}+\Spmatrix^{(1,C')}_{q \to g q} \, ,
\eeqn
with
\beqn
\nonumber \Spmatrix^{(1,A')}_{q \to g q} &=& -\frac{g_s {\left(g^{\rm scalar}_s\right)}^2 \mu^{3 \epsilon}C_F}{s_{12}^2} \, \SUNT^a \,  \bar{u}(p_2)\slashed{\epsilon}{\left(p_1\right)}\pslashed_{12}\hat{\gamma}^{\nu}\gamma^{\alpha}\hat{\gamma}^{\rho}.u(\tilde{P})
\\ &\times&  \int_q \, \frac{{\left(p_{12}-q\right)}_{\alpha} \left(-{\eta^{\epsilon}_{\nu \rho}} \right)}{q^2 t_{12q}} \, ,
\\ \nonumber \Spmatrix^{(1,B')}_{q \to g q} &=& \frac{g_s {\left(g^{\rm scalar}_s\right)}^2 \mu^{3 \epsilon}(C_A-2C_F)}{2s_{12}} \, \SUNT^a \, \bar{u}(p_2)\hat{\gamma}^{\rho}\gamma^{\alpha}\slashed{\epsilon}{\left(p_1\right)}\gamma^{\beta}\hat{\gamma}^{\nu}u(\tilde{P})
\\ &\times& \int_q \, \frac{{\left(p_{12}-q\right)}_{\beta} {\left(p_2-q\right)}_{\alpha} \left(- \eta^{\epsilon}_{\nu \rho}\right)}{q^2 t_{2q} t_{12q}} \, ,
\\ \nonumber \Spmatrix^{(1,C')}_{q \to g q} &=& - \frac{{\left(g^{\rm scalar}_s\right)}^3 \mu^{3 \epsilon}C_A}{2s_{12}} \, \SUNT^a \, \epsilon_{\mu}{\left(p_1\right)} \bar{u}(p_2)\hat{\gamma}^{\nu}\gamma^{\alpha}\hat{\gamma}^{\beta}u(\tilde{P}) 
\\ &\times& \int_q \, \frac{{\left(p_2-q\right)}_{\alpha} V^{\rm Cin}_{3g}\left(p_1,q,-p_1-q;\mu,\nu_1,\mu_1\right) \eta^{\epsilon}_{\nu \nu_1} \eta^{\epsilon}_{\beta \mu_1}}{q^2 s_{1q} t_{2q}} \, ,
\eeqn
where we used the Feynman's rules previously obtained at Lagrangian level (see Section 2). Now let's simplify this expressions using some properties of Dirac matrices in $D$-dimensions: we want to show explicitly that it is possible to use the effective Feynman rules introduced in the end of Section 2. Focusing in $\Spmatrix^{(1,A')}_{q \to g q}$, note that this contribution depend on
\beqn
\nonumber {\rm Int}^{A'} &=& \int_{q} \, \frac{\eta^{\epsilon}_{\nu \rho} \, \hat{\gamma}^{\nu}\left(\slashed{p}_{12}-\slashed{q}\right)\hat{\gamma}^{\rho}}{q^2 t_{12q}} \, ,
\\ &=& \int_{q}  \frac{1}{q^2 t_{12q}} \, \hat{\gamma}_{\rho}\slashed{p}_{12}\hat{\gamma}^{\rho} + \int_{q}  \frac{q_{\alpha}}{q^2 t_{12q}} \, \hat{\gamma}_{\rho}\gamma^{\alpha}\hat{\gamma}^{\rho} \, .
\label{IntegralAprime2}
\eeqn
However, since $p_{12}$ is a physical momenta then $\slashed{p}_{12}=(p_{12})_{\sigma} \tilde{\gamma}^{\sigma}$. Using that $\left\{\tilde{\gamma}^{\alpha},\hat{\gamma}^{\sigma}\right\}=0$ and $\hat{\gamma}^{\rho}\hat{\gamma}_{\rho}=-2\epsilon {\rm Id}$, then we have
\beqn
{\rm Int}^{A'} &=&  \int_{q}  \frac{1}{q^2 t_{12q}} \, \left(2\epsilon \, \slashed{p}_{12}\right) + \left(A(s_{12}) \, (p_{12})_{\alpha}\right) \, \left(\hat{\gamma}^{\rho}{\gamma}^{\alpha}\hat{\gamma}^{\rho}\right)  \, ,
\eeqn
where we used Passarino-Veltman (PV) decomposition to write the vector-type integral in the second term. Due to the fact that vector-type integrals only depend on physical vectors then we can repeat the procedure performed in the first term and we obtain
\beqn
\nonumber {\rm Int}^{A'} &=&  \int_{q}  \frac{1}{q^2 t_{12q}} \, \left(2\epsilon \, \slashed{p}_{12}\right) + \int_{q}  \frac{q_{\alpha}}{q^2 t_{12q}} \, \left(2 \epsilon \tilde{\gamma}^{\alpha}\right)
\\ &=& 2\epsilon \, \int_{q} \, \frac{ \, \left(\slashed{p}_{12}-\slashed{q}\right)}{q^2 t_{12q}} \, ,
\eeqn
which is equivalent to use the effective scalar rules discussed in the last part of Section 2. Note that we have not used the fact that $\DDirac=4$, which implies this result immediately. 

The situation is analogous when we move to $\Spmatrix_{q\to gq}^{(1,C')}$, but some subtleties appear when treating $\Spmatrix_{q\to gq}^{(1,B')}$. That contribution depends on
\beqn
\nonumber {\rm Int}^{B'} &=& \int_{q} \, \frac{\hat{\gamma}^{\rho}(\pslashed_{2}-\slashed{q})\gamma^{\mu}(\pslashed_{12}-\slashed{q})\hat{\gamma}^{\nu} \eta^{\epsilon}_{\nu\rho}}{q^2 t_{2q} t_{12q}} 
\\ \nonumber &=& \int_{q} \, \frac{1}{q^2 t_{2q} t_{12q}} \, \hat{\gamma}^{\rho}\pslashed_{2}\gamma^{\mu}\pslashed_{12}\hat{\gamma}_{\rho} - \int_{q} \, \frac{q_{\alpha}}{q^2 t_{2q} t_{12q}} \, \hat{\gamma}^{\rho}\left(\pslashed_{2}\gamma^{\mu}\gamma^{\alpha}+\gamma^{\alpha}\gamma^{\mu}\pslashed_{12}\right)\hat{\gamma}_{\rho}
\\ &+& \int_{q} \, \frac{q_{\alpha}q_{\beta}}{q^2 t_{2q} t_{12q}} \, \hat{\gamma}^{\rho}\gamma^{\alpha}\gamma^{\mu}\gamma^{\beta}\hat{\gamma}_{\rho} \, ,
\label{INTBprime1}
\eeqn
where $\left\{\rho,\nu \right\}$ run over the non-physical dimensions. Depending on the number of external gluon polarizations, $\mu$ can live in $4$ ($n_g=2$) or in $\DST$-dimensions ($n_g=2-2\epsilon$). Using PV decomposition, the tensor-type integrals can be expanded as
\beqn
\int_{q} \, \frac{q_{\alpha}}{q^2 t_{2q} t_{12q}} &=& \sum_{i} A_{i}(s_{12}) (p_{i})_{\alpha} \, ,
\\ \int_{q} \, \frac{q_{\alpha}q_{\beta}}{q^2 t_{2q} t_{12q}} &=& \sum_{i,j} A_{ij}(s_{12}) (p_{i})_{\alpha}(p_{j})_{\beta} \, + B(s_{12}) \eta^{\DST}_{\alpha\beta} \, ,
\eeqn
with the inclusion of a term proportional to the $\DST$-dimensional metric tensor. Replacing these expansions in Eq. \ref{INTBprime1} and using that $p_{12}$ and $p_2$ are 4-vectors, we obtain
\beqn
\nonumber {\rm Int}^{B'} &=& A_{0}(s_{12}) \pslashed_{2}\hat{\gamma}^{\rho}\gamma^{\mu}\hat{\gamma}_{\rho}\pslashed_{12} -  \sum_{i} A_{i}(s_{12}) \, \left(\pslashed_{2}\hat{\gamma}^{\rho}\gamma^{\mu}\hat{\gamma}_{\rho}\pslashed_{i}+\pslashed_{i}\hat{\gamma}^{\rho}\gamma^{\mu}\hat{\gamma}_{\rho}\pslashed_{12}\right)
\\ \nonumber &+& \sum_{i,j} A_{ij}(s_{12}) \, \pslashed_{i}\hat{\gamma}^{\rho}\gamma^{\mu}\hat{\gamma}_{\rho}\pslashed_{j} \, + B(s_{12}) \, \hat{\gamma}^{\rho}\gamma^{\alpha}\gamma^{\mu}\gamma_{\alpha}\hat{\gamma}_{\rho} \, ,
\label{INTBprime2}
\eeqn
since $\hat{\gamma}^{\rho}\pslashed_{i}=-\pslashed_{i}\hat{\gamma}^{\rho}$ and $A_{ij}=A_{ji}$ due to symmetry properties. On the other hand,
\beqn
\gamma^{\alpha}\gamma^{\mu}\gamma_{\alpha} &=& (2-\DDirac)\gamma^{\mu} \, ,
\\ \hat{\gamma}^{\rho}\hat{\gamma}_{\rho} &=& (\DST-4) {\rm Id} \, ,
\\ \hat{\gamma}^{\rho}\gamma^{\mu}\hat{\gamma}_{\rho} &=& (4-\DST) \gamma^{\mu} + 2 \hat{\gamma}^{\mu} \, ,
\eeqn
where we used that Dirac's algebra dimension is $\DDirac$ and it is equal to the number of gamma matrices available. So, after applying these properties and the fact that
\beqn
\hat{\gamma}^{\rho}\gamma^{\alpha}\gamma^{\mu}\gamma_{\alpha}\hat{\gamma}_{\rho} &=& \gamma^{\alpha}\hat{\gamma}^{\rho}\gamma^{\mu}\hat{\gamma}_{\rho}\gamma_{\alpha}+2 \left(\gamma^{\mu}\hat{\gamma}^{\rho}\hat{\gamma}_{\rho}- \hat{\gamma}^{\rho}\hat{\gamma}_{\rho}\gamma^{\mu} \right) = \gamma^{\alpha}\hat{\gamma}^{\rho}\gamma^{\mu}\hat{\gamma}_{\rho}\gamma_{\alpha} \, ,
\eeqn
we can rewrite ${\rm Int}^{B'}$ as
\beqn
\nonumber {\rm Int}^{B'} &=& (4-\DST) \left[A_{0}(s_{12}) \pslashed_{2}\gamma^{\mu}\pslashed_{12} -  \sum_{i} A_{i}(s_{12}) \, \left(\pslashed_{2}\gamma^{\mu}\pslashed_{i}+\pslashed_{i}\gamma^{\mu}\pslashed_{12}\right) + \sum_{i,j} A_{ij}(s_{12}) \, \pslashed_{i}\gamma^{\mu}\pslashed_{j} \right. 
\\ \nonumber &+& \left. B(s_{12}) \, \gamma^{\alpha}\gamma^{\mu}\gamma_{\alpha}\right] +2 \left[A_{0}(s_{12}) \pslashed_{2}\hat{\gamma}^{\mu}\pslashed_{12} -  \sum_{i} A_{i}(s_{12}) \, \left(\pslashed_{2}\hat{\gamma}^{\mu}\pslashed_{i}+\pslashed_{i}\hat{\gamma}^{\mu}\pslashed_{12}\right) \right.
\\ &+& \left. \sum_{i,j} A_{ij}(s_{12}) \, \pslashed_{i}\hat{\gamma}^{\mu}\pslashed_{j} \, + B(s_{12}) \, \gamma^{\alpha}\hat{\gamma}^{\mu}\gamma_{\alpha}\right] \, ,
\label{INTBprime3}
\eeqn
where we can always express $4-\DST=\eta^{\epsilon}_{\rho\nu}(-(\eta^{\epsilon})^{\rho\nu})$. Note that the metric tensor inside the parenthesis comes from commuting and symmetrizing the product $\hat{\gamma}^{\rho}\hat{\gamma}^{\nu}$. Also, the contributions involving $\hat{\gamma}^{\mu}$ violate helicity conservation, so they vanish when we restrict external particles to have helicity configurations compatible with standard QCD interactions. 

Summarizing these observations, we conclude that the replacement $-\eta^{\epsilon}_{\rho\nu} \hat{\gamma}^{\rho}\gamma^{\alpha}\gamma^{\mu}\gamma^{\beta}\hat{\gamma}^{\nu} \to (-2\epsilon) \gamma^{\alpha}\gamma^{\mu}\gamma^{\beta}$ is valid. Thus we get an effective Feynman rule for scalar-gluons interaction with fermions, which consists in considering them as scalar particles with propagator $\frac{-2 \imath \epsilon}{p^2+\imath 0}$ (see Eq. \ref{PropagadorGluonEscalar}) and remove the corresponding Dirac matrix in the vertex. On the other hand it is useful to remember that in usual DREG schemes, if scalar-gluons are introduced then we have to set $D_{\rm Dirac}=4$. But this limit has to be taken after replacing integrals. In other words, it is possible that some new terms (i.e. not present in the expressions when using effective Feynman rules for scalar-gluon) survive when applying directly Lagrangian level Feynman rules. But this terms are always proportional to integrals which vanish in the limit $\DDirac \to 4$. This situation occurs in $\Spmatrix_{q\to gq}^{(1,B')}$ because there is a term proportional to $\gamma^{\alpha}\gamma^{\mu}\gamma_{\alpha}=-2(1-\epsilon)\gamma^{\mu}$ (see Eq. \ref{INTBprime3}).

So, after this discussion, we can rewrite the SCA-nHV contributions as
\beqn
\Spmatrix^{(1,A')}_{q \to g q} &=& \frac{2 \, g^3_s \mu^{3 \epsilon} \epsilon C_F}{s_{12}^2} \, \SUNT^a \,  \bar{u}(p_2)\slashed{\epsilon}{\left(p_1\right)}\pslashed_{12}\gamma^{\alpha}u(\tilde{P}) \, \int_q \, \frac{{\left(p_{12}-q\right)}_{\alpha}}{q^2 t_{12q}} \, ,
\\ \Spmatrix^{(1,B')}_{q \to g q} &=& \frac{g^3_s  \mu^{3 \epsilon} \epsilon (2C_F-C_A)}{s_{12}} \, \SUNT^a \, \bar{u}(p_2)\gamma^{\alpha}\slashed{\epsilon}{\left(p_1\right)}\gamma^{\beta}u(\tilde{P}) \, \int_q \, \frac{{\left(p_{12}-q\right)}_{\beta} {\left(p_2-q\right)}_{\alpha}}{q^2 t_{2q} t_{12q}} \, ,
\\ \Spmatrix^{(1,C')}_{q \to g q} &=& \frac{g^3_s  \mu^{3 \epsilon} \epsilon C_A}{s_{12}} \, \SUNT^a \, \epsilon_{\mu}{\left(p_1\right)} \bar{u}(p_2)\gamma^{\alpha}u(\tilde{P}) \, \int_q \, \frac{{\left(p_2-q\right)}_{\alpha} {\left(2 q+p_1 \right)}^{\mu}}{q^2 s_{1q} t_{2q}} \, ,
\eeqn
where we used the effective Feynman rules for scalar-gluons setting $D_{\rm Dirac}=4$.

Finally, we want to make a brief comment about SCA-HV components. When working in HSA/HSB schemes it is possible that STD contributions mix helicity-preserving and helicity-violating terms, whose origin is the contraction of $4$-dimensional metric tensors (coming from the gluon propagator) with $\DST$-dimensional structures. We discuss this point in the next subsections, using the results for $\Spmatrix_{q\to g q}$ to give an explicit example.

%*******************************************************************************
\subsection{Amplitude level results}
%*******************************************************************************
Following with the study of $q \to gq$ splitting amplitude, we performed the computation without specifying the polarization of the involved particles. This implies having larger spinorial structures and more complex tensor-type integrals, but this will allow us to compute contributions to the NLO Altarelli-Parisi kernel in an easier way. 

Let's start with the NLO standard-QCD contribution to the splitting matrix. After writing explicitly the corresponding Feynman diagrams and replacing the involved loop-integrals, we find that
\beqn
\nonumber \Spmatrix^{(1,{\rm STD})}_{q \to g q} &=& \frac{\CG g_s^3 \mu^{\epsilon}}{2 s_{12} \epsilon^2} \, {\left(\frac{-s_{12}-\imath 0}{\mu^2}\right)}^{-\epsilon} \SUNT^a \left[\vphantom{\frac{1}{nP}} C^{({\rm STD},1)}_{q \to gq} \, \bar{u}(p_2)\slashed{\epsilon}(p_1)u(\tilde{P}) \right.
\\  &+& \left. C^{({\rm STD},2)}_{q \to gq} \, \frac{1}{nP}\bar{u}(p_2)\slashed{n}u(\tilde{P}) p_2\cdot \epsilon(p_1) + \vphantom{\frac{1}{nP}} \delta_{\alpha,1} \, C^{({\rm STD},3)}_{q \to gq} \, \bar{u}(p_2)\hat{\gamma}^{\mu}u(\tilde{P}) \hat{\epsilon}_{\mu}(p_1) \right] \, ,
\label{SplittingGeneralqgq}
\eeqn
where the coefficients $C^{({\rm STD},i)}_{q \to gq}$ are given by
\beqn
\nonumber C^{({\rm STD},1)}_{q \to gq} &=&  2(C_A-2C_F) {}_2 F_1 \left(1,-\epsilon,1-\epsilon,\frac{z_1}{z_1-1}\right) -2 C_A {}_2 F_1 \left(1,-\epsilon,1-\epsilon,\frac{z_1-1}{z_1}\right)
\\ \nonumber &-& 2 \frac{C_A \left(\epsilon(\delta\epsilon^2+\epsilon-3)+1\right)-C_F\left(\delta \epsilon^3+3\epsilon^2-6\epsilon+2\right)}{(\epsilon-1)(2 \epsilon-1)}
\\ &+&  (1-\alpha_R) \delta \epsilon^2 \frac{C_A \left(2\epsilon+1+\alpha_R\right)-2C_F\epsilon}{(\epsilon-1)(2\epsilon-1)}\, ,
\\ \nonumber C^{({\rm STD},2)}_{q \to gq} &=&  \frac{2\epsilon^2 (C_A-C_F)(\delta \epsilon-1)}{(\epsilon-1)(2\epsilon-1)} + \frac{\delta (1-\alpha_R) \epsilon}{2(1-z_1)^2(\epsilon-1)(2\epsilon-1)} \left[ \vphantom{\frac{a}{b}} 2(1-z_1)^2 \epsilon^2 \left(2C_F-C_A(\alpha_R+2)\right) \right.
\\ \nonumber &+& \left. C_A(1-z_1)^2\epsilon \, {}_2 F_1 \left(1,1-\epsilon,2-2\epsilon,\frac{1}{z_1}\right) +C_A z_1 (\epsilon-1) {}_2 F_1 \left(1,-\epsilon,1-\epsilon,\frac{z_1-1}{z_1}\right) \right.
\\ &+& \left. C_A \left((z^2_1-4 z_1+2)\epsilon + z_1 \right) \vphantom{\frac{a}{b}}\right]\, ,
\\ \nonumber C^{({\rm STD},3)}_{q \to gq} &=& 2 (1-\alpha_R) C_A \left[{}_2 F_1 \left(1,-\epsilon,1-\epsilon,\frac{z_1-1}{z_1}\right)+\frac{(1-z_1)\epsilon}{z_1(2\epsilon-1)^2}{}_2 F_1\left(1,1-\epsilon,2-2\epsilon,\frac{1}{z_1}\right)\right]
\\ &+&  \frac{(1-\alpha_R) \left[2 C_F(1-2\epsilon)\epsilon-C_A\left(\epsilon((1+\epsilon\delta)(1-\alpha_R)-6\epsilon+7)-4\right)\right]}{(\epsilon-1)(2\epsilon-1)}\, ,
\eeqn
where $\delta$ controls Dirac's algebra dimension and we left $\alpha_R$ as a free parameter. Note that there is a term that explicitly involves an helicity-violating interaction. It is proportional to $1-\alpha_R$ and only contributes when we work in HSA scheme ($\alpha=1$) because external gluons must have $2-2\epsilon$ polarizations in order to allow for this kind of interactions. Also, it is worth noting that modifying $\alpha_R$ only introduces ${\cal O}(\epsilon^2)$ differences in coefficients $C^{({\rm STD},1)}_{q \to gq}$ and $C^{({\rm STD},2)}_{q \to gq}$. However, expanding $C^{({\rm STD},3)}_{q \to gq}$ we find
\beq
C^{({\rm STD},3)}_{q \to gq} = 6 (1-\alpha_R) C_A + {\cal O}(\epsilon) \, ,
\eeq
which implies that $\Spmatrix^{(1,{\rm STD})}_{q \to g q}$ acquires an additional contribution to the double $\epsilon$ pole which is proportional to $\delta_{\alpha 1}(1-\alpha_R)$.

If we want to check our calculations, we can set $\alpha_R=1$ to recover well known results in FDH ($\delta=0$) and CDR/HV schemes ($\delta=1$). In particular, when using CDR we have to assume that $\epsilon_{\mu}(p_1)$ is a $\DST$-dimensional vector, while in the remaining schemes $\epsilon_{\mu}(p_1)$ is associated to a $4$-dimensional space. It is important to appreciate that we used the properties $\epsilon(p_1) \cdot n =0$ (related to the definition of the null-vector $n$) and $\epsilon(p_1) \cdot p_1 =0$ (because the outgoing gluon is a physical massless vector particle, with transverse polarization) to simplify the expressions.

Following with the study of different contributions to the splitting amplitude, we can compute $\Spmatrix^{(1,{\rm SCA-nHV})}$. After replacing integrals and performing some simplifications, it can be expressed as
\beqn
\nonumber \Spmatrix^{(1,{\rm SCA-nHV})}_{q \to g q} &=& \CG {\left(\frac{-s_{12}-\imath 0}{\mu^2}\right)}^{-\epsilon} \SUNT^a \, \frac{g_s^3 \mu^{\epsilon}}{s_{12}}  \, \frac{\epsilon  \left(C_F-C_A\right)}{(\epsilon -1) (2 \epsilon -1)}  \, \left[\bar{u}(p_2)\slashed{\epsilon}(p_1)u(\tilde{P}) \vphantom{\frac{1}{nP}} \right.
\\ &-& \left. \frac{1}{nP} \bar{u}(p_2)\slashed{n}u(\tilde{P}) \, p_2\cdot \epsilon(p_1) \right]\, ,
\label{SplittingqgqSCALARgeneral}
\eeqn
where we consider a $4$-dimensional Dirac's algebra. Note that this expression is simpler than the STD contribution presented before. This is due to the absence of two-gamma matrices in the spinorial chain, which were replaced by a $\epsilon$-dimensional metric, and the simplification of some gluon-propagators. Also, it is worth noting that SCA-nHV terms are finite in the limit $\epsilon \to 0$, so they can be added to the other contributions without modifying the divergent structure. This allows us to interpret the addition to the SCA-nHV terms to the splitting as a DREG scheme choice. Moreover, note that from Eqs. \ref{SplittingGeneralqgq} and \ref{SplittingqgqSCALARgeneral} we can recover the relation
\beqn
\Spmatrix^{(1,{\rm STD},HV)}_{q\to gq} = \Spmatrix^{(1,{\rm STD},FDH)}_{q\to gq} + \Spmatrix^{(1,{\rm SCA-nHV})}_{q\to gq} \, ,
\label{RelacionSPLITTINGSqgq}
\eeqn
which tells us that HV results can be obtained from FDH ones by just adding SCA-nHV contributions. This is a really interesting property, because sometimes it is easier to perform the computation using $4$-dimensional algebra. Moreover, this relation is still valid when we set the polarization of external particles to the possible $4$-dimensional physical values. And, in that situation, we can take advantage of working in FDH scheme because we can apply a wide range of novel techniques, such as the helicity method. 

%************************************************************************************************************************************%
\subsection{Scheme dependence and divergent structure}
Following with the analysis of our results, we can test the decomposition suggested in Eq. \ref{FormulaCatani1}. First, we assume that $\alpha=0$ (i.e. we neglect HSA scheme) and use only STD diagrams. If we expand in series around $\epsilon=0$ and rearrange divergent contributions, we find
\beqn
\Spmatrix^{(1, {\rm STD})}_{q\to g q} &=& \Spmatrix^{(1)}_{H,q\to g q} + \IC_{C,q \to g q} \left(p_1,p_2;\tilde{P} \right) \Spmatrix^{(0)}_{q\to g q} \, ,
\label{FormulaCatani1qgq}
\eeqn
with
\beqn
\nonumber \IC_{C,q \to g q} \left(p_1,p_2;\tilde{P} \right) &=& \frac{\CG g_s^2}{\epsilon ^2} \left(\frac{-s_{12}-\imath 0}{\mu^2}\right)^{-\epsilon} \left[\left(C_A-2 C_F\right) \left(\, _2F_1\left(1,-\epsilon ;1-\epsilon ;\frac{z_1}{z_1-1}\right)-1\right) \right. 
\\ &-& \left. C_A \, _2F_1\left(1,-\epsilon ;1-\epsilon ;\frac{z_1-1}{z_1}\right)\right] \, ,
\eeqn
\beqn
\nonumber \Spmatrix^{(1)}_{H,q\to gq} &=& \CG {\left(\frac{-s_{12}-\imath 0}{\mu^2}\right)}^{-\epsilon} \SUNT^a \,  \frac{g_s^3 \mu^{\epsilon}}{s_{12}} \, \left[\left(C_A\frac{2(1-\delta \epsilon)+\delta(1-\alpha_R)(1+2\epsilon+\alpha_R)}{2 (\epsilon-1) (2 \epsilon-1)}  \right.\right.
\\ \nonumber &-& \left.\left. C_F\frac{1-\alpha_R \delta  \epsilon}{(2\epsilon-1)(\epsilon-1)}\right)  \bar{u}(p_2)\slashed{\epsilon}(p_1)u(\tilde{P})  \right.
\\ &+& \left. \frac{C^{({\rm STD},2)}_{q \to gq}}{2\epsilon^2} \frac{1}{nP}\bar{u}(p_2)\slashed{n}u(\tilde{P}) \, p_2 \cdot \epsilon(p_1) \right] \, ,
\eeqn
where we left $\delta$ and $\alpha_R$ as free parameters. The structure of $\IC_{C,q \to g q}$ exactly agrees with the expected singular behavior of unrenormalized splitting amplitudes. However, some discrepancies appear in the finite contribution. According to Ref. \cite{Catani:2011st}, $\Spmatrix^{(1)}_H$ only contains rational functions of $z$ and $\epsilon$. This is completely true when $\alpha_R=1$, since it reduces to
\beqn
C^{({\rm STD},2)}_{q \to gq} (\alpha_R=1)&=& 2\epsilon^2 \frac{(C_A-C_F) (\delta  \epsilon-1)}{(\epsilon-1) (2 \epsilon-1)} \, ,
\eeqn
and the finite remainder becomes
\beqn
\nonumber \Spmatrix^{(1)}_{H,q\to gq} (\alpha_R=1)&=& \CG {\left(\frac{-s_{12}-\imath 0}{\mu^2}\right)}^{-\epsilon} \SUNT^a \,  \frac{g_s^3 \mu^{\epsilon}}{s_{12}} \frac{(C_F-C_A)(\delta\epsilon-1)}{(\epsilon-1)(2\epsilon-1)} \left[\vphantom{\frac{1}{nP}}\bar{u}(p_2)\slashed{\epsilon}(p_1)u(\tilde{P}) \right.
\\  &-& \left. \frac{1}{nP} \, \bar{u}(p_2)\slashed{n}u(\tilde{P}) \, p_2\cdot \epsilon(p_1)\right] \, .
\eeqn
But when considering $\alpha_R=0$, this contribution involves a non-vanishing combination of hypergeometric functions, which can not be expressed using only rational terms. So, when we work in HSB scheme, $\Spmatrix^{(1)}_{H}$ is no longer a pure rational function.

The situation becomes worse if we choose to work in HSA scheme, setting $\alpha=1$ and $\alpha_R=0$. In that case, it is not possible to cast $\Spmatrix^{(1,{\rm STD})}$ in the form expressed in Eq. \ref{FormulaCatani1qgq} because the divergent structure verifies
\beqn
\nonumber \Spmatrix^{(1,{\rm STD})}_{q\to g q}(HSA) &=& \CG g_s^2 \left(\frac{-s_{12}-\imath 0}{\mu^2}\right)^{-\epsilon} \left[\left(-\frac{C_A}{\epsilon ^2}+\frac{C_A \log(z_1)+(2 C_F-C_A) \log(1-z_1)}{\epsilon }\right) \Spmatrix^{(0)}_{q \to gq} \right.
\\ \nonumber &+&  \left.\left(\frac{3 C_A}{\epsilon ^2}+\frac{2 (2 C_A+C_F)-C_A (1-z_1) \log(-1+z_1)-C_A (1+z_1) \log(z_1)}{2 \epsilon }\right) \right. 
\\ &\times& \left. \frac{\SUNT^a g_s \mu^{\epsilon}}{s_{12}}\bar{u}(p_2)\hat{\gamma}^{\mu}u(\tilde{P}) \hat{\epsilon}_{\mu}(p_1) \, + {\cal O}(\epsilon^0)\right] \, ,
\eeqn
which involves additional $\epsilon$ poles that can not be absorbed in any term proportional to the LO splitting amplitude. This indicates that something else has to be added when performing computations inside HSA scheme (or, conversely, that the definition of HSA scheme must be different). In fact, we need to take into account all the scalar-gluon contributions, both SCA-nHV and SCA-HV. To understand this, we remind the reader that HS schemes assume $\DDirac=4-2\epsilon=\DST$ (i.e. $\delta=1$). Because gluon polarization vectors arise after solving Euler-Lagrange equations in a $\DDirac$-dimensional space, there must be $2-2\epsilon$ degrees of freedom coming from gluons. But in HS schemes, we decompose $\DST$-dimensional gluons into $4$-dimensional vectors (i.e. vector gluons) and $\DST-4$ scalar particles, which forces us to include both vector and scalar gluons simultaneously in our computations. In the case of HSB ($\alpha=0=\alpha_R$), external gluons are always $4$-dimensional particles but we must consider both vector and scalar virtual gluons. Since they have the same kind of couplings, to take into account both contributions we just have to add the propagators, which leads to
\beqn
\nonumber D^{(\alpha_R=0)}_G(k,\mu,\nu) + D_S(k,\mu,\nu) &=& \frac{\imath}{k^2+\imath 0} \left( \left(-\eta^{4}_{\mu \nu} + \frac{n_{\mu}k_{\nu}+n_{\nu}k_{\mu}}{nk}\right) + \left(-\eta^{\epsilon}_{\mu \nu}\right) \right)
\\ &=& \imath \frac{d^{\DST}_{\mu \nu}(k,n)}{k^2+\imath 0} =  D_G^{(\alpha_R=1)}(k,\mu,\nu) \, .
\eeqn
This relation tells us that the consistent version of HSB is the HV scheme ($\delta=1$ and $\alpha=0$). On the other hand, in HSA scheme we must allow the presence of scalar-gluons as external particles. Again, this is equivalent to add the same kind of diagrams but decomposing the outgoing gluon polarization vector as $\epsilon_{\mu}=\tilde{\epsilon}_{\mu}+\hat{\epsilon}_{\mu}$. In other words, if we add all the contributions required to cure the inconsistencies of HSB, we just end in CDR scheme ($\delta=1$ and $\alpha=1$). We will emphasize this point in the following subsection, when computing Altarelli-Parisi kernels.

In summary, after analyzing the scheme dependence of our results for $q \to g q$ splitting amplitude and comparing them with Catani's formula (Eq. \ref{FormulaCatani1qgq}), we conclude that HSA/HSB configurations are not suitable choices for performing calculations. Instead, we will use CDR, HV and FDH schemes, with the possibility of changing the number of fermion polarizations (playing with the parameters $\beta$ and $\beta_R$ previously defined).

%************************************************************************************************************************************%
\subsection{NLO corrections to AP kernels}
Having LO and NLO contributions to the splitting matrix we can obtain the NLO correction to the Altarelli-Parisi (AP) kernel $q \to g q$. In order to do that, we use the expansion
\beqn
\textit{\textbf{P}}_{q\to g q} &=& \frac{s_{12}}{2 \mu^{2\epsilon}} \left[ \left(\Spmatrix^{(0)}_{q \to g q}\right)^{\dagger}\Spmatrix^{(0)}_{q \to g q} + 2 {\rm Re} \left(\left(\Spmatrix^{(0)}_{q \to g q}\right)^{\dagger}\Spmatrix^{(1)}_{q \to g q} \right) \right] + {\cal O}(\alpha_s^3)\, ,
\eeqn 
where we must consider the regulator $\epsilon$ as a complex-valued parameter. If we sum over the physical polarization states of outgoing particles, sum over colors (averaging the incoming ones) and project over the helicity-space of incoming particles, we obtain the polarized AP kernels. Also, it is possible to sum and average over the physical polarizations of the parent parton, which leads to the definition of the unpolarized AP kernels.\footnote{In Ref. \cite{Catani:1996pk}, a distinction is made between \textit{unpolarized} (i.e. averaged over initial polarization states) and \textit{azimuthally averaged} AP kernels. Here we present only polarized and unpolarized, since we can perform the azimuthal average starting from the polarized kernels.}

As expected, the sum over polarizations depend on the scheme being used. If we consider FDH or HV, external particles have physical $4$-dimensional polarizations, but when we set in CDR, they live in a $\DST$-dimensional space. So, in the last scenario, a scalar-gluon can be considered as an external particle, which implies that we must also consider spin-flip contributions at amplitude level. If we compute STD contributions to $\Spmatrix_{q \to g q}$, we can obtain AP kernels in any scheme. It is important to note that, when considering CDR scheme, spin-flip contributions are hidden inside the definition of the $\DST$-dimensional polarization vector, as we saw in Eq. \ref{SplittingLOqgqSEPARACION}. So, we do not need to include explicitly external scalar-gluons, but we can use them to give a physical interpretation to some contributions.

After this brief discussion, let's show explicit results. Starting at LO, we get
\beqn
\left\langle s \right|\hat{P}^{(0)}_{q \to g q}(z_1,k_{\bot})\left|s'\right\rangle &=& C_F \delta_{s,s'} \frac{g_s^2}{z_1} \left(1+ (1-z_1)^2 - \alpha \delta \epsilon z_1^2 \right) \, ,
\\ P^{(0)}_{q\to g q} &=& C_F \frac{g_s^2}{z_1} \, \left(1+ (1-z_1)^2 - \alpha \delta \epsilon z_1^2 \right) \, ,
\eeqn
for the polarized and unpolarized kernels, respectively. Note that when summing over external fermions polarizations, we get a global factor ${\rm Tr}({\rm Id})=4-4\epsilon \beta$ multiplying our results, but it cancels with the average factor. So, $q \to g q$ AP kernels are independent of the number of fermion polarizations. Also, we can prove that
\beqn
\left\langle s \right|\hat{P}_{q \to g q}(z_1,k_{\bot})\left|s'\right\rangle &=& \delta_{s,s'} P_{q \to g q} \, ,
\eeqn
since the kernel is diagonal in helicity space. For this reason, we only present the NLO correction to the unpolarized kernel, which is given by
\beqn
\nonumber P^{(1)}_{q\to g q} &=&\frac{\CG g_s^2}{\epsilon ^2}\left(\frac{-s_{12}-\imath 0}{\mu^2}\right)^{-\epsilon} \left[ P^{(0)}_{q\to gq} \left(\frac{(C_F-C_A)\left(\epsilon(\delta\epsilon^2+\epsilon-3)+1\right)}{(\epsilon -1) (2 \epsilon -1)}   \right.\right.
\\ \nonumber &+&  \left.\left. \left(C_A-2 C_F\right) \, _2F_1\left(1,-\epsilon;1-\epsilon;\frac{z_1}{z_1-1}\right)- C_A \, _2F_1\left(1,-\epsilon;1-\epsilon;\frac{z_1-1}{z_1}\right) + C_F \right) \right.
\\ &+&\left. \frac{g_s^2 C_F}{z_1}\, \frac{({z_1}-2) ({z_1}-1) \epsilon ^2 (\delta  \epsilon -1) \left(C_A-C_F\right)}{(\epsilon -1) (2 \epsilon -1)}\right] \, + {\rm c.c.}\, ,
\label{KernelNLOqgq}
\eeqn
where $\alpha=1$ in CDR and $\alpha=0$ in FDH/HV schemes. As expected, we can appreciate that NLO corrections are independent of $\beta_R$ and $\beta$. On the other hand, it is important to take into account that we must consider only the real part of the r.h.s.

\begin{figure}[htb]
	\centering
		\includegraphics[width=0.80\textwidth]{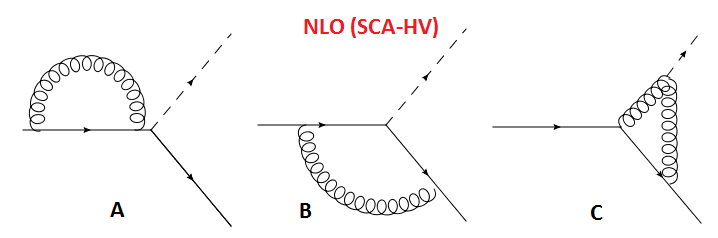}
	\caption{Feynman diagrams associated with the external scalar-gluon contributions to $q(\tilde{P}) \to \phi(p_1) q(p_2)$ at NLO.}
	\label{fig:NLODiagramsScalar1BIS}
\end{figure}

To conclude this section, let's make a remark about the role of scalar-gluons when performing computations in CDR scheme. As we mentioned in the beginning of this section, we can decompose a $\DST$-dimensional gluon as a $4$-dimensional vector gluon and $\DST-4$ scalar particles. Using the LO scalar-gluon contribution (see Eq. \ref{SplittingLOqgqSEPARACION}) and computing the associated unpolarized AP kernel we obtain
\beqn
\nonumber P^{(0)}_{q \to \phi q} &=& \frac{g_s^2 C_F}{4(1-\beta \epsilon)} {\rm Tr}\left[\pslashed_2\hat{\gamma}^{\mu}\slashed{\tilde{P}}\hat{\gamma}^{\nu} \right] \left(\sum_{\rm scalars} \hat{\epsilon}_{\mu}(p_1) \hat{\epsilon}^{*}_{\nu}(p_1)\right)
\\ &=& -{g_s^2 \epsilon C_F \delta z_1} \, ,
\eeqn 
where $\phi$ denotes external scalar-gluons and we use the replacement suggested in Eq. \ref{SumaPolarizacionesSCALARES}. To obtain the NLO correction to this result, it is necessary to take into account some SCA-HV diagrams and compute the corresponding splitting matrix. Since we are decomposing only external gluons, the required contributions can be recovered from $\Spmatrix^{(1,{\rm STD})}$ by just making the replacement $\epsilon_{\mu}(p_1)\to \hat{\epsilon}_{\mu}(p_1)$. So, we can write
\beqn
\nonumber \Spmatrix^{(1,{\rm SCA-HV})}_{q \to \phi q} &=& \left(\Spmatrix^{(1,A)}_{q \to g q}+\Spmatrix^{(1,B)}_{q \to g q}+\Spmatrix^{(1,C)}_{q \to g q} \right)_{\epsilon \to \hat{\epsilon}, \delta \to 1, \alpha_R \to 1}
\\ &=& \frac{\CG g_s^3 \mu^{\epsilon}}{2 s_{12} \epsilon^2} \, {\left(\frac{-s_{12}-\imath 0}{\mu^2}\right)}^{-\epsilon} \SUNT^a C^{({\rm STD},1)}_{q \to gq}(\alpha_R=1,\delta=1) \, \bar{u}(p_2)\hat{\gamma}^{\mu}u(\tilde{P}) \hat{\epsilon}_{\mu}(p_1) \, ,
\eeqn
with the corresponding Feynman diagrams shown in Fig. \ref{fig:NLODiagramsScalar1BIS}. After summing over external particles polarizations and averaging, we get
\beqn
\nonumber P^{(1)}_{q \to \phi q} &=& \frac{\CG g_s^4 z_1 C_F}{\epsilon} \left(\frac{-s_{12}-\imath 0}{\mu^2}\right)^{-\epsilon } \left[C_A \, _2F_1\left(1,-\epsilon ;1-\epsilon ;\frac{z_1-1}{z_1}\right) -C_F\right.
\\ &-& \left. \left(C_A-2 C_F\right) \, _2F_1\left(1,-\epsilon ;1-\epsilon ;\frac{z_1}{z_1-1}\right)+\frac{(\epsilon  (\epsilon +2)-1) \left(C_A-C_F\right)}{2 \epsilon -1}\right] \, + {\rm c.c.} \, .
\eeqn
We can appreciate that 
\beqn
P^{CDR}_{q \to g q} &=& P^{HV}_{q \to g q} + P_{q \to \phi q} \, ,
\eeqn
which reflects the fact that additional gluon polarizations can be interpreted as scalar particles, and, in consequence, that it is possible to recover CDR results working with external $4$-dimensional gluons and adding the remaining degrees of freedom treating them as scalar-particles. Of course, this separation has to be performed with each external gluon to be consistent, which makes a bit cumbersome to carry out this analysis in general.

%%%%%%%%%%%%%%%%%%%%%%%%%%%%%%%%%%%%%%%%%%%%%%%%%%%%%%%%%%%%%%%%
\section{The $g\to q \bar{q}$ splitting matrix}
%%%%%%%%%%%%%%%%%%%%%%%%%%%%%%%%%%%%%%%%%%%%%%%%%%%%%%%%%%%%%%%%
In the previous section we treated in great detail the splitting amplitude $q \to g q$. Here we focus in the process $g \to q \bar{q}$, which is closely related to the first one. However, due to the fact that it is initiated by a vector particle, there are some differences.

As a starting point, we write
\beq
\Spmatrix_{g \to q 	\bar{q}} = \Spmatrix^{(0)}_{g \to q 	\bar{q}} + \Spmatrix^{(1)}_{g \to q 	\bar{q}} \ ,
\eeq
where the LO contribution is
\beqn
\Spmatrix^{(0)}_{g \to q 	\bar{q}} &=&  \frac{g_s \mu^{\epsilon}}{s_{12}} \SUNT^a \bar{u}(p_1)\slashed{\epsilon}(\tilde{P})v(p_2) \ ,
\eeqn
where $p_i$ is the physical momentum of particle $i$ and we associate the massless vector $\tilde{P}$ to the incoming gluon in the collinear limit, in spite of having a momenta $p_{12}$ which verifies $p^2_{12}=s_{12}$.

\begin{figure}[htb]
	\centering
		\includegraphics[width=0.80\textwidth]{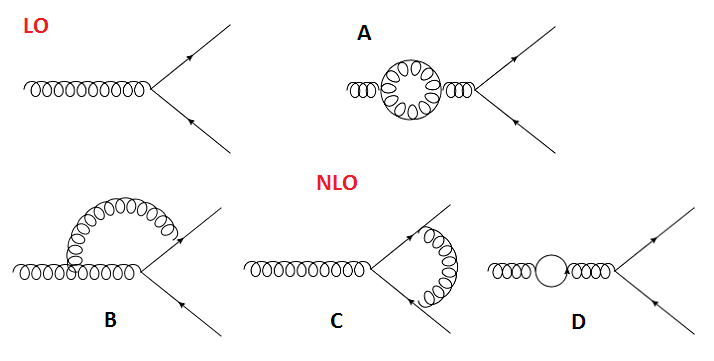}
	\caption{Feynman diagrams associated with $g(\tilde{P}) \to q(p_1) \bar{q}(p_2)$ at NLO. Here the incoming gluon is off-shell and its virtuality is $(p_{12})^2=s_{12}$. Only STD contributions are drawn here.}
	\label{fig:NLODiagrams2}
\end{figure}

The NLO standard-QCD contribution can be expanded as
\beqn
\Spmatrix_{g\to q \bar{q}}^{(1,{\rm STD})} &=& \Spmatrix_{g\to q \bar{q}}^{(1,A)} + \Spmatrix_{g\to q \bar{q}}^{(1,B)} +\Spmatrix_{g\to q \bar{q}}^{(1,C)}+\Spmatrix_{g\to q \bar{q}}^{(1,D)} \, ,
\eeqn
where $\Spmatrix_{g\to q \bar{q}}^{(1,i)}$ refers to the diagram $i\in \left\{A,B,C,D \right\}$, as shown in Fig. \ref{fig:NLODiagrams2}. Note that diagrams A and D expands the self-energy correction to the incoming gluon with a tiny virtuality $s_{12}$. For that reason, we can rewrite their contribution as
\beqn
\Spmatrix_{g\to q \bar{q}}^{(1,A)}+\Spmatrix_{g\to q \bar{q}}^{(1,D)} &=& \Pi(p_{12}^2) \, \Spmatrix_{g\to q \bar{q}}^{(0)} \, ,
\label{SplittinggqqbSELFENERGY}
\eeqn
where $\Pi(p_{12}^2)$ can be extracted from $\Pi_{\mu \nu}(p_{12})$ after contracting with two gluon polarization vectors $\epsilon^{*}_{\mu}(\tilde{P})\epsilon_{\nu}(\tilde{P})$. (See Appendix B for further details on the computation of $\Pi(s_{12})$ and $\Sigma(s_{12})$).

For this process, there are four possible SCA-nHV diagrams which contribute to the amplitude. Following Fig. \ref{fig:NLODiagramsScalar2} and using Feynman rules at Lagrangian level, we have 
\beqn
\Spmatrix^{(1,{\rm SCA-nHV})}_{g \to q \bar{q}} &=& \Spmatrix_{g\to q \bar{q}}^{(1,A')} +\Spmatrix_{g\to q \bar{q}}^{(1,A'')} + \Spmatrix_{g\to q \bar{q}}^{(1,B')} + \Spmatrix_{g\to q \bar{q}}^{(1,C')} \, ,
\eeqn
with
\beqn
\Spmatrix_{g\to q \bar{q}}^{(1,A')} &=& -\frac{g^3_s \mu^{3 \epsilon}}{s^2_{12}} \,C_A \SUNT^a \, \epsilon_{\mu}(\tilde{P}) \bar{u}(p_1)\gamma^{\nu}v(p_2) \, d_{\nu \nu_1}\left(p_{12},n\right)
\\ \nonumber &\times& \int_q \, \frac{\left(-\eta^{\epsilon}_{\rho \rho_1}\right)d_{\sigma \sigma_1}\left(p_{12}-q,n\right)V^{\rm Cin}_{3g}\left(-p_{12},q,p_{12}-q;\mu,\rho,\sigma\right)}{q^2 t_{12q}}
\\ \nonumber &\times& V^{\rm Cin}_{3g}\left(-q,p_{12},q-p_{12};\rho_1,\nu_1,\sigma_1\right)\, ,
\\ \Spmatrix_{g\to q \bar{q}}^{(1,A'')} &=& -\frac{g^3_s \mu^{3 \epsilon}}{2 s^2_{12}} \,C_A \SUNT^a \, \epsilon_{\mu}(\tilde{P}) \bar{u}(p_1)\gamma^{\nu}v(p_2) \, d_{\nu \nu_1}\left(p_{12},n\right)
\\ \nonumber &\times& \int_q \, \frac{\eta^{\epsilon}_{\rho \rho_1}\eta^{\epsilon}_{\sigma \sigma_1}V^{\rm Cin}_{3g}\left(-p_{12},q,p_{12}-q;\mu,\rho,\sigma\right)}{q^2 t_{12q}} V^{\rm Cin}_{3g}\left(-q,p_{12},q-p_{12};\rho_1,\nu_1,\sigma_1\right) \, ,
\\ \Spmatrix_{g\to q \bar{q}}^{(1,B')} &=& -\frac{g^3_s \mu^{3 \epsilon}}{2 s_{12}} \, C_A \SUNT^a \, \epsilon_{\mu}(\tilde{P}) \bar{u}(p_1)\hat{\gamma}^{\rho_1}\gamma^{\alpha}\hat{\gamma}^{\sigma_1}v(p_2) \, \eta^{\epsilon}_{\rho_1 \rho}\eta^{\epsilon}_{\sigma_1 \sigma} \, ,
\\ \nonumber &\times& \int_q \, \frac{{\left(p_1-q\right)}_{\alpha} V^{\rm Cin}_{3g}\left(-p_{12},q,p_{12}-q;\mu,\rho,\sigma\right)}{q^2 t_{1q} t_{12q}} \, ,
\\ \Spmatrix_{g\to q \bar{q}}^{(1,C')} &=& \frac{g^3_s \mu^{3 \epsilon}}{2 s_{12}}\, (C_A-2C_F) \SUNT^a \bar{u}(p_1)\hat{\gamma}^{\rho}\gamma^{\alpha}\slashed{\epsilon}(\tilde{P})\gamma^{\beta}\hat{\gamma}^{\sigma}v(p_2) \, \int_q \, \frac{q_{\alpha} {\left(q-p_{12}\right)}_{\beta} \left(-\eta^{\epsilon}_{\rho \sigma}\right)}{q^2 t_{1q} t_{12q}} \,  .
\eeqn

When we discussed the structure of the contributions to $q \to g q$ splitting amplitude, we mention the possibility of having $q_{\epsilon}^2$-type integrals. Here we face the problem explicitly when analyzing $\Spmatrix^{(1,A')}_{g \to q \bar{q}}$. If we expand the triple gluon vertex, we find
\beqn
\Spmatrix_{g\to q \bar{q}}^{(1,A')} &=& \frac{g^3_s \mu^{3 \epsilon}}{s^2_{12}} \,C_A \SUNT^a \, \epsilon_{\mu}(\tilde{P}) \bar{u}(p_1)\gamma^{\nu}v(p_2) \, d_{\nu \sigma_1}\left(p_{12},n\right) \eta^{\epsilon}_{\rho \rho_1} \,\int_q \, \frac{q_{\rho}q_{\rho_1} \, d_{\mu \sigma_1}\left(p_{12}-q,n\right)}{q^2 t_{12q}} \, .
\eeqn
Since the scalar contribution is computed setting $D_{\rm Dirac}=4$, we can write the involved integral as
\beqn
\nonumber {\rm Int}^{A''}=\int_q \, \frac{q_{\rho}q_{\rho_1} \, d_{\mu \sigma_1}\left(p_{12}-q,n\right)}{q^2 t_{12q}} &=& F_1(k_i\cdot k_j) \eta^{4}_{\rho \rho_1}\eta^{4}_{\mu \sigma_1} \, + \sum_{P} F_2(k_i\cdot k_j,P) \eta^{4}_{a_1 a_2} (k_i)_{a_3} (k_j)_{a_4}
\\ &+& \sum_{P,Q} F_3(k_i\cdot k_j,P,Q) (k_{i_1})_{a_1} (k_{i_2})_{a_2}(k_{i_3})_{a_3}(k_{i_4})_{a_4} \, ,
\eeqn
where $P$ is a permutation of Lorentz indices $\left\{\rho,\rho_1,\mu,\sigma \right\}$, $k_i \in \left\{p_{12},n \right\}$ and $Q$ is a ordering of $\left\{k_i\right\}$. The important fact here is that ${\rm Int}^{A''}$ only has $4$-dimensional components, which implies $\eta^{4}_{\alpha \beta} (\eta^{\epsilon})^{\alpha \beta}=0$. Thus, $\Spmatrix_{g\to q \bar{q}}^{(1,A')}=0$ when using a standard scheme for scalar-gluon contributions. The remaining terms of the splitting matrix can be written as
\beqn
\Spmatrix_{g\to q \bar{q}}^{(1,A'')} &=& \frac{g^3_s \mu^{3 \epsilon}}{s^2_{12}} \epsilon \,C_A \SUNT^a \, \epsilon_{\mu}(\tilde{P}) \bar{u}(p_1)\gamma^{\nu}v(p_2) \, d_{\nu \nu_1}\left(p_{12},n\right) \, \int_q \, \frac{{\left(2q-p_{12}\right)}_{\mu}{\left(2q-p_{12}\right)}_{\nu_1}}{q^2 t_{12q}} \, ,
\\ \Spmatrix_{g\to q \bar{q}}^{(1,B')} &=& -\frac{g^3_s \mu^{3 \epsilon}}{s_{12}} \epsilon\, C_A \SUNT^a \, \epsilon_{\mu}(\tilde{P}) \bar{u}(p_1)\gamma^{\alpha}v(p_2) \, \int_q \, \frac{{\left(p_1-q\right)}_{\alpha} {\left(p_{12}-q\right)}_{\mu}}{q^2 t_{1q} t_{12q}} \, ,
\\ \Spmatrix_{g\to q \bar{q}}^{(1,C')} &=& -\frac{g^3_s \mu^{3 \epsilon}}{s_{12}}\epsilon\, (C_A-2C_F) \SUNT^a \bar{u}(p_1)\gamma^{\alpha}\slashed{\epsilon}(\tilde{P})\gamma^{\beta}v(p_2) \, \int_q \, \frac{q_{\alpha} {\left(q-p_{12}\right)}_{\beta}}{q^2 t_{1q} t_{12q}} \,  ,
\eeqn
where we used the same argument presented in the previous section to make the replacements $\hat{\gamma}^{a}\gamma^c\hat{\gamma}^{b} \to -(\eta^{\epsilon})^{ab} \gamma^c$ and $\hat{\gamma}^{a}\gamma^c\gamma^d\gamma^e\hat{\gamma}^{b} \to -(\eta^{\epsilon})^{ab} \gamma^c\gamma^d\gamma^e$.

Related with the scalar-gluon contributions, here we saw an important fact. Although many diagrams can be constructed by using the effective rules, some of them are going to be zero due to the presence of only $q_{\epsilon}^2$-integrals. This integrals appear when a transverse index contracts with the loop-momentum $q$. So, to avoid them, transverse indices should form \textit{closed chains}, that is
\beq
(\eta^{\epsilon})_{a_1 a_2} (\eta^{\epsilon})^{a_2 a_3} \ldots (\eta^{\epsilon})^{a_n a_1} = (\eta^{\epsilon})^{a_1}_{a_1} \, , 
\eeq
which is equivalent to say that each chain is going to be proportional to the trace of the transverse metric tensor (${\rm Tr}\left[\eta^{\epsilon}\right]=-2\epsilon=\left(\eta^{\epsilon}\right)^{\mu}_{\mu}$).

\begin{figure}[htb]
	\centering
		\includegraphics[width=0.70\textwidth]{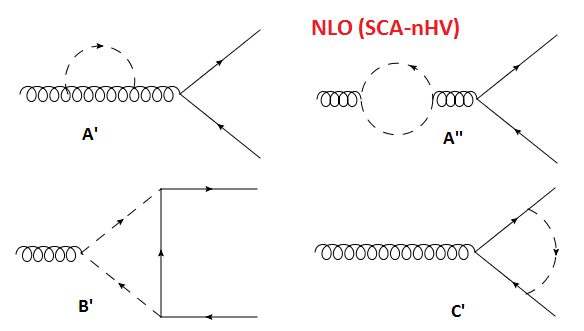}
	\caption{Feynman diagrams associated with SCA-nHV contribution to $g(\tilde{P}) \to q(p_1) \bar{q}(p_2)$ at NLO.}
	\label{fig:NLODiagramsScalar2}
\end{figure}

\subsection{Amplitude level results}
Before showing the explicit results for the $g \to q \bar{q}$ splitting matrix, let's work out the possible spinorial structures which are going to appear. First of all, LO contribution is proportional to $\bar{u}(p_1)\slashed{\epsilon}(\tilde{P})v(p_2)$ and there will be a term proportional to this in $\Spmatrix^{(1)}_{g \to q \bar{q}}$. Due to the symmetry $p_1 \leftrightarrow p_2$, the properties
\beqn
\bar{u}(p_1)\pslashed_{12}v(p_2)&=&0
\\ p_{12}\cdot \epsilon(\tilde{P}) &=&  n\cdot \epsilon(\tilde{P}) = \tilde{P}\cdot \epsilon(\tilde{P})=0 \, ,
\eeqn
and the presence of only two physical vectors ($p_{12}$ and $n$), we can only have one additional spinor-chain with one gamma matrix inside: $\bar{u}(p_1)\slashed{n}v(p_2) p_1\cdot \epsilon(\tilde{P})$. Although there can be spinor-chains of up to five gamma-matrices, Dirac's algebra and the previous properties allow to reduce them to combinations of $\bar{u}(p_1)\slashed{\epsilon}(\tilde{P})v(p_2)$ and $\bar{u}(p_1)\slashed{n}v(p_2) p_1\cdot \epsilon(\tilde{P})$. For these reasons, after replacing Feynman integrals in the expressions for $\Spmatrix^{(1,i)}_{g \to q \bar{q}}$, we get
\beqn
\nonumber \Spmatrix^{(1,{\rm STD})}_{g \to q \bar{q}} &=& \frac{\CG  g_s^3 \mu^{\epsilon} \SUNT^a}{\epsilon^2 s_{12}} {\left(\frac{-s_{12}-\imath 0}{\mu^2}\right)}^{-\epsilon}\,  \left[C^{({\rm STD},1)}_{g \to q \bar{q}} \vphantom{\frac{1}{nP}} \, \bar{u}(p_1)\slashed{\epsilon}(\tilde{P})v(p_2) \right.
\\ &+& \left. C^{({\rm STD},2)}_{g \to q \bar{q}} \,\frac{1}{nP}\bar{u}(p_1)\slashed{n}v(p_2) p_1\cdot \epsilon(\tilde{P}) \right] \, ,
\label{SplittingGeneralgqqb}
\eeqn
for the NLO standard contribution, where the coefficients $C^{({\rm STD},i)}_{g \to q \bar{q}}$ are given by
\beqn
\nonumber C^{({\rm STD},1)}_{g \to q \bar{q}} &=& N_f\, \frac{2 (\epsilon-1 ) \epsilon(1-\beta_R\epsilon) }{4 (\epsilon-2)\epsilon- 3}+C_F \, \frac{ \epsilon  \left(3-(2+\delta ) \epsilon +2 \delta  \epsilon ^2\right)-2}{(\epsilon-1) (2\epsilon-1)} 
\\ \nonumber &+& C_A \, \left( \frac{3+\epsilon ^2 (2 (\epsilon-2 )+\delta  (1+2 (\epsilon-2) \epsilon ))}{(\epsilon-1) (3-2\epsilon) (2\epsilon-1)}- \, _2F_1\left(1,-\epsilon;1-\epsilon;\frac{z_1-1}{z_1}\right) \right.
\\ &-& \left. _2F_1\left(1,-\epsilon;1-\epsilon;\frac{z_1}{z_1-1}\right) +2 \right) \, ,
\\ \nonumber C^{({\rm STD},2)}_{g \to q \bar{q}} &=& 0\, ,
\eeqn
where we set $\alpha_R=1$ since we will not use HSA/HSB schemes here. It is interesting to note that the full NLO correction to the splitting matrix is proportional to $\Spmatrix^{(0)}_{g\to q \bar{q}}$. Besides that, we can appreciate that $C^{({\rm STD},1)}_{g \to q \bar{q}}$ is symmetric when interchanging particles $1$ and $2$.

Again, when using $\alpha=1$ we have to assume that $\mu$ is a $\DST$-dimensional Lorentz index, while in the remaining schemes $\mu$ is associated to a $4$-dimensional space. Moreover, if we rearrange the contributions to $\Spmatrix^{(1)}_{g \to q \bar{q}}$ in the last scenario, we find that it verifies
\beqn
\Spmatrix^{(1)}_{g\to q \bar{q}} &=& \Spmatrix^{(1)}_{H,g\to q \bar{q}} + \IC_{C,g\to q \bar{q}} \left(p_1,p_2;\tilde{P} \right) \Spmatrix^{(0)}_{g\to q \bar{q}} \, ,
\label{FormulaCatani1gqqb}
\eeqn
with
\beqn
\nonumber \IC_{C,g \to q \bar{q}} \left(p_1,p_2;\tilde{P} \right) &=& \frac{\CG g_s^2}{\epsilon ^2} \left(\frac{-s_{12}-\imath 0}{\mu^2}\right)^{-\epsilon} \left[3 C_A- (3 \epsilon +2) C_F+2 \epsilon  b_0 \vphantom{\frac{z_1}{z_1-1}} \right.
\\ &-& \left. C_A \left(\, _2F_1\left(1,-\epsilon ;1-\epsilon ;\frac{z_1-1}{z_1}\right)+\, _2F_1\left(1,-\epsilon ;1-\epsilon ;\frac{z_1}{z_1-1}\right)\right)\right] \, ,
\\ \nonumber \Spmatrix^{(1)}_{H,g\to q \bar{q}} &=& \CG g_s^2 \left(\frac{-s_{12}-\imath 0}{\mu^{2}}\right)^{-\epsilon} \left[C_A \left(\frac{2-3 \delta }{6(3-2 \epsilon) }+\frac{1-\delta }{\epsilon -1}+\frac{\delta -18}{2(2\epsilon -1)}\right)  \right.
\\ &+& \left. C_F \left(\frac{\delta -1}{\epsilon -1}+\frac{8}{2 \epsilon -1}\right) + N_f \frac{6 \beta_R (1-\epsilon)+8 \epsilon -10}{3 (4 (\epsilon -2) \epsilon +3)} \right] \, \Spmatrix^{(0)}_{g \to q \bar{q}} \, ,
\eeqn
as expected according to Eqs. \ref{FormulaCatani1} and \ref{FormulaCatani2}. In contrast to the $q \to gq$ splitting, $\Spmatrix^{(1)}_{g \to q \bar{q}}$ depends on $\beta_R$. However, this parameter seems to define a well-behaved scheme since it respects the universal divergent structure of splitting amplitudes and the finite remainder is kept composed only by rational functions.

On the other hand, for the scalar-gluon contribution we have
\beqn
\Spmatrix^{(1,{\rm SCA-nHV})}_{g \to q \bar{q}} &=& \CG g_s^2  {\left(\frac{-s_{12}-\imath 0}{\mu^2}\right)}^{-\epsilon}    \, \frac{(2 (2-\epsilon) \epsilon -1) C_A+(4 (\epsilon -2) \epsilon +3) C_F}{(\epsilon -1) (2 \epsilon -3) (2 \epsilon -1)} \, \Spmatrix^{(0)}_{g \to q \bar{q}} \, ,
\label{SplittinggqqbSCALARgeneral}
\eeqn
and we can recover the relation
\beqn
\Spmatrix^{(1,{\rm STD},HV)}_{g\to q \bar{q}} = \Spmatrix^{(1,{\rm STD},FDH)}_{g\to q \bar{q}} + \Spmatrix^{(1,{\rm SCA-nHV})}_{g\to q \bar{q}} \, ,
\eeqn
which tells us, again, that HV results can be recovered from FDH ones by just adding SCA-nHV contributions.

\subsection{NLO corrections to AP kernels}
Finally we can compute the contributions to both polarized and unpolarized AP kernel. For the LO contribution we get
\beqn
\left\langle \mu \right|\hat{P}^{(0)}_{g \to q \bar{q}}(z_1,k_{\bot})\left|\nu\right\rangle &=& -g_s^2 (1-\beta \epsilon) \, T_R \, \left((\eta^{\DDirac})^{\mu\nu}+\frac{4 (z_1-1) z_1}{k_{\bot}^2} k_{\bot}^{\mu} k_{\bot}^{\nu}\right) \,
\\ P^{(0)}_{g\to q \bar{q}} &=& \frac{g_s^2 (1-\beta \epsilon) \, T_R}{ (1-\alpha  \epsilon)} ((1-z_1)^2 + z_1^2 -\alpha  \delta  \epsilon) \, ,
\label{KernelLOgqqb}
\eeqn
where we can appreciate that the results depend explicitly on $\beta$ (i.e. the number of external fermions polarizations). Due to the fact that $\Spmatrix^{(1)}_{g \to q \bar{q}}$ is proportional to LO, NLO corrections to AP kernels can be written as
\beqn
P^{(1)}_{g\to q \bar{q}} &=&  \frac{\CG}{\epsilon ^2} g_s^2 \left(\frac{-s_{12}-\imath 0}{\mu^2}\right)^{-\epsilon}  \,  C^{{\rm STD},1}_{g \to q \bar{q}} P^{(0)}_{g\to q \bar{q}} \, + {\rm c.c.} \, ,
\label{KernelNLOgqqb}
\eeqn
where we kept only the real part of the r.h.s. We can appreciate that this expressions depends on both $\beta$ and $\beta_R$, and it is not possible to cancel this dependence by setting $\beta=\beta_R$. But it is interesting to appreciate that the additional factors in Eq. \ref{KernelLOgqqb} disappear in TSC scheme.

%%%%%%%%%%%%%%%%%%%%%%%%%%%%%%%%%%%%%%%%%%%%%%%%%%%%%%%%%%%%%%%%
\section{The $g \to gg$ splitting matrix}
%%%%%%%%%%%%%%%%%%%%%%%%%%%%%%%%%%%%%%%%%%%%%%%%%%%%%%%%%%%%%%%%
Finally, we arrive to the $g \to gg$ splitting amplitude. It is worth noticing that this case involves dealing with many properties of polarization vectors, but it has the advantage of being free of spinor chains. For that reason, here we deal only with scalar products which are well-defined in DREG for every value of $D$.

As done with the previous configurations, the splitting matrix can be decomposed as
\beq
\Spmatrix_{g \to g g} = \Spmatrix^{(0)}_{g \to gg} + \Spmatrix^{(1)}_{g \to gg} \ ,
\eeq
where the LO contribution is
\beqn
\nonumber \Spmatrix^{(0)}_{g \to gg} &=&  \frac{2 g_s \mu^{\epsilon}}{s_{12}} \SUNT^a(A) \left( p_1 \cdot \epsilon(\tilde{P}) \, \epsilon(p_1)\cdot \epsilon(p_2) - p_1 \cdot \epsilon(p_2)\, \epsilon(p_1)\cdot \epsilon(\tilde{P}) \right.
\\ &+& \left. p_2 \cdot \epsilon(p_1)\, \epsilon(p_2)\cdot \epsilon(\tilde{P})  \right) \ ,
\eeqn
where $p_i$ is the physical momentum of particle $i$ and $\left(\SUNT^a(A)\right)_{bc}= \imath f_{abc}$ are the generators of $SU(3)_C$ in the adjoint representation.

\begin{figure}[htb]
	\centering
		\includegraphics[width=0.90\textwidth]{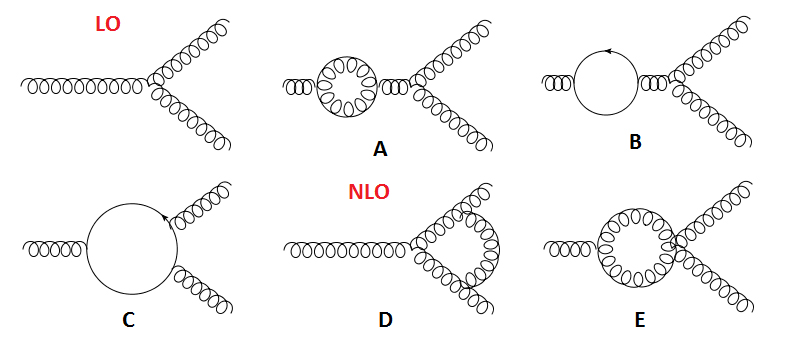}
	\caption{Feynman diagrams associated with the standard QCD contribution to $g(\tilde{P}) \to q(p_1) g(p_2)$ at NLO. Here the incoming gluon is off-shell and its virtuality is $(p_{12})^2=s_{12}$.}
	\label{fig:NLODiagrams3}
\end{figure}

The NLO standard-QCD contribution can be expanded as
\beqn
\Spmatrix_{g\to gg}^{(1,{\rm STD})} &=& \Spmatrix_{g\to gg}^{(1,A)} + \Spmatrix_{g\to gg}^{(1,B)} +\Spmatrix_{g\to gg}^{(1,C)}+\Spmatrix_{g\to gg}^{(1,D)}+\Spmatrix_{g\to gg}^{(1,E)} \, ,
\eeqn
being $\Spmatrix_{g\to g g}^{(1,i)}$ associated with diagram $i\in \left\{A,B,C,D,E \right\}$, as shown in Fig. \ref{fig:NLODiagrams3}. We have to remark that due to symmetry properties, diagrams C and D only describe the associated topology. In other words, there are two diagrams C (and D), which are obtained from the displayed graph by interchanging particles $1$ and $2$; $\Spmatrix^{(1,C)}$ and $\Spmatrix^{(1,D)}$ include the sum over all the possible relabellings of final-state particles associated with the process.

On the other hand, diagrams A and B expands the self-energy correction to the incoming gluon with a tiny virtuality $s_{12}$. As we have done in the $g \to q \bar{q}$ splitting, we can rewrite their contribution as
\beqn
\Spmatrix_{g\to g g}^{(1,A)}+\Spmatrix_{g\to g g}^{(1,B)} &=& \Pi(p_{12}^2) \, \Spmatrix_{g\to g g}^{(0)} \, ,
\label{SplittinggggSELFENERGY}
\eeqn
with $\Pi(p_{12}^2)$ given in Appendix B.

\begin{figure}[htb]
	\centering
		\includegraphics[width=0.80\textwidth]{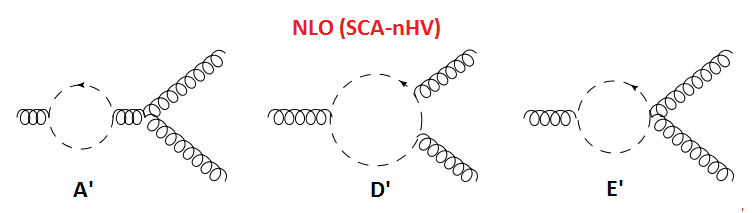}
	\caption{Feynman diagrams associated with the scalar-gluon contribution to $g(\tilde{P}) \to g(p_1) g(p_2)$ at NLO. We only consider diagrams which contribute non-trivially to the splitting amplitude.}
	\label{fig:NLODiagramsScalar3}
\end{figure}

When dealing with the scalar-gluon contribution, we find many possible diagrams. However, as we have seen in the previous computations (explicitly in $\Spmatrix^{(1)}_{g \to q \bar{q}}$), the only non-trivial terms arise from taking the trace of transverse metrics. In other words, transverse indices have to form a closed chain and be completely contracted with metric tensors; otherwise, we will have $q_{\epsilon}^2$-integrals, which are set to zero when $D_{\rm Dirac}=4$. So, following Fig. \ref{fig:NLODiagramsScalar3} and using effective Feynman rules for scalar-gluons, the SCA-nHV contribution can be written as 
\beqn
\Spmatrix^{(1,{\rm SCA-nHV})}_{g \to gg} &=& \Spmatrix_{g\to gg}^{(1,A')} +\Spmatrix_{g\to gg}^{(1,D')} + \Spmatrix_{g\to gg}^{(1,E')} \, ,
\eeqn
with
\beqn
\nonumber \Spmatrix_{g\to gg}^{(1,A')} &=& -\frac{g^3_s \mu^{3 \epsilon} \epsilon}{s^2_{12}} \,C_A \SUNT^a(A) \, \epsilon_{\mu}(\tilde{P}) \epsilon_{\nu}(p_1)\epsilon_{\rho}(p_2) \, d_{\alpha \alpha_1}\left(p_{12},n\right) \, V^{\rm Cin}_{3g}\left(-p_{12},p_1,p_2;\alpha_1,\nu,\rho\right)
\\ &\times& \int_q \, \frac{\left(2q-p_{12}\right)_{\mu}\left(2q-p_{12}\right)_{\alpha}}{q^2 t_{12q}} \, ,
\\ \nonumber \Spmatrix_{g\to gg}^{(1,D')} &=& -\frac{g^3_s \mu^{3 \epsilon} \epsilon}{2 s_{12}} \, C_A \SUNT^a(A) \, \epsilon_{\mu}(\tilde{P}) \epsilon_{\nu}(p_1) \epsilon_{\rho}(p_2) \, \int_q \, \frac{\left(2q-p_{12}\right)_{\mu}}{q^2 t_{12q}}
\\ &\times& \left[ \frac{\left(2q-p_1\right)_{\nu}\left(2q-2p_1-p_2\right)_{\rho}}{t_{1q}} - \frac{\left(2q-p_2\right)_{\rho}\left(2q-2p_2-p_1\right)_{\nu}}{t_{2q}}\right] \, ,
\\ \Spmatrix_{g\to gg}^{(1,E')} &=& 0 \, .
\eeqn
It is important to note that $\Spmatrix_{g\to gg}^{(1,E')}$ is zero due to color properties. In fact, we get
\beqn
f_{ade} \left(f_{bex}f_{cdx}+f_{bdx}f_{cex}\right)&=&0 \, ,
\eeqn
where we have contracted the effective 2scalar-2gluon vertex with a factor $f_{ade}$ coming from the triple-gluon interaction.

\subsection{Amplitude level results}
As a first step, let's study the possible structure of the splitting matrix. In this process we have three physical momenta ($p_1$, $p_2$ and $\tilde{P}$, or equivalently, $n$) and three physical on-shell polarizations vectors. Since external legs are massless on-shell particles we have the constraints
\beqn
\tilde{P}\cdot\epsilon(\tilde{P}) &=& 0 = n \cdot \epsilon(\tilde{P}) \Rightarrow p_{12}\cdot(\tilde{P}) =0 \, ,
\\ p_i \cdot \epsilon(p_i) &=& 0 = n \cdot\epsilon(p_i) \ \ , \ i\in\left\{1,2\right\} \ \, ,
\eeqn
where we have forced all the polarization vectors to vanish when contracted with the null-vector $n$, relying in the gauge invariance. So, we have the following non-zero scalar products:
\beq
\left\{p_1\cdot \epsilon(p_2) \, , p_2\cdot \epsilon(p_1) \, ,(p_1-p_2)\cdot \epsilon(\tilde{P})\right\} \, ,
\eeq
and
\beq
\left\{\epsilon(p_1)\cdot \epsilon(p_2) \, ,\epsilon(p_1)\cdot \epsilon(\tilde{P}) \, ,\epsilon(p_2)\cdot \epsilon(\tilde{P}) \right\}\, ,
\eeq
where we are using $p_1\cdot\epsilon(\tilde{P})=-p_2\cdot\epsilon(\tilde{P})$. Now we have to form all the possible structures that involve the three polarization vectors and that are compatible with the symmetry of the system when interchanging particles $1$ and $2$. Thus we get
\beqn
E_1 &=& \epsilon(p_1)\cdot\epsilon(p_2) \, p_1\cdot\epsilon(\tilde{P}) \, ,
\\ E^{\pm}_2 &=& p_2\cdot\epsilon(p_1)\, \epsilon(p_2)\cdot\epsilon(\tilde{P}) \pm p_1\cdot\epsilon(p_2)\, \epsilon(p_1)\cdot\epsilon(\tilde{P}) \, ,
\\ E_3 &=& p_1\cdot\epsilon(p_2) \, p_2\cdot\epsilon(p_1) \, p_1\cdot\epsilon(\tilde{P}) \, ,
\eeqn
and notice that $E^{+}_2$ is symmetric while $E_1$,$E^{-}_2$,$E_3$ are antisymmetric. After replacing Feynman integrals in the expressions for $\Spmatrix^{(1,i)}_{g \to gg}$ and summing all the contributions, we realize that only two structures survive: $E_1+E^{-}_2$ (this is proportional to LO splitting) and $E_1-\frac{2}{s_{12}}E_3$. So, we can write
\beqn
\nonumber \Spmatrix^{(1,{\rm STD})}_{g \to gg} &=& \frac{\CG g_s^3 \mu^{\epsilon} \SUNT^a(A)}{\epsilon^2 s_{12}} {\left(\frac{-s_{12}-\imath 0}{\mu^2}\right)}^{-\epsilon}\,  
\\ \nonumber &\times& \left[\vphantom{\frac{1}{s_{12}}}C^{({\rm STD},1)}_{g \to gg} \left(\epsilon(p_1)\cdot\epsilon(p_2) \, p_1\cdot\epsilon(\tilde{P})+ p_2\cdot\epsilon(p_1)\, \epsilon(p_2)\cdot\epsilon(\tilde{P}) - p_1\cdot\epsilon(p_2)\, \epsilon(p_1)\cdot\epsilon(\tilde{P})\right) \right.
\\ &+& \left. C^{({\rm STD},2)}_{g \to gg} \, p_1\cdot\epsilon(\tilde{P})\left(\epsilon(p_1)\cdot\epsilon(p_2) -\frac{2}{s_{12}}p_1\cdot\epsilon(p_2)\,p_2\cdot\epsilon(p_1) \right)  \right] \, ,
\label{SplittingGeneralggg}
\eeqn
for the NLO standard contribution, where the coefficients $C^{({\rm STD},i)}_{g \to gg}$ are given by
\beqn
\nonumber C^{({\rm STD},1)}_{g \to gg} &=& 2 C_A \left[1- \, _2F_1 \left(1,-\epsilon;1-\epsilon;\frac{z_1}{z_1-1}\right) \right.
\\ &-& \left.  \, _2F_1 \left(1,-\epsilon;1-\epsilon;\frac{z_1-1}{z_1}\right)  \right] \, ,
\\ \nonumber C^{({\rm STD},2)}_{g \to gg} &=&  \frac{2 \epsilon^2 \left((\delta\epsilon -1) C_A+N_f(1-\beta_R \epsilon)\right)}{(\epsilon-1)(2\epsilon-1)(2\epsilon-3)} \, ,
\eeqn
where we set $\alpha_R=1$ to exclude HSA/HSB schemes.

Following Eq. \ref{FormulaCatani1}, $\Spmatrix^{(1)}_{g \to gg}$ can be rewritten as
\beqn
\Spmatrix^{(1)}_{g\to gg} &=& \Spmatrix^{(1)}_{H,g\to gg} + \IC_{C,g\to gg} \left(p_1,p_2;\tilde{P} \right) \Spmatrix^{(0)}_{g\to gg} \, ,
\label{FormulaCatani1ggg}
\eeqn
with
\beqn
\nonumber \IC_{C,g\to gg}\left(p_1,p_2;\tilde{P} \right) &=& \frac{\CG g_s^2}{\epsilon^2} \left(\frac{-s_{12}-\imath 0}{\mu^{2}}\right)^{-\epsilon} \, C_A \left(1- \, _2F_1\left(1,-\epsilon ;1-\epsilon ;\frac{z_1-1}{z_1}\right) \right.
\\ &-& \left. \, _2F_1\left(1,-\epsilon ;1-\epsilon ;\frac{z_1}{z_1-1}\right)\right)
\\ \nonumber \Spmatrix^{(1)}_{H,g\to gg} &=& \CG \left(\frac{-s_{12}-\imath 0}{\mu^{2}}\right)^{-\epsilon} \SUNT^a(A) \frac{2 g_s^{3}\mu^{\epsilon}}{s_{12}}\frac{C_A (\delta  \epsilon -1)+N_f(1-\beta_R \epsilon)}{ (\epsilon -1) (2 \epsilon -3) (2 \epsilon -1)} \, 
\\ &\times& p_1\cdot\epsilon(\tilde{P}) \, \left(\epsilon(p_1)\cdot\epsilon(p_2) -\frac{2}{s_{12}}p_1\cdot\epsilon(p_2)\,p_2\cdot\epsilon(p_1) \right) \, ,
\eeqn
as expected. Moving to the scalar-gluon contribution we obtain
\beqn
\nonumber \Spmatrix^{(1,{\rm SCA-nHV})}_{g \to gg} &=& \frac{\CG g_s^3 \mu^{\epsilon} \epsilon C_A \SUNT^a(A)}{s_{12} (\epsilon -1) (2 \epsilon -3) (2 \epsilon -1)}\left(\frac{-s_{12}-\imath 0}{\mu^2}\right)^{-\epsilon} \, p_1\cdot\epsilon(\tilde{P})
\\ &\times& \left(\epsilon(p_1)\cdot\epsilon(p_2) -\frac{2}{s_{12}}p_1\cdot\epsilon(p_2)\,p_2\cdot\epsilon(p_1) \right) \, ,
\label{SplittinggggSCALARgeneral}
\eeqn
and comparing it with STD contributions in different schemes we get
\beqn
\Spmatrix^{(1,{\rm STD},HV)}_{g\to gg} = \Spmatrix^{(1,{\rm STD},FDH)}_{g\to gg} + \Spmatrix^{(1,{\rm SCA-nHV})}_{g\to gg} \, ,
\eeqn
which agrees with the relation found for $q \to gq$ and $g \to q \bar{q}$ splittings.

\subsection{NLO corrections to AP kernel}
Finally we can compute the contributions to the Altarelli-Parisi kernels. At LO we have
\beqn
\nonumber \left\langle \mu \right|\hat{P}^{(0)}_{g \to gg}(z_1,k_{\bot})\left|\nu\right\rangle &=& -\frac{2 g_s^2 C_A}{z_1(1-z_1)} \left[(1-2 (1-z_1) z_1) \left((1-\alpha) (\eta^{4})^{\mu\nu}+\alpha  (\eta^{\DST})^{\mu\nu}\right) \vphantom{\frac{a}{k_{\bot}^2}}\right.
\\ &+& \left. 2 \frac{(1-z_1)^2 z_1^2}{k_{\bot}^2} k_{\bot}^{\mu} k_{\bot}^{\nu} \left(1-\alpha \delta  \epsilon\right)\right] \, ,
\\ P^{(0)}_{g\to gg} &=& \frac{2 g_s^2 (1-(1-z_1)z_1)^2 C_A \left(1-\alpha \delta  \epsilon\right)}{(1-z_1) z_1 (1- \alpha  \epsilon )} \, ,
\eeqn
for the polarized and unpolarized kernels, respectively. Note that when we set $\alpha=1$, then external gluons have $2-2\epsilon$ polarizations and they are treated like $\DST$-dimensional vectors. So, we must set $\delta=1$, which allows us to cancel the $\alpha$ dependence in the unpolarized kernel.

Moving to NLO, we obtain
\beqn
\nonumber \left\langle \mu \right|\hat{P}^{(1)}_{g \to gg}(z_1,k_{\bot})\left|\nu\right\rangle &=& \CG g_s^2 \left(\frac{-s_{12}-\imath 0}{\mu^2}\right)^{-\epsilon}  \left[\frac{C_A}{\epsilon^2} \left(1-\, _2F_1\left(1,-\epsilon ;1-\epsilon ;\frac{z_1-1}{z_1}\right) \right. \right.
\\ \nonumber &-& \left. \left. \, _2F_1\left(1,-\epsilon ;1-\epsilon ;\frac{z_1}{z_1-1}\right)\right)  \, \left\langle \mu \right|\hat{P}^{(0)}_{g \to gg}(z_1,k_{\bot})\left|\nu\right\rangle \right.
\\ \nonumber &+& \left. \frac{2 g_s^2 C_A \left(1-2 \alpha \delta  (1-z_1) z_1 \epsilon\right) \left(C_A (\delta  \epsilon -1)+N_f(1-\beta_R\epsilon)\right)}{s_{12} (1-z_1) z_1 (\epsilon -1) (2 \epsilon -3) (2 \epsilon -1)} k_{\bot}^{\mu} k_{\bot}^{\nu}\right]
\\ \, &+& {\rm c.c.} \, ,
\eeqn
\beqn
\nonumber P^{(1)}_{g\to gg} &=&  \CG g_s^2 \left(\frac{-s_{12}-\imath 0}{\mu^2}\right)^{-\epsilon } \frac{C_A}{\epsilon^2} \left[\frac{g_s^2 \epsilon^2  \left(1-2 \alpha \delta  \epsilon z_1(1-z_1)\right) \left(C_A (\delta  \epsilon -1)+N_f(1-\beta_R \epsilon)\right)}{(1-\alpha\epsilon)(\epsilon -1) (2 \epsilon -3) (2 \epsilon -1)} \right.
\\ &+& \left. \left(1- \, _2F_1\left(1,-\epsilon ;1-\epsilon ;\frac{z_1-1}{z_1}\right)-\, _2F_1\left(1,-\epsilon ;1-\epsilon ;\frac{z_1}{z_1-1}\right)\right) P^{(0)}_{g \to gg}\right] \, + {\rm c.c.}\, ,
\label{KernelNLOggg}
\eeqn
where $\alpha=1$ in CDR and $\alpha=0$ in FDH/HV schemes. It is worth noticing that the polarized kernel contains some terms proportional to $\tilde{P}^{\mu}$ and $n^{\mu}$, but since $\tilde{P} \cdot \epsilon(\tilde{P}) = n\cdot \epsilon(\tilde{P})=0$ we neglect them to simplify the result.

%%%%%%%%%%%%%%%%%%%%%%%%%%%%%%%%%%%%%%%%%%%%%%%%%%%%%%%%%%%%%%%%
\section{Splittings matrices involving photons}
%%%%%%%%%%%%%%%%%%%%%%%%%%%%%%%%%%%%%%%%%%%%%%%%%%%%%%%%%%%%%%%%
Let's consider an extension of massless QCD with the inclusion of a QED photon. This is a natural step when we want to study photon-production in the context of hadron colliders, since photons represent a very clean signal in the detector and QCD corrections can not be ignored. This model can be described by extending the gauge group to ${\rm SU}(3)_{C}\times{\rm U}(1)_{E}$ which involves adding a new vector field $A_{\mu}$. The associated $D$-dimensional Lagrangian reads
\beqn
{\cal L}_{\rm QCD+QED}&=& {\cal L}_{\rm QCD} - \sum_Q \, g_e \mu^{\epsilon} E_Q \delta_{ij} \, \bar{\Psi}^i_Q\gamma^{\mu}\Psi^j_Q \, A_{\mu} -\frac{1}{4} F^{\mu\nu}F_{\mu\nu} \,  ,
\eeqn
where $\left\{i,j\right\}$ are color indices, $g_e$ is the electromagnetic coupling (i.e. the absolute value of electron charge), $E_Q$ is the charge of quark's flavor $Q$ ($E_{u,c,t}=2/3$ and $E_{d,s,b}=-1/3$) and $F_{\mu \nu}=\partial_{\mu}A_{\nu}-\partial_{\nu}A_{\mu}$ is the gauge-field strength tensor for the Abelian group ${\rm U}(1)_E$. From the interaction term, we can deduce that the Feynman rule for the quark-photon-quark vertex is $- \imath g_e \mu^{\epsilon} E_Q \, \gamma^{\mu}$ and it is proportional to the identity matrix $\IDC$ in the color space. Since quarks belong to the fundamental representation of ${\rm SU}(3)_C$, then ${\rm Tr}({\rm Id}_C)=N_C=C_A$ which is going to be important when computing AP kernels. 

In the next subsections, we show the associated splitting functions at NLO in the QCD coupling constant $\alpha_s$: $\Spmatrix_{q \to \gamma q}$ and $\Spmatrix_{\gamma \to q \bar{q}}$. It is worth noticing that processes involving two photons and one gluon (i.e $\gamma \to \gamma g$ or $g \to \gamma \gamma$) vanish due to color conservation, because they are proportional to ${\rm Tr}(\SUNT^a(F))=0$. On the other hand, there are not splittings with one photon and two gluons, because they involve a fermion loop with three vectors attached to it and, after summing all diagrams, we arrive to an expression which is again globally proportional to ${\rm Tr}(\SUNT^a(F))=0$. 

It is worth noticing that we can check the divergent structure of splitting matrices involving photons using a formula similar to Eq. \ref{FormulaCatani1}. For $1\to 2$ processes, any splitting can be written as
\beqn
\Spmatrix^{(1)}\left(p_1,p_2;\tilde{P}\right) &=& \Spmatrix^{(1)}_{H} \left(p_1,p_2;\tilde{P}\right) + \IC^{\gamma}_{C} \left(p_1,p_2;\tilde{P} \right) \Spmatrix^{(0)} \left(p_1,p_2;\tilde{P}\right) \, ,
\label{FormulaCataniFotones}
\eeqn
with $\Spmatrix^{(1)}_{H}$ finite in the limit $\epsilon\to 0$ and containing only rational functions of $p_1$, $p_2$ and $\tilde{P}$, and
\beqn
\nonumber \IC^{\gamma}_{C} \left(p_1,p_2;\tilde{P} \right) &=& \CG g_s^{2} \left(\frac{-s_{12}-\imath 0}{\mu^2} \right)^{-\epsilon}
\\ \nonumber &\times& \left\{ \frac{1}{\epsilon^2}\left(C_{12}-C_1-C_2 \right) + \frac{1}{\epsilon} \left(\gamma_{12}-\gamma_1-\gamma_2 \right) \right.
\\ &-& \left. \frac{1}{\epsilon} \left[\left(C_{12}+C_1-C_2\right)f(\epsilon,z_1)+\left(C_{12}+C_2-C_1\right)f(\epsilon,1-z_1) \right] \right\} \, ,
\label{FormulaCataniFotones2}
\eeqn
associated with the divergent contributions. Note that Eq. \ref{FormulaCataniFotones2} is very similar to Eq. \ref{FormulaCatani2}, with the exception of single pole proportional to $b_0$. Explicitly, 
\beqn
\IC^{\gamma}_{C} \left(p_1,p_2;\tilde{P} \right) &=& \IC_{C} \left(p_1,p_2;\tilde{P} \right) - \CG g_s^{2} \left(\frac{-s_{12}-\imath 0}{\mu^2} \right)^{-\epsilon} \frac{b_0}{\epsilon} \, ,
\eeqn
which is related to the fact that this kind of splitting only involves two colored-particles and we have to remove the single $\epsilon$-pole coming from the renormalization of QCD coupling. Also, we have to take into account that $C_{\gamma}=0=\gamma_{\gamma}$ because photons do not carry color.

Finally, since we are interested in studying the scheme dependence of splitting amplitudes, we are treating external photons and gluons in the same way. In other words, we can adapt the conventions shown in Section 2 for gluons to obtain
\beqn
n_{\gamma} &=& 2 - 2 \alpha \epsilon \, ,
\\ \sum_{\rm phys. pol.} \epsilon_{\mu}(p) \epsilon^{*}_{\nu}(p) &=& - \left(\eta^{4}_{\mu \nu} + \alpha \eta^{\DST-4}_{\mu\nu} \right) + \frac{p_{\mu}n_{\nu}+p_{\nu}n_{\mu}}{n\cdot p} \, , 
\eeqn
where $\epsilon(p)$ denotes the polarization vector associated to photons. The advantage of choosing this gauge is that it allows us to make a straightforward reduction from the pure QCD splittings, since this implies that $\epsilon(p) \cdot n =0$ also for photons.

%*******************************************************************************
\subsection{$q \to \gamma q$}
This process can be considered as an Abelianization of $q \to g q$, because it is not possible to have a triple-gluon vertex contribution. So, having performed a detailed study of $q \to g q$ in previous sections, we are able to extract some important results for $q \to \gamma q$ without doing a full computation again. 

\begin{figure}[htb]
	\centering
		\includegraphics[width=0.80\textwidth]{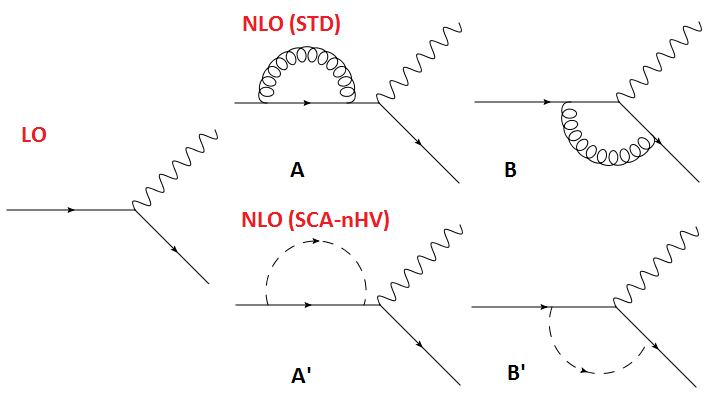}
	\caption{Feynman diagrams associated with $q(\tilde{P}) \to \gamma(p_1) q(p_2)$ at LO and NLO. We include also the SCA-nHV contributions.}
	\label{fig:NLOPhotonDiagrams1}
\end{figure}

First of all, the list of possible Feynman diagrams up to ${\cal O}\left(\alpha_s^2\right)$ is shown in Fig. \ref{fig:NLOPhotonDiagrams1}. Note that they are essentially the same that we used for $q\to g q$ (see Figs. \ref{fig:NLODiagrams1} and \ref{fig:NLODiagramsScalar1}), except for the diagrams that include a triple-gluon vertex. At LO we have
\beqn
\Spmatrix^{(0)}_{q \to \gamma q} &=& \frac{g_e E_Q \mu^{\epsilon}}{s_{12}} \IDC \,  \bar{u}(p_2)\slashed{\epsilon}{\left(p_1\right)}u(\tilde{P}) \ ,
\eeqn
while the NLO standard-QCD corrections can be written as
\beqn
\nonumber \Spmatrix^{(1,{\rm STD})}_{q \to \gamma q} &=& \CG g_s^2 {\left(\frac{-s_{12}-\imath 0}{\mu^2}\right)}^{-\epsilon} \, \IDC \frac{g_e E_Q \mu^{\epsilon} C_F}{s_{12} \epsilon^2}\, \left[\vphantom{\frac{1}{nP}} C^{({\rm STD},1)}_{q \to \gamma q} \, \bar{u}(p_2)\slashed{\epsilon}(p_1)u(\tilde{P}) \right.
\\ &+& \left. C^{({\rm STD},2)}_{q \to \gamma q} \, \frac{1}{nP}\bar{u}(p_2)\slashed{n}u(\tilde{P}) p_2\cdot \epsilon(p_1) \right] \, ,
\label{SplittingGeneralqgammaq}
\eeqn
where the coefficients $C^{({\rm STD},i)}_{q \to \gamma q}$ are given by
\beqn
C^{({\rm STD},1)}_{q \to \gamma q}&=&\frac{\epsilon^2(\delta  \epsilon -1)}{(\epsilon-1)(2\epsilon-1)}+2 -2 \, _2F_1\left(1,-\epsilon;1-\epsilon;\frac{z_1}{z_1-1}\right) \, ,
\\ C^{({\rm STD},2)}_{q \to \gamma q}&=& \frac{1-\delta  \epsilon}{(\epsilon-1)(2\epsilon-1)} \, .
\eeqn
Analogously, for the NLO scalar-gluon contribution we have
\beqn
\nonumber \Spmatrix^{(1,{\rm SCA-nHV})}_{q \to \gamma q} &=& \CG g_s^2  \left(\frac{-s_{12}-\imath 0}{\mu^2}\right)^{-\epsilon} \, \IDC \frac{\epsilon g_e E_Q \mu^{\epsilon} \, C_F}{s_{12}(2  \epsilon-1)(\epsilon-1)} \left[ \bar{u}(p_2)\slashed{\epsilon}(p_1)u(\tilde{P})  \right.
\\ &-& \left. \frac{1}{nP}\bar{u}(p_2)\slashed{n}u(\tilde{P}) p_2\cdot \epsilon(p_1) \right]\, ,
\eeqn
and it is straightforward to verify that
\beqn
\Spmatrix^{(1,{\rm STD},HV)}_{q\to \gamma q} = \Spmatrix^{(1,{\rm STD},FDH)}_{q\to \gamma q} + \Spmatrix^{(1,{\rm SCA-nHV})}_{q\to \gamma q} \, ,
\eeqn
which shows that the cancellation of scalar degrees of freedom occurs separately in Abelian and non-Abelian vertices. 

As a consistency check, following Eq. \ref{FormulaCataniFotones}, we rewrite $\Spmatrix^{(1)}_{q \to \gamma q}$ as
\beqn
\Spmatrix^{(1)}_{q \to \gamma q} &=& \Spmatrix^{(1)}_{H,q \to \gamma q} + \IC^{\gamma}_{C,q \to \gamma q} \left(p_1,p_2;\tilde{P} \right) \Spmatrix^{(0)}_{q \to \gamma q} \, ,
\label{FormulaCatani2qgammaq}
\eeqn
with
\beqn
\IC^{\gamma}_{C,q \to \gamma q} &=& \CG g_s^2 \left(\frac{-s_{12}-\imath 0}{\mu^{2}}\right)^{-\epsilon} \frac{2 C_F}{\epsilon^2} \left(1-\, _2 F_1\left(1,-\epsilon ;1-\epsilon ;\frac{z_1}{z_1-1}\right)\right) \, ,
\\ \nonumber \Spmatrix^{(1)}_{H,q \to \gamma q} &=&  \CG g_s^2 \left(\frac{-s_{12}-\imath 0}{\mu^{2}}\right)^{-\epsilon} \, \IDC \frac{g_e E_Q \mu^{\epsilon} C_F(\delta\epsilon-1)}{s_{12}(2\epsilon-1)(\epsilon-1)}  \, \left[\bar{u}(p_2)\slashed{\epsilon}(p_1)u(\tilde{P}) \right.
\\ &-& \left. \frac{1}{nP}\bar{u}(p_2)\slashed{n}u(\tilde{P}) p_2\cdot \epsilon(p_1) \right] \, ,
\eeqn
where we see that the divergent part (which contains $\epsilon$-poles and branch-cuts) is isolated into $\IC^{\gamma}_C$, while $\Spmatrix^{(1)}_H$ only contains rational functions and is finite in the limit $\epsilon\to 0$. Moreover, the new spinor chain which appears in the NLO computation is entirely contained in $\Spmatrix^{(1)}_H$.

Finally, we can compute the corresponding contributions to the Altarelli-Parisi kernel. Since it is a quark initiated process, the polarized kernel verifies
\beqn
\left\langle s \right|\hat{P}_{q \to \gamma q}(z_1,k_{\bot})\left|s'\right\rangle &=& \delta_{s,s'} P_{q \to \gamma q} \, ,
\eeqn
due to helicity conservation. So, the unpolarized kernel at LO is given by
\beqn
P^{(0)}_{q \to \gamma q} &=& g_e^2 E_Q^2 \, \frac{1+(1-z_1)^2-\alpha\delta\epsilon z_1}{z_1} \, ,
\eeqn
where we can appreciate that the result is independent of the number of fermion polarizations. On the other hand, at NLO we have
\beqn
\nonumber P^{(1)}_{q \to \gamma q} &=&\CG  g_s^2 \, {\left(\frac{-s_{12}-\imath 0}{\mu^2}\right)}^{-\epsilon} \frac{C_F}{\epsilon^2} \left[\left(\frac{2+\epsilon  (\epsilon  (3+\delta  \epsilon )-6)}{(\epsilon -1) (2 \epsilon -1)} - 2 \, _2F_1\left(1,-\epsilon;1-\epsilon;\frac{z_1}{z_1-1}\right)\right) \right.
\\ &\times& \left. P^{(0)}_{q \to \gamma q} +\frac{g_e^2 E_Q^2 \epsilon^2(z_1-1)(z_1-2)(1-\delta\epsilon)}{z_1(2\epsilon-1)(\epsilon-1)} \right]\, + {\rm c.c.} \, ,
\eeqn
where $\alpha$ is a parameter that allows us to change between CDR ($\alpha=1$) and HV/FDH ($\alpha=0$) schemes.

%*******************************************************************************
\subsection{$\gamma \to q \bar{q}$}
\begin{figure}[htb]
	\centering
		\includegraphics[width=0.80\textwidth]{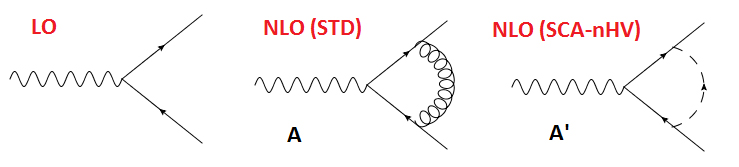}
	\caption{Feynman diagrams associated with $\gamma(\tilde{P}) \to q(p_1) \bar{q}(p_2)$ at LO and NLO. We include also the SCA-nHV amplitudes.}
	\label{fig:NLOPhotonDiagrams2}
\end{figure}

Finally, we arrive to $\Spmatrix_{\gamma \to q \bar{q}}$. Starting with $g \to q \bar{q}$, we have to replace the incoming leg with a photon, which forces us to eliminate self-energy correction (diagrams A and D in Fig. \ref{fig:NLODiagrams2}) and other term which includes a triple-gluon vertex. So, up to ${\cal O}\left(\alpha_s^2\right)$, we only have the diagrams shown in Fig. \ref{fig:NLOPhotonDiagrams2}. The LO contribution reads
\beqn
\Spmatrix^{(0)}_{\gamma \to q 	\bar{q}} &=&  \frac{g_e E_Q \mu^{\epsilon}}{s_{12}} \IDC \, \bar{u}(p_1)\slashed{\epsilon}(\tilde{P})v(p_2) \ ,
\eeqn
while standard NLO correction is
\beqn
\nonumber \Spmatrix^{(1,{\rm STD})}_{\gamma \to q \bar{q}} &=& \CG g_s^2 {\left(\frac{-s_{12}-\imath 0}{\mu^2}\right)}^{-\epsilon} \, \IDC \frac{g_e E_Q \mu^{\epsilon}}{\epsilon^2 s_{12}} \,  \left[C^{({\rm STD},1)}_{\gamma \to q \bar{q}} \vphantom{\frac{1}{nP}} \, \bar{u}(p_1)\slashed{\epsilon}(\tilde{P})v(p_2) \right.
\\ &+& \left. C^{({\rm STD},2)}_{\gamma \to q \bar{q}} \,\frac{1}{nP}\bar{u}(p_1)\slashed{n}v(p_2) p_1\cdot \epsilon(\tilde{P}) \right] \, ,
\label{SplittingGeneralgammaqqb}
\eeqn
with
\beqn
C^{({\rm STD},1)}_{\gamma \to q \bar{q}}&=& C_F \frac{\epsilon  (3-\epsilon  (2-\delta(2 \epsilon -1)))-2}{(\epsilon -1) (2 \epsilon -1)} \, ,
\\ C^{({\rm STD},2)}_{\gamma \to q \bar{q}}&=& 0 \, .
\eeqn
On the other hand, for the NLO scalar-gluon contribution we have
\beqn
\Spmatrix^{(1,{\rm SCA-nHV})}_{\gamma \to q \bar{q}} &=& \CG g_s^2 \left(\frac{-s_{12}-\imath 0}{\mu^2}\right)^{-\epsilon} \, \frac{C_F}{\epsilon -1}  \Spmatrix^{(0)}_{\gamma \to q \bar{q}} \, ,
\eeqn
and, again, we find that the relation
\beqn
\Spmatrix^{(1,{\rm STD},HV)}_{\gamma \to q \bar{q}} = \Spmatrix^{(1,{\rm STD},FDH)}_{\gamma \to q \bar{q}} + \Spmatrix^{(1,{\rm SCA-nHV})}_{\gamma \to q \bar{q}} \, ,
\eeqn
is fulfilled.

Testing the divergent structure of $\Spmatrix^{(1)}_{\gamma \to q \bar{q}}$ we find that
\beqn
\Spmatrix^{(1)}_{\gamma\to q \bar{q}} &=& \Spmatrix^{(1)}_{H,\gamma\to q \bar{q}} + \IC^{\gamma}_{C,\gamma\to q \bar{q}} \left(p_1,p_2;\tilde{P} \right) \Spmatrix^{(0)}_{\gamma \to q \bar{q}} \, ,
\label{FormulaCatani2gammaqqb}
\eeqn
with
\beqn
\IC^{\gamma}_{C,\gamma\to q \bar{q}} &=& \CG g_s^2 \left(\frac{-s_{12}-\imath 0}{\mu^{2}}\right)^{-\epsilon} C_F \left(-\frac{2}{\epsilon^2}-\frac{3}{\epsilon}\right) \, ,
\\ \Spmatrix^{(1)}_{H,\gamma \to q \bar{q}} &=&  \CG g_s^2 \left(\frac{-s_{12}-\imath 0}{\mu^{2}}\right)^{-\epsilon} C_F \frac{2 (3+\delta ) \epsilon -7-\delta }{(\epsilon-1) (2 \epsilon-1)}\, \Spmatrix^{(0)}_{\gamma \to q \bar{q}}\, ,
\eeqn
as expected according to Eq. \ref{FormulaCataniFotones}.

Finally, the corresponding contributions to the Altarelli-Parisi kernels are
\beqn
\left\langle \mu \right|\hat{P}^{(0)}_{\gamma \to q \bar{q}}(z_1,k_{\bot})\left|\nu\right\rangle &=& - g_e^2 E_Q^2 C_A (1-\beta \epsilon) \left((\eta^{\DDirac})^{\mu\nu}+\frac{4 (z_1-1) z_1}{k_{\bot}^2} k_{\bot}^{\mu} k_{\bot}^{\nu}\right) \, , \ \hphantom{a}
\\ P^{(0)}_{\gamma \to q \bar{q}} &=& \frac{g_e^2 E_Q^2 C_A (1-2(1-z_1)z_1-\alpha  \delta \epsilon )(1-\beta \epsilon)}{1-\alpha\epsilon} \, ,
\eeqn
for the LO terms and
\beqn
P^{(1)}_{\gamma \to q \bar{q}} &=& \CG g_s^2 \left(\frac{-s_{12}-\imath 0}{\mu^2}\right)^{-\epsilon} \frac{C_F}{\epsilon^2} \frac{\epsilon  \left(2 \delta  \epsilon ^2-(\delta +2) \epsilon +3\right)-2}{(\epsilon -1) (2 \epsilon -1)} P^{(0)}_{\gamma \to q \bar{q}} \, + {\rm c.c.} \, ,
\eeqn
for the unpolarized NLO correction. The NLO polarized kernel can be expressed as
\beqn
\left\langle \mu \right|\hat{P}^{(1)}_{\gamma \to q \bar{q}}(z_1,k_{\bot})\left|\nu\right\rangle &=& \frac{P^{(1)}_{\gamma \to q \bar{q}}}{P^{(0)}_{\gamma \to q \bar{q}}} \, \left\langle \mu \right|\hat{P}^{(0)}_{\gamma \to q \bar{q}}(z_1,k_{\bot})\left|\nu\right\rangle \,
\eeqn
because $\Spmatrix^{(1)}_{\gamma \to q \bar{q}}$ is proportional to $\Spmatrix^{(1)}_{\gamma \to q \bar{q}}$.

%%%%%%%%%%%%%%%%%%%%%%%%%%%%%%%%%%%%%%%%%%%%%%%%%%%%%%%%%%%%%%%%%%%%%%%  
\section{Conclusions}
%%%%%%%%%%%%%%%%%%%%%%%%%%%%%%%%%%%%%%%%%%%%%%%%%%%%%%%%%%%%%%%%%%%%%%%  
In this work we have studied the double collinear limit and we computed the associated splitting matrices at NLO in $\alpha_s$ for both pure QCD and QCD plus photon-quark interactions. As a first consistency check, we have verified that the divergent structure of splitting matrices agrees with the general form shown in the literature (for example, in Refs. \cite{Catani:1998bh, Catani:2011st, Catani:2012iw}). Moreover, we found that the scheme dependence can be predicted up to ${\cal O}\left(\epsilon^0\right)$ using Eq. 11 in Ref. \cite{Catani:2003vu}. Also, we compared our results for usual DREG schemes with those available in Ref. \cite{Kosower:1999rx}, and, again, we found an agreement.

Besides the comparison of explicit results, we shown that FDH and HV schemes can be related at the amplitude level by introducing scalar-gluons. In fact, we verified that the relation
\beqn
\Spmatrix^{(1,{\rm STD},HV)} &=& \Spmatrix^{(1,{\rm STD},FDH)}+ \Spmatrix^{(1,{\rm SCA-nHV})} \, ,
\label{RELACIONFINAL}
\eeqn
is always fulfilled. Moreover, if we only consider fixed helicity configurations allowed by standard $4$-dimensional QCD interactions, then we can extend the validity of Eq. \ref{RELACIONFINAL} to include the CDR scheme. This is an important fact because it allows us to perform the same computation following two different paths, each of them having advantages in certain situations. For example, if we want to compute fixed-polarization splitting amplitudes (or matrices) in the CDR/HV scheme, it is more suitable to work with the r.h.s. of Eq. \ref{RELACIONFINAL}, because we settle $D_{\rm Dirac}=4$ and many useful identities can be used. In particular, we can use Fierz identities to contract spinor chains and reduce them to bispinor products. The improvement in the treatment of results can be much better when more particles are involved (for instance, when studying the multiple-collinear limit).

On the other hand, if we want to compute Altarelli-Parisi kernel corrections, it is better to use the l.h.s. of Eq. \ref{RELACIONFINAL} and work with $D_{\rm Dirac}=4-2\epsilon$. The reason is that when we close spinor chains and sum over polarizations, we get rid of spinors and obtain traces which involves Dirac's matrices. Since the relations that we use to solve Dirac's traces are valid with any value of $D_{\rm Dirac}$, then we can simplify them and the final result only contains scalar products. Also, we do not have to compute each helicity configuration separately, which makes the computation straightforward. This can be considered a great advantage, even if this procedure involves dealing with tensor type integrals which can have up to three free Lorentz indices. (In Appendix A we collect all the integrals required for the double collinear limit).

In the context of AP kernels, we also showed that it is possible to relate CDR and HV computations by just taking into account external scalar gluons. In fact, for the $q \to g q$ process, we find
\beqn
P^{CDR}_{q \to g q} &=& P^{HV}_{q \to g q} + P_{q \to \phi q} \, ,
\label{RELACIONFINAL2}
\eeqn
which is a complement to Eq. \ref{RELACIONFINAL} at the squared-amplitude level. Of course this relation can be extended to more general processes: we just have to decompose external gluons into $4$-dimensional vectors plus scalar particles and compute each contribution separately.

Finally, let's make some comments about the alternative schemes studied in this article. In Section 2 we introduced some parameters that allowed us to control Dirac's algebra dimension ($\delta$), the number of gluon polarizations ($\alpha_R$ and $\alpha$ for internal and external particles, respectively) and the number of fermion polarizations ($\beta_R$ for internal fermions and $\beta$ for external ones). By examining the behavior of $\Spmatrix^{(1)}_{q \to g q}$ with different parameter's values and comparing the divergent structure predicted by Eqs. \ref{FormulaCatani1} and \ref{FormulaCatani2}, we conclude that the hybrid-schemes (i.e. $\alpha_R=0$) are not consistent unless we include the corresponding scalar-gluon contributions. But, after adding these contributions, we get the same results provided by HV and CDR schemes. In other words, we show that the consistent version of HSA and HSB schemes are CDR and HV, respectively.

As anticipated in Section 2, FDH and TSC schemes are compatible with the supersymmetric Ward identity, even at one-loop level. In Ref. \cite{Catani:1996pk}, it was shown that tree-level Altarelli-Parisi kernels computed in FDH and TSC schemes fulfilled this identity, i.e.   
\beqn
P_{g\to g g}(z) +P_{g\to q \bar{q}}(z) &=& P_{q\to q g}(1-z)+P_{q\to q g}(z) \, ,
\label{SUPERWARD}
\eeqn
given that we set $C_A=C_F=T_R=N_f$. In this situation, if we identify quarks and gluinos then QCD is similar to ${\cal N}=1$ super Yang-Mills theory. From a physical point of view, this is possible because we consider the same number of bosonic and fermionic degrees of freedom. However, from Eqs. \ref{KernelNLOqgq}, \ref{KernelNLOgqqb} and \ref{KernelNLOggg} we can explicitly show that Eq. \ref{SUPERWARD} is verified at one-loop level, for both FDH and TSC schemes. This result makes TSC an interesting choice, since it has a very symmetric and \textit{democratic} way of treating all the particles involved in the computation.

It is interesting to appreciate that we performed the computations following a path that allowed us to keep track of Lorentz indices and metric tensors. In other words, we replaced integrals in $\Spmatrix^{(1)}$ before contracting with $\Spmatrix^{(0)}$ and summing over polarizations. This involved dealing with tensor-type integrals, which makes the calculation more complicated. If we were only interested in obtaining NLO corrections to AP-kernels, we could have first performed the contraction, and then replace the corresponding scalar integrals. However, scalar $\qeps$-integrals could appear in all schemes, with the exception of CDR. In spite of that, this approach is better suited when considering multiple-collinear splitting amplitudes, because tensor-type integrals become very lengthy and complicated when increasing the number of external physical momenta.

The natural next-step of this work is to extend the analysis to cover the multiple-collinear limit, and the possibility of computing them using recursion-relations \cite{Catani:2012ri}, even at loop-level.

%%%%%%%%%%%%%%%%%%%%%%%%%%%%%%%%%%%%%%%%%%%%%%%%%%%%%%%%%%%%%%%%%%%%%%%  
  \subsection*{Acknowledgments}
We are very grateful to Stefano Catani, for taking a carefull look to this paper and making extremely valuable comments about it. This work is partially supported by UBACYT, CONICET, ANPCyT, the
Research Executive Agency (REA) of the European Union under
the Grant Agreement number PITN-GA-2010-264564 (LHCPhenoNet),
by the Spanish Government and EU ERDF funds
(grants FPA2007-60323, FPA2011-23778 and CSD2007-00042
Consolider Project CPAN) and by GV (PROMETEUII/2013/007).
%%%%%%%%%%%%%%%%%%%%%%%%%%%%%%%%%%%%%%%%%%%%%%%%%%%%%%%%%%%%%%%%%%%%%%%  

%%%%%%%%%%%%%%%%%%%%%%%%%%%%%%%%%%%%%%%%%%%%%%%%%%%%%%%%%%%%%%%%%%%%%%%
\section*{Appendix A: Loop integrals in the light-cone gauge}
Here we show the list of Feynman integrals used to perform the computations of standard double-collinear splitting functions.
First of all, following Ref. \cite{Kosower:1999rx}, we introduce the auxiliary functions
\beqn
\nonumber f_1 (z)&=&\frac{2 \CG}{\epsilon ^2}\left(-\Gamma(1-\epsilon)\Gamma(1+\epsilon)z^{-1-\epsilon }(1-z)^{\epsilon }-\frac{1}{z}+\frac{(1-z)^{\epsilon }}{z} \, _2 F_1(\epsilon ,\epsilon;1+\epsilon;z)\right)
\\ &=&  - \frac{2 \CG}{\epsilon ^2 z} \, _2 F_1 \left(1 ,-\epsilon;1-\epsilon;\frac{z-1}{z}\right) \, ,
\\ f_2&=&-\frac{\CG}{\epsilon ^2} \, ,
\eeqn
where $z \in \left[0,1\right]$, since it is a partonic momentum fraction. On the other hand, due to the fact that double-collinear limit only involves $1\to 2$ processes, we will have bubble and triangle integrals.

Let's start with scalar integrals. We have three different types of bubbles
\beqn
I_1 &=& \int_q \frac{1}{q^2(q-p_{12})^2} =  \frac{f_2 \epsilon  (-s_{12}-\imath 0)^{-\epsilon }}{2 \epsilon-1 } \, ,
\\ I_2 &=& \int_q \frac{1}{q^2(q-p_{12})^2 nq} =  \frac{f_2 (-s_{12}-\imath 0)^{-\epsilon }}{nP} \, ,
\\ I_3 &=& \int_q \frac{1}{q^2(q-p_{12})^2 n\cdot(q-p_1)} =  \frac{\CG (-s_{12}-\imath 0)^{-\epsilon } }{nP z_1 (1-2 \epsilon ) \epsilon } \, _2F_1\left(1,1-\epsilon ;2-2 \epsilon ;\frac{1}{z_1}\right) \, ,
\eeqn
and three triangle integrals
\beqn
I_4 &=& \int_q \frac{1}{q^2(q-p_1)^2(q-p_{12})^2} = -\frac{f_2}{s_{12}} (-s_{12}-\imath 0)^{-\epsilon} \, ,
\\ I_5 &=& \int_q \frac{1}{q^2(q-p_1)^2(q-p_{12})^2 nq} =  \frac{f_1(z_1) (-s_{12}-\imath 0)^{-\epsilon }}{s_{12} nP} \, ,
\\ I_6 &=& \int_q \frac{1}{q^2(q+p_1)^2(q-p_2)^2 nq} =  \frac{(z_1 (2-4 \epsilon )+2 \epsilon -1) \, I_1 + \epsilon nP \, I_3}{nP s_{12} (z_1-1) z_1 (2 \epsilon +1)} \, ,
\eeqn
where $p_i$ are the four-momenta associated with the outgoing massless particles $i$, $p_{12}=p_1+p_2$ is the incoming particle momentum, which satisfies $p^2_{12}=s_{12}$, and $nP=n \cdot p_{12}=n\cdot \tilde{P}$. It is important to note that more scalar integrals are required for the computations performed in this work, but we can recover them from these results by just changing variables or relabeling momenta. Moreover, when using conventional schemes (FDH, HV and CDR), contributions proportional to $I_3$ and $I_6$ vanish.

Since in intermediate steps we left many Lorentz indices uncontracted, we also required tensor-type integrals with up to three free indices. To get them, we used Passarino-Veltman decomposition and the Mathematica package {\sc FIRE} \cite{Smirnov:2008iw, Smirnov:2013dia} to reduce scalar integrals. The required bubble integrals were
\beqn
I_7(\mu) &=& \int_q \frac{q^{\mu}}{q^2(q-p_{12})^2} = -\frac{f_2 (-s_{12}-\imath 0)^{-\epsilon }}{2(1-2 \epsilon) } p_{12}^{\mu} \, ,
\\ I_8(\mu,\nu) &=& \int_q \frac{q^{\mu}q^{\nu}}{q^2(q-p_{12})^2} = \frac{2-\epsilon}{2(3-2\epsilon)} I_1 \left(p_{12}^{\mu} p_{12}^{\nu}-\frac{s_{12}}{4-2\epsilon}\eta^{\mu\nu}\right) \, ,
\\ I_9(\mu) &=& \int_q \frac{q^{\mu}}{q^2(q-p_{12})^2 nq} = \frac{\epsilon f_2 (-s_{12}-\imath 0)^{-\epsilon } }{nP (2 \epsilon -1)} \, \left(p_{12}^{\mu}-\frac{s_{12}}{2 nP \epsilon} n^{\mu}\right) \, ,
\\ \nonumber I_{10}(\mu,\nu) &=& \int_q \frac{q^{\mu}q^{\nu}}{q^2(q-p_{12})^2 nq} = \frac{\epsilon f_2 (-s_{12}-\imath 0)^{-\epsilon } }{4 nP (\epsilon -1) (2 \epsilon -1)} \left(s_{12}  \eta^{\mu\nu}+p_{12}^{\mu} \left(2(\epsilon -1) p_{12}^{\nu}-\frac{s_{12}}{nP} n^{\nu}\right) \right.
\\ &+& \left. \frac{s_{12}}{nP} n^{\mu} \left(\frac{s_{12}}{\epsilon nP} n^{\nu}-p_{12}^{\nu}\right)\right) \, ,
\\ \nonumber I_{11}(\mu) &=& \int_q \frac{q^{\mu}}{(q+p_1)^2(q-p_2)^2 nq} = \frac{1}{2 nP^2} \left[s_{12} n^{\mu} \left(I_3 nP (1-2z_1)-2 I_1\right) \right.
\\ &+& \left. 2 nP p_{12}^{\mu} \left(I_1+I_3 nP (1-z_1)\right)+2 I_3 nP^2 p_2^{\mu}\right] \, ,
\eeqn
\beqn
\nonumber I_{12}(\mu,\nu) &=& \int_q \frac{q^{\mu}q^{\nu}}{(q+p_1)^2(q-p_2)^2 nq} = s_{12} \eta^{\mu\nu} \frac{  (1-2 z_1) I_1+2 I_3 nP (1-z_1) z_1}{4 nP(\epsilon -1)}
\\ \nonumber &+&\frac{p_2^{\mu} p_2^{\nu}}{2 nP} \left((2 z_1+1) I_1+2 I_3 nP z_1^2\right)+\frac{p_1^{\mu} p_1^{\nu}}{2nP} \left((2 z_1-3) I_1+2 I_3 nP (1-z_1)^2\right)
\\ \nonumber&+&\frac{s_{12}^2 n^{\mu} n^{\nu}}{4 nP^3 (\epsilon-1)}  \left((2 z_1-1) (2 \epsilon -3) I_1+I_3 nP (2 (z_1-1) z_1 (2 \epsilon -3)+\epsilon -1)\right) 
\\ \nonumber&+&\frac{s_{12}\left(p_2^{\nu} n^{\mu}+p_2^{\mu} n^{\nu}\right) \left((z_1 (6-4 \epsilon )-1) I_1+2 I_3 nP z_1 (z_1 (3-2 \epsilon )+\epsilon -2)\right)}{4 nP(\epsilon -1)}
\\ \nonumber &+&\frac{s_{12} \left(p_1^{\nu} n^{\mu}+p_1^{\mu} n^{\nu}\right) \left((z_1 (6-4 \epsilon )+4 \epsilon -5) I_1-2 I_3 nP (z_1-1) (z_1 (2 \epsilon -3)-\epsilon +1)\right)}{4 nP(\epsilon -1)}
\\ &+&\frac{p_2^{\mu} p_1^{\nu}+p_1^{\mu} p_2^{\nu}}{2 nP} \left((2 z_1-1) I_1+2 I_3 nP (z_1-1) z_1\right) \, .
\eeqn
Tensor-type triangle integrals used in this this work were
\beqn
I_{13}(\mu) &=& \int_q \frac{q^{\mu}}{q^2(q-p_1)^2(q-p_{12})^2} = -\frac{f_2 (-s_{12}-\imath 0)^{-\epsilon} \left((\epsilon -1) p_1^{\mu}+\epsilon  p_2^{\mu}\right)}{(2 \epsilon -1)s_{12}} \, ,
\\ \nonumber I_{14}(\mu,\nu) &=& \int_q \frac{q^{\mu}q^{\nu}}{q^2(q-p_1)^2(q-p_{12})^2} = \frac{f_2 (-s_{12}-\imath 0)^{-\epsilon}}{4(1-\epsilon)(2\epsilon-1)} \left(\frac{2 \epsilon}{s_{12}}  p_2^{\mu} \left((\epsilon -2) p_1^{\nu}+(\epsilon -1) p_2^{\nu}\right) \right.
\\ &+& \left. \frac{2 (\epsilon -2)}{s_{12}} p_1^{\mu} \left((\epsilon -1) p_1^{\nu}+\epsilon p_2^{\nu}\right)+\epsilon \eta^{\mu\nu}\right) \, ,
\eeqn
\beqn
\nonumber I_{15}(\mu,\nu,\rho) &=& \int_q \frac{q^{\mu}q^{\nu}q^{\rho}}{q^2(q-p_1)^2(q-p_{12})^2} = \frac{f_2 (-s_{12}-\imath 0)^{-\epsilon}}{4 (1-\epsilon) (2 \epsilon -3) (2 \epsilon -1)}
\\ \nonumber &\times& \left[\epsilon  \left(p_1^{{\rho}} (\epsilon-2) \left(\eta^{{\mu}{\nu}}+2 \frac{(\epsilon-1)(\epsilon-3)}{s_{12} (\epsilon -2) } p_2^{{\mu}} p_2^{{\nu}}\right) \right.\right.
\\ \nonumber &+& \left. \left. p_1^{{\nu}}(\epsilon-2) \left(\eta^{{\mu}{\rho}}+2 \frac{(\epsilon-3)}{s_{12}  } p_2^{{\mu}} p_1^{{\rho}}+2 \frac{(\epsilon-1)(\epsilon-3)}{s_{12} (\epsilon -2) } p_2^{{\mu}} p_2^{{\rho}}\right) \right. \right.
\\ \nonumber &+& \left.\left. (\epsilon -1) \left(\left(p_2^{{\rho}} \eta^{{\mu}{\nu}}+p_2^{{\nu}} \eta^{{\mu}{\rho}}\right)+p_2^{{\mu}} \left(\eta^{{\nu}{\rho}}+2 \frac{(\epsilon -2)}{s_{12}} p_2^{{\nu}} p_2^{{\rho}}\right)\right)\right) \right.
\\ \nonumber &+& \left. p_1^{{\mu}} (\epsilon-2)\left(\epsilon  \left( \eta^{{\nu}{\rho}}+2 \frac{(\epsilon-3)}{s_{12}} p_2^{{\nu}} p_1^{{\rho}}+2 \frac{(\epsilon-1)(\epsilon-3)}{s_{12} (\epsilon -2)} p_2^{{\nu}} p_2^{{\rho}}\right) \right. \right.
\\ &+& \left.\left. 2 \frac{(\epsilon-3)}{s_{12}} p_1^{{\nu}} \left((\epsilon -1) p_1^{{\rho}}+\epsilon  p_2^{{\rho}}\right)\right)\right] \, ,
\eeqn
\beqn
\nonumber I_{16}(\mu) &=& \int_q \frac{q^{\mu}}{q^2(q-p_1)^2(q-p_{12})^2 nq} = \frac{(-s_{12}-\imath 0)^{-\epsilon} }{2 nP s_{12} (1-z_1)} \left(p_2^{\mu} \frac{z_1 f_1(z_1)-2f_2}{1-z_1}  \right.
\\ &-& \left. f_1(z_1) p_1^{\mu}  -\frac{s_{12}(f_1(z_1)-2 f_2)}{2 nP(1-z_1)} n^{\mu}\right) \, ,
\\ \nonumber I_{17}(\mu,\nu) &=& \int_q \frac{q^{\mu}q^{\nu}}{q^2(q-p_1)^2(q-p_{12})^2 nq} = -\frac{(-s_{12}-\imath 0)^{-\epsilon}}{s_{12} nP} \left[\vphantom{\frac{a^2}{b^2}}f_{5,aa}(z_1)p_1^{\mu}p_1^{\nu} \right.
\\ \nonumber &+& \left. f_{5,ab}(z_1)\frac{p_1^{\mu}p_2^{\nu}+p_2^{\mu}p_1^{\nu}}{2}+f_{5,bb}(z_1)p_2^{\mu}p_2^{\nu} +\left(\frac{s_{12}}{2 nP}\right)^{2}f_{5,qq}(z_1)n^{\mu}n^{\nu} \right.
\\ &+& \left. s_{12} f_{5,aq}(z_1)\frac{p_1^{\mu}n^{\nu}+n^{\mu}p_1^{\nu}}{2 nP} + s_{12} f_{5,bq}(z_1)\frac{p_2^{\mu}n^{\nu}+n^{\mu}p_2^{\nu}}{2 nP}+s_{12} f_{5,g}\eta^{\mu\nu}\right]\, ,
\eeqn
\beqn
\nonumber I_{18}(\mu) &=& \int_q \frac{q^{\mu}}{q^2(q+p_1)^2(q-p_2)^2 nq} =  \frac{(1-2 \epsilon)I_1}{2 nP s_{12} (1-z_1) z_1 \epsilon } \left((1-z_1) p_1^{\mu}+z_1 p_2^{\mu}-\frac{s_{12}}{2\, nP} n^{\mu}\right)
\\ &+& \frac{I_3}{2 s_{12} (1-z_1) z_1} \left((1-z_1) p_1^{\mu}-z_1 p_2^{\mu}+\frac{s_{12} (1-2 z_1)}{2\, nP } n^{\mu}\right)\, ,
\\ \nonumber I_{19}(\mu,\nu) &=& \int_q \frac{q^{\mu}q^{\nu}}{q^2(q+p_1)^2(q-p_2)^2 nq} = \frac{1}{4 nP \, s_{12} z_1 (1-z1)} \, \left[\vphantom{\frac{a}{a}} f_{19,aa}(z_1,s_{12},nP)  p_1^{\mu}p_1^{\nu} \right.
\\ \nonumber &+& \left. f_{19,ab}(z_1,s_{12},nP)  \frac{p_1^{\mu}p_2^{\nu}+p_2^{\mu}p_1^{\nu}}{2} + f_{19,bb}(z_1,s_{12},nP) p_2^{\mu}p_2^{\nu} \right.
\\ \nonumber &+& \left. f_{19,qq}(z_1,s_{12},nP) n^{\mu}n^{\nu}+ f_{19,aq}(z_1,s_{12},nP) \frac{p_1^{\mu}n^{\nu}+n^{\mu}p_1^{\nu}}{2}\right.
\\ &+& \left. f_{19,bq}(z_1,s_{12},nP) \frac{p_2^{\mu}n^{\nu}+n^{\mu}p_2^{\nu}}{2} + f_{19,g}(z_1,s_{12},nP) \eta^{\mu\nu}\right]  \, ,
\eeqn
where coefficients $\left\{f_{5,ij},f_{5,g}\right\}$ are given in Ref. \cite{Kosower:1999rx} and
\beqn
\nonumber f_{19,aa}(z_1,s_{12},nP) &=& \frac{1-{z_1}}{{z_1} (2 \epsilon -1)} \left[I_1 nP (2 {z_1} \epsilon +\epsilon -1)+{\bar{I}_{13}} s_{12} (\epsilon -1)+nP {z_1} (2 I_3 nP ({z_1}-1) \epsilon \right.
\\ &+& \left. I_4 s_{12} (\epsilon -1))\right]\, ,
\\ f_{19,ab}(z_1,s_{12},nP) &=& \frac{2 \epsilon  (nP (-2 I_1 {z_1}+I_1+{z_1} (I_4 s_{12}-2 I_3 nP ({z_1}-1)))+{\bar{I}_{13}} s_{12})}{2 \epsilon -1} \, ,
\eeqn
\beqn  
\nonumber f_{19,bb}(z_1,s_{12},nP) &=& \frac{{z_1} }{(1-{z_1}) (2 \epsilon -1)} \left[I_1 nP ((2 {z_1}-3) \epsilon +1)+{\bar{I}_{13}} s_{12} (\epsilon -1) \right.
\\ &+& \left. nP {z_1} (2 I_3 nP ({z_1}-1) \epsilon +I_4 s_{12} (\epsilon -1))\right]\, ,
\\ \nonumber f_{19,qq}(z_1,s_{12},nP) &=& \frac{s_{12}^2}{4 nP^2 (1-{z_1}) {z_1} (2 \epsilon -1)} \left[I_1 nP (2 {z_1}-1) (2 ({z_1}-1) {z_1} (2 \epsilon -1)-\epsilon +1) \right.
\\ \nonumber &+& \left. s_{12} (\epsilon -1) ({\bar{I}_{13}}+I_4 nP {z_1})+2 I_3 nP^2 ({z_1}-1) {z_1} \left(2 ({z_1}-1) \right.\right.
\\ &\times& \left.\left. {z_1} (2 \epsilon -1)+\epsilon \right)\right] \, ,
\\ \nonumber f_{19,aq}(z_1,s_{12},nP) &=& -\frac{s_{12} }{nP {z_1} (2 \epsilon -1)} \left[I_1 nP (2 {z_1} (-2 {z_1} \epsilon +{z_1}+\epsilon )+\epsilon -1) \right.
\\ &+& \left. s_{12} (\epsilon -1) ({\bar{I}_{13}}+I_4 nP {z_1}) + 2 I_3 nP^2 ({z_1}-1) {z_1} (-2 {z_1} \epsilon +{z_1}+\epsilon )\right] \, ,
\\ \nonumber f_{19,bq}(z_1,s_{12},nP) &=& \frac{s_{12}}{nP ({z_1}-1) (2 \epsilon -1)} \left[I_1 nP (2 {z_1} ({z_1} (2 \epsilon -1)-3 \epsilon +2)+\epsilon -1) \right.
\\ \nonumber &+& \left. {\bar{I}_{13}} s_{12} (\epsilon -1)+nP {z_1} (2 I_3 nP ({z_1}-1) (2 {z_1} \epsilon -{z_1}-\epsilon +1) \right.
\\ &+& \left. I_4 s_{12} (\epsilon -1)) \right] \, ,
\\ f_{19,g}(z_1,s_{12},nP) &=& \frac{s_{12} (nP (-2 I_1 {z_1}+I_1+{z_1} (I_4 s_{12}-2 I_3 nP ({z_1}-1)))+{\bar{I}_{13}} s_{12})}{2(2 \epsilon -1)}   \, ,
\eeqn
with $\bar{I}_{13}=I_{13}(\alpha)n^{\alpha}$.

Finally, let's make a brief comment about $q_{\epsilon}^2$-integrals. They appear if we introduce $4$-dimensional metric tensors when performing the computation with $\DDirac=4-2\epsilon$. This situation is only possible in the context of HSA/HSB schemes, which were defined in Section 2. To compute $q_{\epsilon}^2$-integrals we require tensor-type Feynman integrals with rank greater than 2, and then we have to contract them with a transverse-dimensional metric tensor $\eta^{\epsilon}$. The scalar integrals required in our computations are
\beqn
I^{\epsilon}_1 &=& \int_q \frac{q_{\epsilon}^2}{q^2 (q-p_1)^2(q-p_{12})^2} = \frac{(4-\DDirac)f_2 \epsilon}{4(\epsilon-1)(2\epsilon-1)}(-s_{12}-\imath 0)^{-\epsilon} \, ,
\\ I^{\epsilon}_2 &=& \int_q \frac{q_{\epsilon}^2}{q^2 (q-p_1)^2(q-p_{12})^2 nq} = \frac{(4-\DDirac)f_{5,g}(z_1)}{nP}(-s_{12}-\imath 0)^{-\epsilon} \, ,
\\ \nonumber I^{\epsilon}_3 &=& \int_q \frac{q_{\epsilon}^2}{q^2 (q+p_1)^2(q-p_2)^2 nq} = \frac{(\DDirac-4)}{4 nP (1-z_1) z_1 (1-2 \epsilon )^2} \left[I_1 (2 z_1-1) (2 \epsilon -1) \right.
\\ &+& \left. I_2 \, nP (\epsilon -z_1)+z_1 (2 \epsilon -1) (2 I_3 \, nP (z_1-1)+I_4 s_{12})\right] \, ,
\eeqn
and, also, we used some vector-type $q_{\epsilon}^2$-integrals
\beqn
\nonumber I^{\epsilon}_4(\mu) &=& \int_q \frac{q^{\mu} q_{\epsilon}^2}{q^2 (q+p_1)^2(q-p_2)^2 n\cdot (q+p_1)}= \frac{\DDirac-4}{8(1-z_1)(2\epsilon-1)} \left[\vphantom{\frac{a}{b}} (2I_2-s_{12}z_1I_5) p_1^{\mu} \right.
\\ &+& \left. \left(\frac{2(1-2z_1)I_2+s_{12}z_1^2 I_5}{z_1-1}+\frac{2I_2}{1-\epsilon}\right) p_2^{\mu} + \left(\frac{z_1(2I_2-s_{12}I_5)}{z_1-1}+\frac{2I_2}{\epsilon-1}\right) \frac{s_{12}n^{\mu}}{2 nP} \right] \, ,
\eeqn
\beqn
\nonumber I^{\epsilon}_5(\mu) &=& \int_q \frac{q^{\mu} q_{\epsilon}^2}{q^2 (q+p_1)^2(q-p_2)^2 nq}= \frac{\DDirac-4}{8(2\epsilon-1)(\epsilon-1)} \left[\vphantom{\frac{a}{b}} 2\left((1-z_1)(2\epsilon-1)I_3-\epsilon I_2\right) p_1^{\mu} \right.
\\ &+& \left. 2\left(z_1(1-2\epsilon)I_3-\epsilon I_2\right) p_2^{\mu} + \left(2\epsilon I_2 +(2z_1-1)(2\epsilon-1)I_3\right) \frac{s_{12}n^{\mu}}{nP} \right] \, .
\eeqn

%%%%%%%%%%%%%%%%%%%%%%%%%%%%%%%%%%%%%%%%%%%%%%%%%%%%%%%%%%%%%%%%%%%%%%%
\section*{Appendix B: Parton self-energies}

\subsection*{Gluon self-energy}
\begin{figure}[htb]
	\centering
		\includegraphics{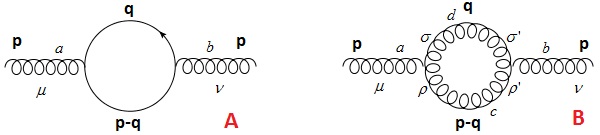}
	\caption{Diagrams contributing to the gluon self-energy $\Pi^{\mu \nu}$ at NLO. We explicitly indicate the conventions used for labeling momenta and color and Lorentz indices.}
	\label{fig:Figura1}
\end{figure}

When computing the gluon self-energy at one-loop level, we find that there are two Feynman diagrams which contribute to $\Pi^{\mu \nu}$. They are shown in Fig. \ref{fig:Figura1}. Using conventional Feynman rules, we define
\beqn
\Pi^{\mu \nu}(p) &=& \Pi^{\mu \nu}_{A}(p) + \Pi^{\mu \nu}_{B}(p) \, ,
\eeqn
with
\beqn
\Pi^{\mu \nu}_{A}(p) &=& \left(g^2_s \mu^{2\epsilon}N_f {\rm Tr}\left[\SUNT^a \SUNT^b \right] \right)  \, \int_q \, \frac{{\rm Tr}\left[\gamma^{\nu}\slashed{q}\gamma^{\mu}(\slashed{q}-\slashed{p}) \right]}{q^2 (q-p)^2} \, ,
\\ \nonumber \Pi^{\mu \nu}_{B}(p) &=& \frac{g^2_s \mu^{2\epsilon}f_{a d c} f_{d b c}}{2} \, \int_q \frac{d_{\sigma \sigma'}(q) d_{\rho \rho'}(p-q)}{q^2 (q-p)^2}
\\ &\times& V^{\rm Cin}_{3g}(-p,q,p-q;\mu,\sigma,\rho)V^{\rm Cin}_{3g}(-q,p,q-p;\sigma',\nu,\rho') \, ,
\eeqn 
where we are using $p$ as the external momenta which verifies $p^2=s$.

After integrating the loop-momentum we arrive to
\beqn
\Pi_A^{\mu \nu}(p) &=& \frac{2 f_2 g_s^2 (\epsilon -1) \epsilon  N_f (1-\beta_R\epsilon) \delta _{ab} \left(s {\eta}^{\mu\nu}-p^{\mu} p^{\nu}\right)}{4 (\epsilon -2) \epsilon +3} \left(\frac{-s-\imath 0}{\mu^2}\right)^{-\epsilon } \, ,
\\ \nonumber \Pi_B^{\mu \nu}(p) &=& \frac{f_2 g_s^2 C_A  \delta _{ab}}{2 {np}^2 (4 (\epsilon -2) \epsilon +3)} \, \left(\frac{-s-\imath 0}{\mu^2}\right)^{-\epsilon } \left({np}^2 s \left(-(D+38) \epsilon +16 \epsilon ^2+24\right) {\eta}^{\mu\nu}+np \, p^{\mu}  \right.
\\ &-& \left. \left((D-2) np \, \epsilon \,  p^{\nu} - 8 s (\epsilon -1) (2 \epsilon -3) n^{\nu}\right)+8 s (\epsilon -1) (2 \epsilon -3) n^{\mu} \left(s n^{\nu}-np \, p^{\nu}\right)\right) \, ,
\eeqn
where $f_2=-\CG/{\epsilon}^2$. Note that there some terms which are proportional to $D$. To understand the origin of these terms, we put a flag multiplying the metric tensor inside $d_{\mu \nu}$ and we follow it until we arrive to the final result. The conclusion is that they are always proportional to the contraction of two gluon propagators, so this $D$ is related to the number of polarizations of internal gluons. Thus, we replace $D \to 4- 2 \delta \epsilon$, with $\delta=0$ in FDH and $\delta=1$ in HV/CDR schemes.

With the aim of simplifying the result, we study separately each tensorial structure and reduce the associated coefficients. Our final result is
\beqn
\nonumber \Pi^{\mu \nu}(p) &=& f_2 g_s^2 \left(\frac{-s-\imath 0}{\mu^2}\right)^{-\epsilon } \delta _{ab} \left[ \frac{C_A (\epsilon  ((\delta +8) \epsilon -21)+12)+2 (\epsilon -1) \epsilon  N_f(1-\beta_R\epsilon)}{4 (\epsilon -2) \epsilon +3} \right.
\\ &\times& \left. \left(s \eta^{\mu\nu}-p^{\mu}p^{\nu}\right) + \frac{4 s (\epsilon -1) C_A}{2 \epsilon -1} \left(\frac{p^{\mu}p^{\nu}}{s}-\frac{n^{\mu}p^{\nu}+n^{\nu}p^{\mu}}{np}+\frac{s}{np^2}n^{\mu}n^{\nu}\right)\right] \, .
\eeqn
Before moving forward, let's define the following factor
\beqn
\Pi(p^2) &=&  -f_2 g_s^2 \left(\frac{-s-\imath 0}{\mu^2}\right)^{-\epsilon } \frac{ C_A (\epsilon  ((\delta +8) \epsilon -21)+12)+2 (\epsilon -1) \epsilon  N_f(1-\beta_R \epsilon)}{4 (\epsilon -2) \epsilon +3} \, ,
\eeqn
which is the same that we introduced in Eq. \ref{SplittinggqqbSELFENERGY}.

To conclude this section, let's mention some properties of $\Pi^{\mu \nu}$. First of all, it satisfies current conservation, that is
\beq
p_{\mu} \Pi^{\mu \nu}(p) = 0 = p_{\nu} \Pi^{\mu \nu}(p) \,.
\eeq
If we contract it with two gluon-propagators we get
\beqn
\nonumber \frac{\imath d_{\mu' \mu}(p)}{s} \left(-\imath \Pi^{\mu \nu}(p) \right) \frac{\imath d_{\nu \nu'}(p)}{s} &=& \imath \delta_{a b} \left[\Pi(p^2)  \left(- \eta_{{\nu'}{\mu'}}+\frac{p_{{\nu'}} {n}_{{\mu'}}+{n}_{{\nu'}} p_{{\mu'}}}{np}\right) \right.
\\ &+& \left. g_s^2 f_2 \left(\frac{-s-\imath 0}{\mu^2}\right)^{-\epsilon } \frac{4 (\epsilon -1) C_A {n}_{{\nu'}} {n}_{{\mu'}}}{np^2 (2 \epsilon -1)} \right] \, ,
\eeqn
and if we only consider the leading contribution in the limit $s\to 0$, we obtain
\beqn
\frac{\imath d_{\mu' \mu}(p)}{s} \left(-\imath \Pi^{\mu \nu}(p) \right) \frac{\imath d_{\nu \nu'}(p)}{s} &\approx& \Pi(p^2) \delta_{ab} \frac{\imath d_{\mu' \nu'}(p)}{s} \, ,
\eeqn
that is proportional to $d_{\mu' \nu'}(p)$.

On the other hand, if we contract $\Pi^{\mu \nu}(p)$ with a propagator and a polarization vector associated with a massless external leg with momentum
\beqn
\tilde{P}^{\mu} &=& p^{\mu} - \frac{s}{2 \, np}n^{\mu} \, ,
\eeqn
then we obtain
\beqn
\frac{\imath d_{\mu' \mu}(p)}{s}\left(-\imath \Pi^{\mu \nu}(p)\right)\, \epsilon_{\nu}(\tilde{P}) &=& \Pi(p^2) \, \epsilon_{\mu'}(\tilde{P}) \, ,
\eeqn
where we have used that $p^{\nu}\epsilon_{\nu}(\tilde{P})=0=n^{\nu}\epsilon_{\nu}(\tilde{P})$ to simplify the expressions. Again, we note that the result is a numerical factor times the polarization vector, which explains why self-energy corrections are proportional to $\Spmatrix^{(0)}$ (see Eqs. \ref{SplittinggqqbSELFENERGY} and \ref{SplittinggggSELFENERGY}). And, moreover, that numerical factor is the same that we found when we contracted $\Pi^{\mu \nu}(p)$ with two gluon propagators.

\subsection*{Quark self-energy}
\begin{figure}[htb]
	\centering
		\includegraphics{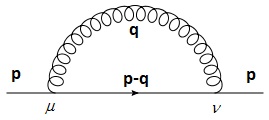}
	\caption{Diagram contributing to the quark self-energy $\Sigma$ at NLO. We explicitly indicate the conventions used for labeling momenta and Lorentz indices.}
	\label{fig:Figura2}
\end{figure}

In this case, there is only one Feynman diagram which contributes to $\Sigma$ and it is shown in Fig. \ref{fig:Figura2}. Using conventional Feynman rules, we define
\beqn
\Sigma_{ij}(p) &=& g^2_s \mu^{2\epsilon} (\SUNT^a \SUNT^a)_{ij} \, \int_q \, \frac{\gamma^{\nu}(\slashed{p}-\slashed{q})\gamma^{\mu}}{q^2 (q-p)^2} \, d_{\mu \nu}(q) \, ,
\eeqn 
where we use the definitions introduced in previous sections.

After integrating the loop-momentum we arrive to
\beqn
\Sigma(p) &=& -\frac{ f_2 g_s^2 \mu^{2\epsilon} C_F ((D-2) np \epsilon  \slashed{p} \, +4 s (\epsilon -1) \slashed{n})}{2 np (2 \epsilon -1)} \, \left(\frac{-s-\imath 0}{\mu^2}\right)^{-\epsilon } \, .
\eeqn
Note that there are some terms which are proportional to $D$. The situation is different from what was happening with the gluon self-energy. In the previous case we can have the contraction of two $d_{\mu \nu}$'s, which originates a terms proportional to the number of gluon polarizations. However, here we only have one gluon propagator. But, working with the Dirac chain we find
\beq
\gamma^{\nu}(\slashed{p}-\slashed{q})\gamma^{\mu} = -\gamma^{\nu}\gamma^{\mu}(\slashed{p}-\slashed{q}) + 2 (p-q)^{\mu} \gamma^{\nu} \, ,
\eeq
and taking into account that $d_{\mu \nu}=d_{\nu \mu}$, we can interchange $\mu-\nu$ indices, and we get
\beqn
\gamma^{\nu}\gamma^{\mu}(\slashed{p}-\slashed{q}) \, d_{\mu \nu}(q) &=& \gamma^{\mu}\gamma^{\nu}(\slashed{p}-\slashed{q}) \, d_{\mu \nu}(q)
\\ \nonumber &=& \frac{1}{2} \left\{\gamma^{\mu},\gamma^{\nu} \right\}(\slashed{p}-\slashed{q}) \, d_{\mu \nu}(q)
\\ \nonumber &=& -(\slashed{p}-\slashed{q}) \, (2-2\epsilon \delta) \, ,
\eeqn
where we used $\eta^{\mu \nu} d_{\mu \nu}(q)=-(2-2 \epsilon \delta)$. 

With the aim of simplifying the result, we study separately each spinorial structure and reduced the associated coefficients. Our final result is
\beqn
\Sigma_{ij}(p) &=& - f_2 g_s^2 \left(\frac{-s-\imath 0}{\mu^2}\right)^{-\epsilon } C_F \left(\frac{\epsilon  (\delta  \epsilon -1) }{1-2 \epsilon} \slashed{p}   -\frac{2 s (\epsilon -1)}{np(1-2\epsilon) } \, \slashed{n} \right) \, .
\eeqn
Now let's study some properties of $\Sigma(p)$. As a first step, if we contract it with two quark propagators, we get
\beqn
\frac{\imath \slashed{p}}{s}\left(-\imath \, \Sigma(p)\right)\frac{\imath \slashed{p}}{s} &=& \imath g_s^2 C_F f_2 \left(\frac{-s-\imath 0}{\mu^2}\right)^{-\epsilon } \left(\frac{(\epsilon  (\delta  \epsilon -5)+4)}{s (2 \epsilon -1)} \, \slashed{p}+\frac{2 (\epsilon -1)}{np (2 \epsilon -1)} \, \slashed{n} \right) \, ,
\eeqn
and if we take only the most divergent part in the limit $s\to 0$, we obtain
\beqn
\frac{\imath \slashed{p}}{s}\left(-\imath \, \Sigma(p)\right)\frac{\imath \slashed{p}}{s} &\approx& \left( g_s^2 f_2 C_F \left(\frac{-s-\imath 0}{\mu^2}\right)^{-\epsilon } \, \frac{\epsilon(5-\delta\epsilon)-4}{1-2 \epsilon} \right) \, \frac{\imath \slashed{p}}{s} \, ,
\eeqn
which is proportional to the quark propagator and motivates the following definition
\beqn
\Sigma(p^2) &=& g_s^2 f_2 C_F \left(\frac{-s-\imath 0}{\mu^2}\right)^{-\epsilon } \, \frac{\epsilon(5-\delta\epsilon)-4}{1-2 \epsilon} \, .
\eeqn
On the other hand, we can contract $\Sigma(p)$ with a quark propagator and a massless spinor $u(\tilde{P})$, and we get
\beqn
\frac{\imath \slashed{p}}{s}\left(-\imath \Sigma(p) \right)u(\tilde{P}) &=& \Sigma(p^2) \, u(\tilde{P}) \, ,
\eeqn
which turns out to be a numerical factor times $u(\tilde{P})$. Moreover, that factor is the same that we found when contracting $\Sigma$ with two propagators, in the limit $s \to 0$. And, again, this explains the result shown in Eq. \ref{SplittingqgqSELFENERGY}.

%%%%%%%%%%%%%%%%%%%%%%%%%%%%%%%%%%%%%%%%%%%%%%%%%%%%%%%%%%%%%%%%%%%%%%%
%%%%%%%%%%%%%%%%%%%%%%%%%%%%%%%%%%%%%%%%%%%%%%%%%%%%%%%%%%%%%%%%%%%%%%%
%\begin{center}
%\large{ \bf References}
%\end{center}

\end{document}